\documentclass[fleqn,usenatbib]{mnras}

\usepackage{newtxtext,newtxmath}

\usepackage[T1]{fontenc}

\DeclareRobustCommand{\VAN}[3]{#2}
\let\VANthebibliography\thebibliography
\def\thebibliography{\DeclareRobustCommand{\VAN}[3]{##3}\VANthebibliography}

\usepackage{graphicx}	
\usepackage{amsmath}	
\usepackage{array}
\usepackage{caption}
\usepackage{float}
\usepackage{subcaption}
\usepackage{placeins}
\usepackage{gensymb}
\usepackage{multirow}
\usepackage{multicol}
\usepackage{xcolor}
\usepackage{hyperref}

\def\h2{H$_2$}

\title[Fragmentation and gas dynamics: G12.42 and G19.88]{ATOMS: ALMA Three-millimeter Observations of Massive Star-forming regions - XII: Fragmentation and multi-scale gas kinematics in protoclusters G12.42+0.50 and G19.88-0.53}

\author[Anindya Saha et al.]
{Anindya Saha,\thanks{E-mail: anindya.s1130@gmail.com (A.S)}$^{\star 1}$
Anandmayee Tej,\thanks{E-mail: tej@iist.ac.in (A.T)}$^{1}$
Hong-Li Liu,\thanks{E-mail: hongliliu2012@gmail.com (H.L.L)}$^{2}$
Tie Liu,$^{3,4}$
Namitha Issac,$^{5}$
Chang Won Lee,$^{6,7}$
\newauthor
Guido Garay, $^{8}$
Paul F. Goldsmith,$^{9}$
Mika Juvela,$^{10}$
Sheng-Li Qin,$^{2}$
Amelia Stutz,$^{11,12}$
Shanghuo Li,$^{7}$
\newauthor
Ke Wang,$^{13,14}$
Tapas Baug,$^{15}$
Leonardo Bronfman,$^{8}$
Feng-Wei Xu,$^{13,14}$
Yong Zhang,$^{16}$
Chakali Eswaraiah,$^{17}$
\\
Affiliations are listed at the end of the paper}

\date{Accepted XXX. Received YYY; in original form ZZZ}

\pubyear{2022}

\begin{document}
\label{firstpage}
\pagerange{\pageref{firstpage}--\pageref{lastpage}}
\maketitle

\begin{abstract}
We present new continuum and molecular line data from the ALMA Three-millimeter Observations of Massive Star-forming regions (ATOMS) survey for the two protoclusters, G12.42+0.50 and G19.88-0.53. The 3~mm continuum maps reveal seven cores in each of the two globally contracting protoclusters. These cores satisfy the radius-mass relation and the surface mass density criteria for high-mass star formation. Similar to their natal clumps, the virial analysis of the cores suggests that they are undergoing gravitational collapse ($\rm \alpha_{vir} << 2$). The clump to core scale fragmentation is investigated and the derived core masses and separations are found to be consistent with thermal Jeans fragmentation.  We detect large-scale filamentary structures with velocity gradients and multiple outflows in both regions. {\it Dendrogram} analysis of the H$^{13}$CO$^{+}$ map identifies several branch and leaf structures with sizes $\sim$ 0.1 and 0.03~pc, respectively. The supersonic gas motion displayed by the branch structures is in agreement with the Larson power-law indicating that the gas kinematics at this spatial scale is driven by turbulence. The transition to transonic/subsonic gas motion is seen to occur at spatial scales of $\sim$0.1~pc indicating the dissipation of turbulence. In agreement with this, the leaf structures reveal gas motions that deviate from the slope of Larson's law. From the large-scale converging filaments to the collapsing cores, the gas dynamics in G12.42+0.50 and G19.88-0.53 show scale-dependent dominance of turbulence and gravity and the combination of these two driving mechanisms needs to be invoked to explain massive star formation in the protoclusters.
\end{abstract}



\section{Introduction} \label{sec:Introduction}
Massive stars ($M_{\star} \gtrsim 8\,\rm M_{\odot}$) dictate the dynamical and chemical evolution of the surrounding interstellar medium (ISM) and the galaxy through their mechanical, radiative, and chemical feedback. However, despite the tremendous theoretical, computational, and observational advances in the last decade \citep[][and references therein]{{2012ApJ...759....9K},{2014prpl.conf..149T},{2018ARA&A..56...41M}}, the formation mechanism of high-mass stars, in particular the processes involved in the initial stages, are still not clearly understood. 
The major issue lies in building a comprehensive multi-wavelength database of this elusive early phase of high-mass stars. Rarity, short evolutionary time scales, formation in clusters, large distances, and high extinction in embedded environment are the factors that pose observational challenges. In recent years, high-sensitivity and high-resolution observations have rendered several statistical and individual case studies possible with facilities like SMA \citep[e.g.][]{{2009ApJ...696..268Z},{2011ApJ...735...64W},{2011ApJ...733...26Z},{2014ApJ...792..116Z},{2015ApJ...805..171L},{2017ApJ...841...97S},{2019A&A...622A..54P},{2019ApJ...886..130L}}, ALMA \citep[e.g.][]{{2019ApJ...886..102S},{2019ApJ...886...36S},{2020MNRAS.496.2790L},{2021ApJ...909..199O},{2021MNRAS.503.4601B},{2021A&A...648A.100B},{HLLiu21},{HLLiu22a},{HLLiu22b}}, and CARMA \citep[e.g.][]{{2011A&A...530A.118P},{2013ApJ...773..123S}}. Additionally, several surveys like SEDIGISM \citep{{2017A&A...601A.124S},{2019A&A...622A.155Y},{2021arXiv211110850Y},{2022A&A...658A..54C}}, GLOSTAR \citep{2021A&A...651A..88N} are also aimed towards addressing various aspects of the formation mechanism and early evolutionary phases of high-mass stars. 

\begin{figure*}
    \centering
    \includegraphics[width=0.85\textwidth]{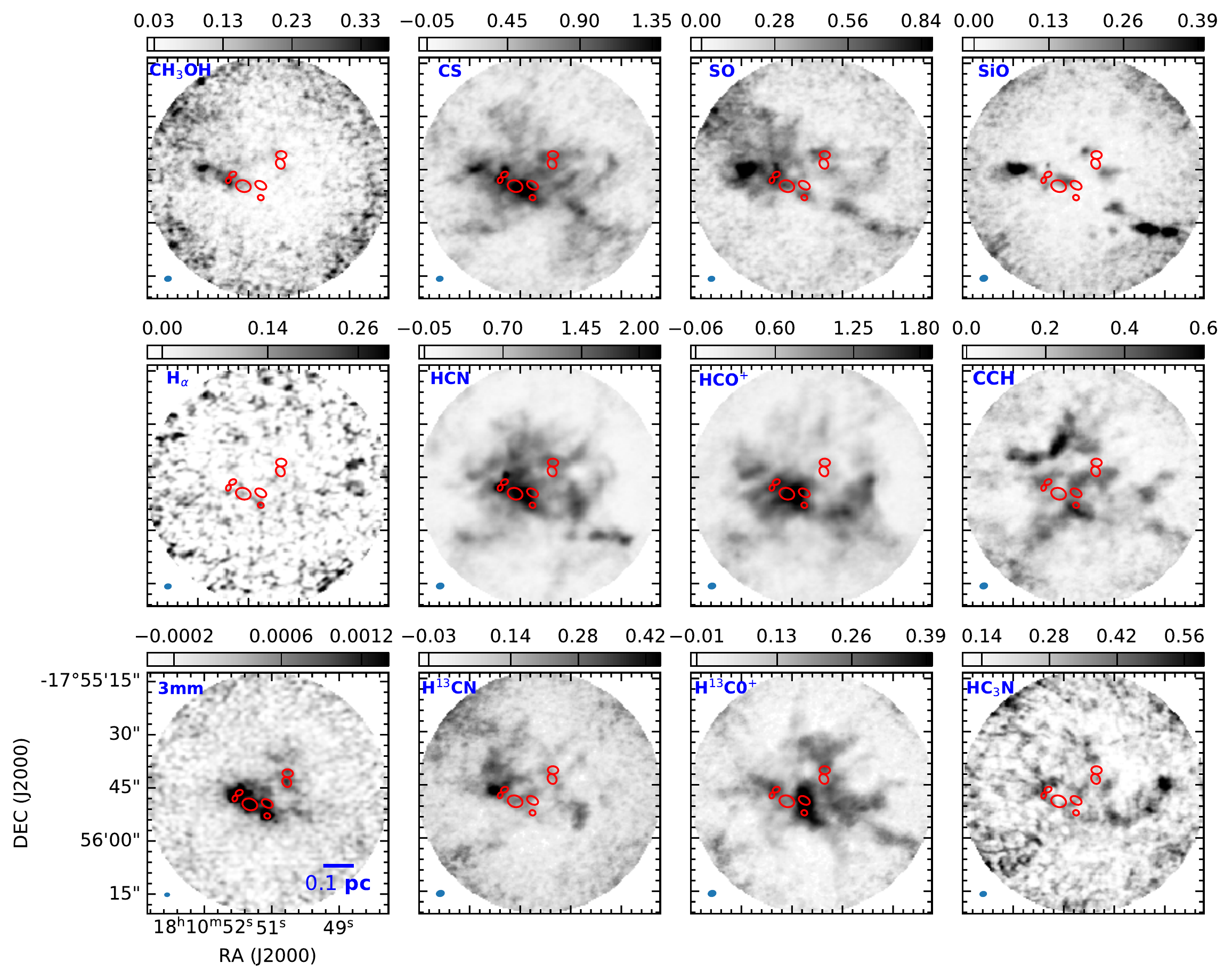}
    \caption{Continuum (3~mm) and moment zero maps of all the molecular transitions observed towards G12.42+0.50 in the ATOMS survey are shown. The continuum map is generated from the high-resolution 12-m data only while the line maps are from the 7-m + 12-m combined data. The maps are integrated over the velocity range [$-32, 58$] $\rm km\,s^{-1}$ which is approximately $\rm V_{LSR} \pm 50\,\rm km\,s^{-1}$ and incorporates all emission features. This enables easy comparison between various molecular lines with the same integrated velocity range. The color bar indicates the flux scale in $\rm Jy\,beam^{-1}\,km\,s^{-1}$ for moment zero maps and $\rm Jy\,beam^{-1}$ for the continuum map. The beam size is indicated at the bottom left in each panel. Red ellipses represent the identified cores (see section \ref{core_identify}).}
    \label{all_mole_mom0_1242}
\end{figure*}
\begin{figure*}
    \centering
    \includegraphics[width=0.85\textwidth]{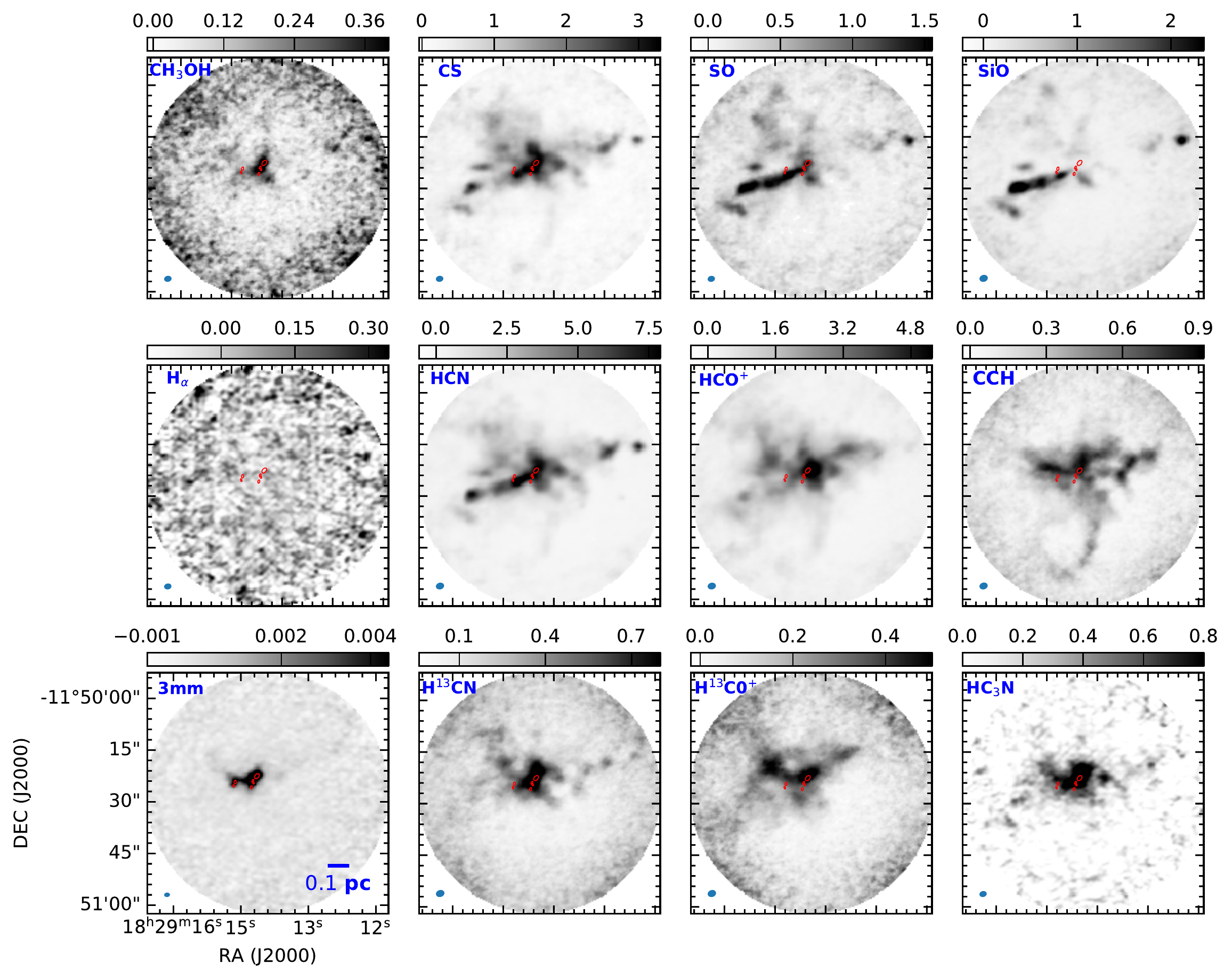}
    \caption{Same as Figure \ref{all_mole_mom0_1242}, but for the protocluster G19.88-0.53. The maps are integrated over the velocity range [$-21, 109$] $\rm km\,s^{-1}$ which is approximately $\rm V_{LSR} \pm 65\,\rm km\,s^{-1}$.}
    \label{all_mole_mom0_1988}
\end{figure*}

In this paper, we present new ALMA data on the massive protoclusters, G12.42+0.50 and G19.88-0.53 (hereafter G12.42 and G19.88, respectively), observed as a part of the ATOMS survey (ALMA Three-millimeter Observations of Massive Star-forming regions survey) \citep{2020MNRAS.496.2790L}. This survey provides ALMA Band 3 observations for both continuum and molecular line emission for 146 active star-forming regions \citep{2004A&A...426...97F} with majority being potential high-mass star-forming regions. The survey is primarily aimed at revealing the spatial distribution of the probed dense gas tracers and deciphering the role of stellar feedback and filaments in the formation of high-mass stars. In the first paper of the ATOMS series, \citet{2020MNRAS.496.2790L} present the survey giving details of the configuration used for the continuum and spectral line observations. The authors have also highlighted the major goals of the survey.
%
\begin{table}
\caption{Basic information of the protoclusters}
\centering
\begin{tabular}{c c c c c}
 \hline
Source & \multicolumn{2}{c}{Coordinates$^a$} & Distance &  $\rm V_{LSR}$\\[0.5ex] 
& RA(J2000) & DEC(J2000) & (kpc) &($\rm km\,s^{-1}$) \\[0.5ex]
\hline
G12.42+0.50 & 18:10:50.6 & -17.55.47.2 & 2.4$^b$ & 18.3$^c$ \\
G19.88-0.53 & 18:29:14.3 & -11.50.27.0 & 3.3$^d$ & 44.0$^e$ \\
\hline
\end{tabular}
\label{ego_basicparam}
\begin{flushleft}
{\bf Note:} $^a$ Coordinates of the associated IRAS point source taken from \citet{2020MNRAS.496.2790L}; $^b$ \citet{{2010ApJ...710..150C},{2012ApJS..202....1H}}; $^c$ \citet{2015MNRAS.451.2507Y}; $^d$ \citet{{2009ApJ...699.1153R},{2014MNRAS.445.1170G}}; $^e$ \citet{{2007ApJ...654..361Q}, {2002ApJ...566..945B}}.
\end{flushleft}
\end{table}
%
G12.42 and G19.88 have been studied by \citet[][hereafter Issac19]{2019MNRAS.485.1775I} and \citet[][hereafter Issac20]{2020MNRAS.497.5454I}, respectively. Both these sources are classified as `extended green objects' (EGOs) by \citet{2008AJ....136.2391C}. These are a class of objects believed to be associated with outflows from massive young stellar objects (MYSOs). Issac19 and Issac20 have also provided a brief overview of recent literature on EGOs. G12.42, located at a distance of 2.4~kpc, is catalogued as a "possible" outflow candidate \citep{2008AJ....136.2391C}. Using the  Giant Metrewave Radio Telescope (GMRT) radio continuum observations at 610 and 1390~MHz, Issac19 suggest the co-existence of an UC H\,{\small II} region and an ionized jet that are likely powered by the MYSO, IRAS 18079-1756. The ionized jet inference is strongly supported by near-infrared (NIR) spectroscopic observations presented by these authors where the detection of the shock-excited lines of $ \rm H_2$ and [Fe\,{\small II}] are reported.
Additionally, these authors show that the observed radio emission is located at the centroid position of the detected wide-angle bipolar CO outflow which also supports the radio thermal jet scenario \citep{{1996ASPC...93....3A},{1997IAUS..182...83R}}. The other protocluster, G19.88, located at a distance of 3.31 kpc, is catalogued as a "likely" outflow candidate and associated with IRAS 18264-1152 \citep{2008AJ....136.2391C}. Issac20 present GMRT observations at the frequencies mentioned above. Their study reveals the presence of an ionized jet which is deciphered to be associated with a massive, dense, and hot ALMA 2.7~mm core powering a bipolar CO outflow. In combination with the ALMA Band 3 and 7 continuum and line emission data, G19.88 is understood to be an active protocluster with high-mass star-forming cores in various evolutionary phases. \citet{HLLiu21} (the third paper in the ATOMS series) catalogued the high mass star-forming cores associated with G12.42 as "unknown cores,” as they do not enshroud hyper/ultra compact H\,{\small II} regions and the spectra lack evidence of complex organic molecules (COM) (e.g., CH$_3$OCHO, CH$_3$CHO, CH$_3$OH, C$_2$H$_5$CN). Whereas, G19.88 is listed as “pure w-cHMC,” which contains high-mass cores with relatively low levels of COM richness and not associated with hyper/ultra compact H\,{\small II} regions.
The details of the two protoclusters are compiled in Table\ref{ego_basicparam}. In this paper, we use the data from the ATOMS survey to conduct a detailed study of the detected 3~mm cores and focus on a multi-spectral line study to understand the kinematics and gas dynamics of the regions associated with G12.42 and G19.88 from clump to core scale. The paper is organised as follows. Section \ref{obs_data} discusses the ALMA observations carried out as a part of the ATOMS survey and the complementary archival data used in this study. Results obtained from the continuum and molecular line analysis are presented in Section \ref{result}. Sections \ref{grav_stab_global_coll} and \ref{frag_Section} address the gravitational stability and fragmentation scenario of the clumps. Section \ref{dyn-SF} delves into detailed analysis and discussion of multi-scale gas kinematics that includes the large-scale filamentary structures, small-scale density structures and the cores. The interplay between gravity and turbulence in driving star formation in the two protoclusters is explored in Section \ref{driving_mech}. Section \ref{conclusion} summarizes the results and subsequent interpretation of this study.

\section{Observations and archival data}
\label{obs_data}
\subsection{ALMA observations}
ALMA data from the ATOMS survey (Project ID: 2019.1.00685; PI:Tie Liu) are used for studying the G12.42 and G19.88 complexes. The 12-m array observations of both complexes were conducted on 1 November 2019. The ACA observations of the same were conducted on the 2 November 2019 and 3 November 2019, with two executions. On-source integration times on each source for 12-m array and ACA array are $\sim$3 minutes and 8 minutes, respectively. 
Calibration of the 12-m array data and ACA data were done separately using the Common Astronomy Software Applications (CASA) package version 5.6 \citep{2007ASPC..376..127M}. Subsequent to this, the visibility data were also combined and imaged in CASA to recover the very extended emission that is missed in the 12-m array observations. All images used in our analysis are primary beam corrected. The high-resolution 3~mm  ($\sim$99.93 GHz) continuum map, with beam size of 1.75~arcsec $\times$ 1.3~arcsec and {\it rms} noise of 0.2~$\rm mJy\,beam^{-1}$, is constructed using data from line-free spectral channels. 
In addition to the continuum data, in the ATOMS survey eight spectral windows (SPWs) were configured to sample eleven major molecular lines that include dense gas tracers (e.g., J = 1-0 of HCO$^{+}$ , HCN and their isotopes), hot core tracers (e.g., CH$_{3}$OH, HC$_{3}$N), shock tracers (e.g., SiO, SO) and ionized gas tracers ($\rm H_{40\alpha}$). The species name, transitions, rest frequencies and basic parameters (e.g., critical density, and upper level energy) of these molecular lines are summarised in Table 2 of \citet{2020MNRAS.496.2790L}.
Details (angular resolution, linear resolution, maximum recoverable scale (MRS), and the {\it rms} level) of the ATOMS data used in this study are compiled in Table \ref{ALMA_obs}. More details can be found in Table 1 of \citet{2020MNRAS.496.2790L}. 
Continuum maps and the velocity integrated intensity maps of all the molecules observed in ATOMS survey are shown in Figure \ref{all_mole_mom0_1242} and \ref{all_mole_mom0_1988}. 

\begin{table}
\setlength{\tabcolsep}{3pt}
\caption{Details of ALMA data}
\centering
\begin{tabular}{l c c}
 \hline
{\it Continuum data (12-m array)}:\\[0.5ex] 
Source & G12.42+0.50 & G19.88-0.53 \\[0.5ex] 
Angular resolution & 1.75$^{''} \times$ 1.3$^{''}$ & 0.46$^{''} \times 0.28^{''}$\\[0.5ex] 
Linear resolution (pc $\times$ pc) & 0.020 $\times$ 0.015 & 0.007 $\times$ 0.004\\[0.5ex]
{\it rms} noise ($\rm mJy\,beam^{-1}$) & 0.2 & 0.12 \\[0.5ex]
Maximum recoverable scale (MRS) & 18.3$^{''}$ & 5.2$^{''}$ \\[0.5ex]
\hline
{\it Line Data (12-m + 7-m arrays)}:\\[0.5ex] 
Transition & $\rm H^{13}CO^+$ (1 - 0)/  & $\rm HCO^+$ (1 - 0) \\[0.5ex]
& SiO (2 - 1) & \\[0.5ex]
Angular resolution & 2.5$^{''} \times$ 2.0$^{''}$ & 2.4$^{''} \times 1.9^{''}$\\[0.5ex]
\multirow{2}{*}{Linear resolution (pc $\times$ pc)}  & 0.029 $\times$ 0.023$^{a}$  & 0.028$\times$ 0.022$^{a}$  \\[0.5ex]
& 0.040 $\times$ 0.032$^{b}$   & 0.038$\times$ 0.030$^{b}$  \\[0.5ex]
Velocity Resolution ($\rm km\,s^{-1}$) & 0.2 & 0.1 \\[0.5ex]
{\it rms} noise ($\rm mJy\,beam^{-1}$) & 8 & 12\\[0.5ex]
Maximum recoverable scale (MRS) & 76.2$^{''}$ & 76.2$^{''}$ \\
\hline
\end{tabular}
\label{ALMA_obs}
\begin{flushleft}
{\bf Note:} $^a$ and $^{b}$ are the linear resolutions corresponding to G12.42+0.50 and G19.88-0.53, respectively.
\end{flushleft}
\end{table}
%
\subsection{Archival ALMA continuum data}
Continuum emission towards G19.88 complex at 2.7~mm ($\sim$ 111.0~GHz) was retrieved from the ALMA archives (Project ID: 2017.1.00377.S; PI:S. Leurini). Observations were conducted in the session 2017 - 2018. These high-resolution ALMA observations were obtained with the 12-m array in the FDM Spectral Mode. The minimum baselines, maximum baselines, and MRS in this observation are 15.07~m, 2386.1~m and 5.2~arcsec, respectively. The 2.7 mm continuum map, with beam size of 0.46~arcsec $\times$ 0.28~arcsec and {\it rms} noise of 0.12~$\rm mJy\,beam^{-1}$, is used to identify and study the cores associated with G19.88. Details of the ALMA archival data used in this study are listed in Table \ref{ALMA_obs}.
\subsection{Spitzer archival data}
To investigate the morphology of the mid-infrared (MIR) emission in regions associated with G12.42+0.50 and G19.888-0.53, we obtained MIR images from the archives of the Galactic Legacy Infrared Midplane Survey Extraordinaire (GLIMPSE) survey of the \textit{Spitzer} Space Telescope. The Infrared Array Camera (IRAC) mounted on the \textit{Spitzer} Space Telescope is capable of simultaneous broad band imaging at 3.6, 4.5, 5.8, and 8.0$\,\mu$m \citep{2004ApJS..154...10F}. We retrieved images having an angular
resolution of $\lesssim$2 arcsec with a pixel size of $\sim$0.6 arcsec in three IRAC bands (3.6, 4.5, and 8.0$\,\mu$m) \citep{2003PASP..115..953B}.
\subsection{JCMT archival data}
The molecular line data for $\rm {}^{13}CO$ (3 - 2) 
transition were obtained from the archives of James Clerk Maxwell Telescope (JCMT) to investigate the filamentary morphology of the star-forming region associated with G12.42. The velocity structure of filaments using JCMT $\rm {}^{12}CO$ (3 - 2) transition presented in Issac19 was utilized to investigate the gas kinematics.
Heterodyne Array Receiver Program (HARP) mounted on the 15~m JCMT telescope comprises of 16 detectors laid out on a $4 \times 4$ grid, with an on-sky beam separation of 30~arcsec. The molecular line observation for  $\rm {}^{13}CO$ (3 - 2) was carried out using HARP at a rest frequency of $\sim$330.6~GHz.
The beam size of JCMT at 345~GHz is 14~arcsec \citep{2009MNRAS.399.1026B}. $\rm {}^{13}CO$ (3 - 2) and $\rm {}^{12}CO$ (3 - 2) data have a channel width of 0.055 and 0.42$\,\rm km\,s^{-1}$, with a {\it rms} level per channel of 1.9 and 1.0~K, respectively.
\begin{figure}
    \centering
    \includegraphics[width=0.47\textwidth]{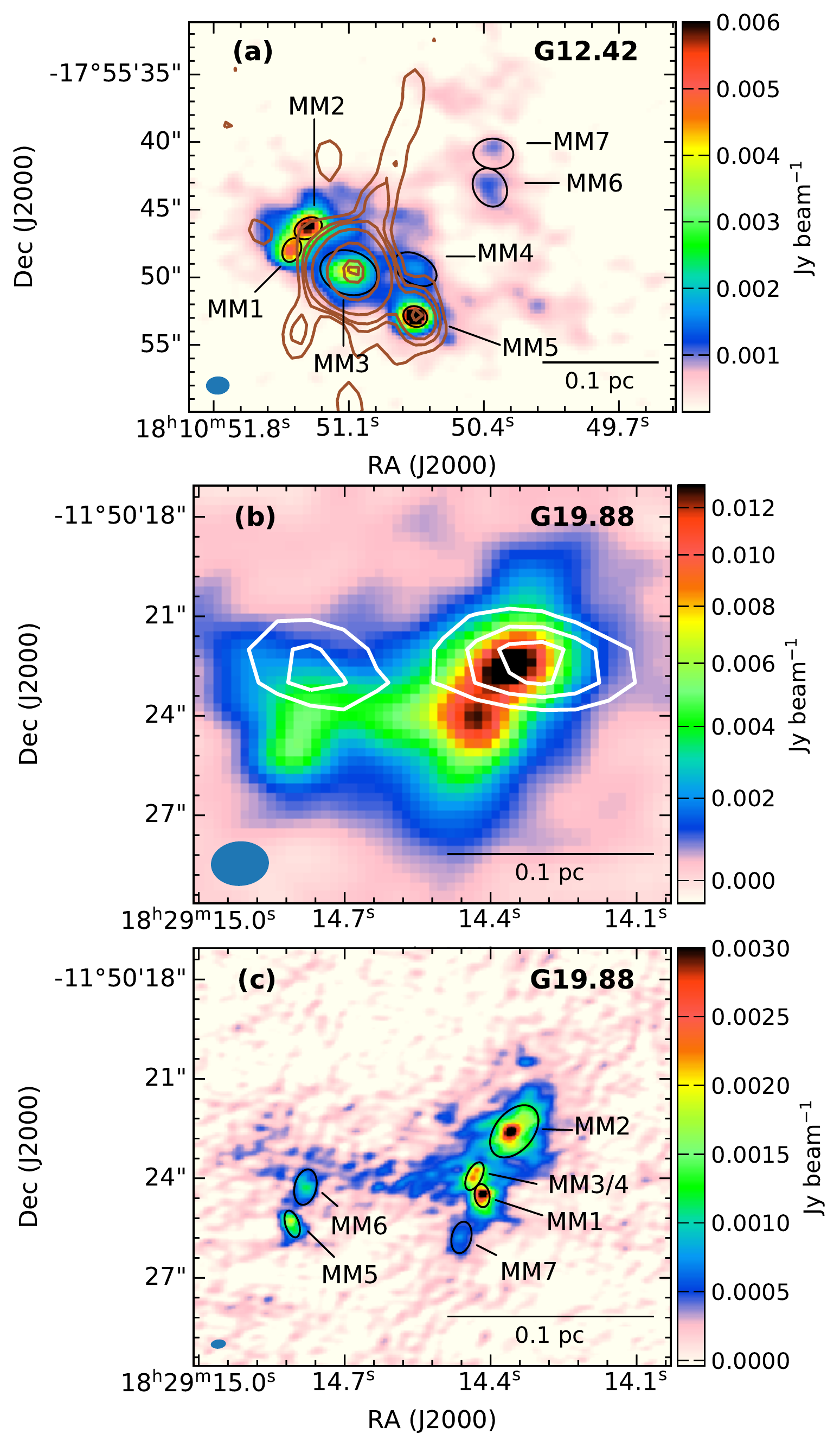}
    \caption{(a) 3~mm continuum map of the region associated with G12.42+0.50 obtained using 12-m array data from ATOMS is shown in colour scale. Core apertures acquired using CASA-\textit{imfit} are shown in black ellipses. The overlaid contours showing the radio emission at 1390 MHz (beam size is 3.0~arcsec $\times$ 2.4~arcsec ) with contour levels at 3, 6, 9, 18, 63, 150 and 172 times $\sigma$ ($\sigma$ $\sim$ 29.7 $\rm \mu Jy\,beam^{-1}$) are from Issac19. (b) Same as figure (a) but for G19.88-0.53. The overlaid contours showing the radio emission at 1390 MHz (beam size is 4.3~arcsec $\times$ 2.7~arcsec ) with contour levels at 3, 4, 5 and 6 times $\sigma$ ($\sigma$ $\sim$ 45 $\rm \mu Jy\,beam^{-1}$) are from Issac20. (c) The colour scale shows the 2.7~mm continuum emission of the region associated with G19.88-0.53 obtained using 12-m array data from the ALMA archive. The black ellipses represent the core apertures acquired using CASA-\textit{imfit}. The beam of the continuum maps are indicated at the bottom left in each figure.}
    \label{continuum_maps}
\end{figure}

\section{Results}
\label{result}
\subsection{Continuum Emission}
We use the observed 3~mm continuum emission maps to identify and study compact cores in the regions associated with the protoclusters under study. The derived properties of these cores enable us to understand their nature and formation. The lower left panel of Figures \ref{all_mole_mom0_1242} and \ref{all_mole_mom0_1988} present the 3~mm continuum map for the entire field of view of the ATOMS survey. An enlarged view of the extended continuum emission morphology and the identified cores in G12.42 and G19.88 are shown in Figure \ref{continuum_maps}.

\subsection*{\it Core identification}
\label{core_identify}
\begin{table*}
\caption{Parameters of detected cores in protocluster G12.42+0.50, using data from 12-m array at 3 mm.}
\centering
\begin{tabular}{c c c c c c c c c c c c}
 \hline
\multirow{2}{*}{Core} & \multicolumn{2}{c}{Peak position}  & \multicolumn{2}{c}{Deconvolved size}  & PA & $F_{\rm int}$ & $F_{\rm peak}$ & $R_{\rm eff}$ & $R_{\rm core}$ & $M^{\rm a}_{\rm core}$ & $\Sigma$ \\ [0.5ex] 
& RA(J2000) & DEC(J2000) & Major($^{''}$) & Minor($^{''}$) & deg & $\rm mJy$ & $\rm mJy\,beam^{-1}$ & ($^{''}$) & $10^{-2}$ pc & ($\rm M_\odot$) & g cm$^{-2}$\\ [0.5ex] 
 \hline
MM1 &18:10:51.39 & -17.55.48.04 & 1.84 & 1.31 & 156 & 10.9 & 5.2 & 0.8 & 0.9 & 24.0 & 19.5 \\
\hline
MM2 &18:10:51.31 &-17.55.46.37 & 2.13 & 1.50 & 116 & 14.9 & 6.3 & 0.9 & 1.0 & 32.9 & 20.2 \\
\hline
MM3 & 18:10:51.10 &-17.55.49.60 & 4.31 & 3.21 & 72 & 22.2 & 3.2 & 1.9 & 2.2 & 48.9 & 6.9  \\
\hline
MM4 & 18:10:50.75 & -17.55.49.40 & 3.4 & 2.17 & 61 & 6.8 & 1.6 & 1.4 & 1.6 & 15.0 & 4.0  \\
\hline
MM5 &18:10:50.75 &-17.55.52.89 &1.79 & 1.48 & 69 & 14.9 & 7.0 & 0.8 & 1.0 & 32.8 & 23.9 \\
\hline
MM6 & 18:10:50.58& -17.55.54.50 & 2.96 & 2.39 & 30 & 4.6 & 1.1 & 1.3 &1.6 & 10.1 & 2.8 \\
\hline
MM7 & 18:10:50.36& -17.55.43.37& 2.97 & 2.22 & 86 & 3.7 & 0.9 & 1.3 & 1.5 & 8.1 & 2.4\\
\hline
\end{tabular}
\label{parameter1242}
\begin{flushleft}
{\bf Note:} $^a$ Temperature of 25~K is considered to calculate the mass of all the cores.
\end{flushleft}
\end{table*}
\begin{table*}
\caption{Parameters of detected cores in the protocluster G19.88-0.53, using 2.7 mm data.}
\begin{tabular}{c c c c c c c c c c c c c}
 \hline
\multirow{2}{*}{Core} & \multicolumn{2}{c}{Peak position}  & \multicolumn{2}{c}{Deconvolved size}  & PA & $F_{\rm int}$ & $F_{\rm peak}$ & $R_{\rm eff}$ & $R_{\rm core}$ & $M_{\rm core}$ & $T^{a}$ & $\Sigma$ \\ [0.5ex] 
& RA(J2000) & DEC(J2000) & Major($^{''}$) & Minor($^{''}$) & deg & $\rm mJy$ & $\rm mJy\,beam^{-1}$ & ($^{''}$) & $10^{-2}$ pc &$\rm M_\odot$ & K & g cm$^{-2}$\\ [0.5ex]

\hline
MM1 &18:29:14.41 & -11.50.24.52 & 0.70 & 0.45 & 8 & 11.3 & 3.0 & 0.3 & 0.5 & 9.3 & 82.7 & 35.7\\
\hline
MM2 &18:29:14.35 &-11.50.22.58 & 1.81& 1.16 & 140 & 39.0 & 2.2 & 0.7 & 1.1 & 22.7 &115.9 & 13.1 \\
\hline
MM3$^{c}$ & 18:29:14.43 &-11.50.24.10 & 0.91 & 0.44 & 155 & 10.5 & 2.3 & 0.3 & 0.5 & 8.7 & 82.7&  26.1\\
\hline
MM4$^{c}$ & 18:29:14.43 & -11.50.23.93& 0.91 & 0.44 & 155 & 10.5 & 2.3 & 0.3 & 0.5 & 8.7 & 82.7 & 26.1\\
\hline
MM5 &18:29:14.80 &-11.50.25.37 &0.84 &0.41 &18 & 6.3 & 1.5 & 0.3 & 0.5 & 9.4 & 47.2 &  32.8\\
\hline
MM6 & 18:29:14.78& -11.50.24.26& 1.09 & 0.65 & 167 & 6.80 & 1.0 & 0.4 & 0.7 & 10.1 & 47.2 & 17.0\\
\hline
MM7 & 18:29:14.46& -11.50.25.78& 0.96 & 0.59 & 167 & 4.0 & 0.7 &  0.4 & 0.6 & 3.3 &  82.7 & 5.8 \\
\hline
\end{tabular}
\label{parameter1988}
\begin{flushleft}
{\bf Note:} $^a$ Temperature corresponding to each core is taken from Issac20, $^c$ MM3 and MM4 are unresolved in the map, hence the quoted parameters refer to the combined region enclosing both the cores.
\end{flushleft}
\end{table*}
Cores in both complexes have been identified in ATOMS III paper \citep{HLLiu21}. These authors have also derived the physical parameters. In addition to this, compact cores in G12.42 (SMA; 1.1~mm) and G19.88 (ALMA; 2.7~mm) have been analyzed in Issac19 and Issac20, respectively. 
\citet{HLLiu21} have detected the most prominent cores (four in G12.42 and two in G19.88) in these protoclusters. Issac19 have identified three SMA cores in G12.42 and Issac20 have detected six ALMA cores in G19.88. Based on these results and a careful visual inspection of the ATOMS continuum map, we are prompted to revisit the core identification before proceeding for further analysis. 
The 3~mm continuum emission towards G12.42 and G19.88 observed with the 12-m array are shown in Figure \ref{continuum_maps}(a) and (b). 
For core detection in G19.88, we use available higher resolution data at 2.7~mm from the ALMA archives. This is the same data set used in Issac20 and the map is shown in Figure \ref{continuum_maps} (c). 

Following the approach introduced in \citet{2020ApJ...896..110L} and outlined in \citet{{2020MNRAS.496.2790L},{HLLiu21},{HLLiu22a}}, cores are extracted by utilizing the \textit{astrodendro} \footnote{Astrodendro is a Python package for computing dendrograms of astronomical data  (\url{http://www.dendrograms.org/})} package and the CASA \textit{imfit} task. The {\it dendrogram} technique \citep{2008ApJ...679.1338R} decomposes the emission into a hierarchy of sub-structures, providing a good representation of the topography of the star-forming region. Dense structures (\textit{leaves} in the terminology of the algorithm) which are not further divisible into smaller structures are considered as potential candidates to become star-forming cores. To identify the leaves, the following parameters are used: (i) $min\_value = 3 \sigma$, which is the minimum threshold to be considered in the dataset and $\sigma$ is the {\it rms} noise; (ii) $min\_delta = \sigma$, the default value in the algorithm which determines how significant a leaf has to be in order to be considered an independent entity; (iii) $min\_npix$ $= N$ pixels, which is the minimum number of pixels needed for a leaf. To ensure cores are resolvable, $N$ is chosen to be equivalent to the synthesized beam
area in units of pixel. The algorithm returns several parameters (i.e., position of the centre, major and minor axis sizes, position angle) of the identified cores. 

Next, the CASA \textit{imfit} task is used with the dendrogram parameter outputs as the corresponding initial estimates of the fit parameters. 
During the fitting process none of the parameters are fixed. A bounding box is used, the size of which is defined based on the retrieved aperture of the leaves in the {\it dendrogram} analysis. We have not considered background subtraction as large-scale extended emission, that can be regarded as the background component, is already filtered-out in the interferometric data.
The derived physical parameters, that includes the peak position, deconvolved major and minor sizes (FWHM$_{\rm maj}$, FWHM$_{\rm min}$), position angle (PA), peak intensity ($F_{\rm peak}$), and integrated flux density ($F_{\rm int}$) are listed in Table \ref{parameter1242} and \ref{parameter1988} for G12.42 and G19.88, respectively. In order to avoid inclusion of spurious cores in our analysis, we retain only those satisfying, $F_{\rm peak} > 5 {\sigma}$. Further, we also reject cores with poorly-fitted shapes by visual inspection of the continuum map overlaid with the identified leaf structures. These usually appear either as filamentary structures or diffuse emission features with an aspect ratio of more than three, between the length of the major and minor axis. The identified cores are overlaid in Figure \ref{continuum_maps} (a) and (c).

For G12.42, we identify seven cores, MM1 - MM7. It is to be noted that in case of this source, cores MM1 and MM2 which are named SMA1 and SMA2 in Issac19, are manually fitted using CASA {\it imfit} since the  dendrogram algorithm resolved it into a single leaf though two distinct cores are seen visually. For G19.88, we also identify seven cores, MM1 - MM7. In their analysis, Issac20, had visually identified six cores and used the CASA {\it imfit} task to determine the parameters. For the common cores, the overall core extraction carried out in this work agrees well with Issac19, Issac20, and \citet{HLLiu21}.

To address the concern of `missing flux' in high-resolution interferometric observations, we compare the retrieved deconvolved sizes of the compact cores with that of the MRS of the two data sets used. Considering the 12-m data used for core identification in G12.42, the deconvolved sizes of the cores lie in the range, 1.6 -- 3.7~arcsec, which is much smaller ($\lesssim20$ per cent) than the MRS of $\sim$18~arcsec. Further, comparing the retrieved fluxes from the 12-m and the combined data, \citet{HLLiu21,HLLiu22a} discuss that the uncertainties on the estimated flux due to `missing flux' is not significant. 
In case of G19.88, we have used the higher resolution archival ALMA data to detect the cores where the MRS quoted is $\sim$5~arcsec and the retrieved sizes lie in the range, 0.6 -- 1.5~arcsec which are again smaller than the MRS. Nevertheless, for the largest core (MM2), we compare the flux retrieved with that from the 12-m ATOMS data and found them to be equal. This indicates that the estimated parameters of the compact cores in G12.42 and G19.88 are not significantly affected and only the diffuse, larger-scale emission between clumps could have been filtered out.

The 3~mm continuum emission can have contributions from both thermal dust emission and free-free emission from ionized gas. Based on the lack of $\rm H_{40\alpha}$ radio recombination line emission observed in the ATOMS survey, \citet{HLLiu21} classified these prominent cores in G12.42 and G19.88 as ones without association with HC or UC HII regions. However, weak centimetre radio emission, interpreted as possible thermal jets, has been detected in cores MM3 and MM5 in G12.42 (Issac19) and core MM2 in G19.88 (Issac20). The radio emission is shown as contours in Figure \ref{continuum_maps}(a) and (b). 
Free-free emission from radio jets become optically thin for frequencies greater than the turnover frequency and the spectral index becomes -0.1 \citep{2018A&ARv..26....3A}.
Without the information of the turnover frequency either from observed radio spectral energy distribution or from theoretical modelling of the jet emission for the three cores discussed above, it is difficult to estimate the contribution of free-free emission at 3~mm . For a crude estimate, we assume the highest frequency data available (5 GHz for G12.42 and 23 GHz for G19.88) as the turnover frequency and extrapolate the measured fluxes to 3~mm (100~GHz) assuming an optically thin spectral index of -0.1. This is reasonably consistent with \citet{2018A&ARv..26....3A} who have adopted a lower limit of 10~GHz for the turnover frequency to estimate the physical parameters of the radio emission from ionized jets. 

For G12.42, 5~GHz continuum flux density is taken from Issac19 and for G19.88, 23~GHz flux density is taken from \citet{2006AJ....131..939Z}. 
Extrapolating gives 20 percent and 5 percent contribution from ionized gas emission to the 3~mm flux estimate of cores MM3 and MM5, respectively, of G12.42 and 2 percent contribution to the 2.7~mm flux estimate for core MM2 of G19.88. For core MM3 of G12.42, the contribution from free-free emission may result in the mass (see next section) being overestimated by a factor of 1.3. 

In summary, free-free emission does not significantly affect the continuum flux-derived parameter estimates (e.g., mass, and density) of the cores studied here.

\begin{figure}
    \centering
    \includegraphics[width=0.44\textwidth]{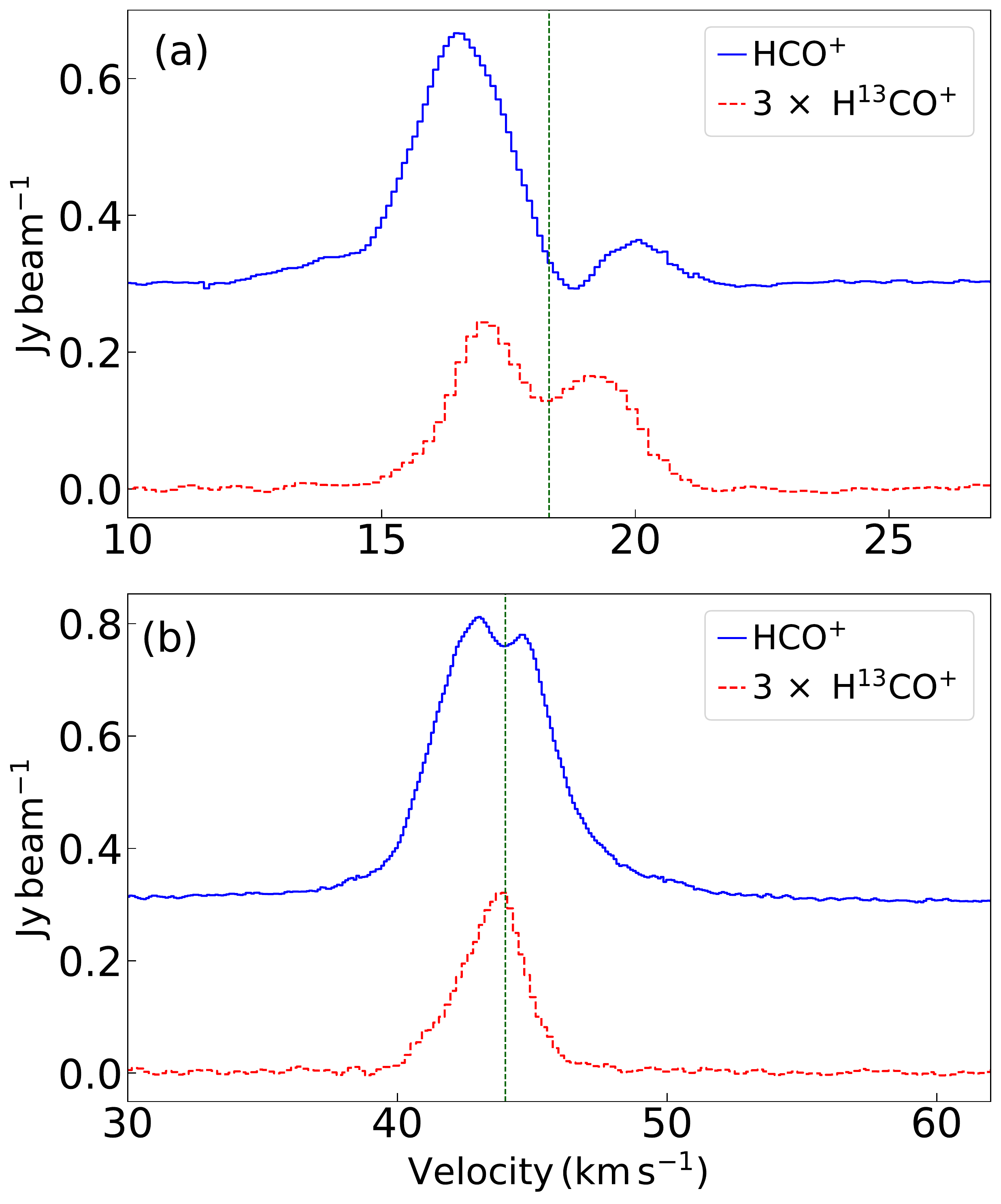}
    \caption{ Clump-averaged spectra of the molecular lines (H$^{13}$CO$^{+}$, HCO$^{+}$) detected towards G12.42+0.50 and G19.88-0.53 using combined 12-m + 7-m data are shown in panel (a) and (b), respectively. The spectra are averaged over a region enclosed by the 3 $\sigma$ contour continuum emission for G12.42 and G19.88, respectively. HCO$^+$ spectra shown in blue (solid) are vertically offset by 0.3~$\rm Jy\,beam^{-1}$ and H$^{13}$CO$^{+}$ spectra shown in red (dashed) are scaled-up by a factor of 3 in both the panels. The vertical dashed line in each panel marks the LSR velocity of 18.3 and 44.0 $\,\rm km\,s^{-1}$ for G12.42 and G19.88, respectively.}
    \label{1242+1988_hco+h13co_3sigma}
\end{figure}
\begin{figure}
    \centering
    \includegraphics[width=0.45\textwidth]{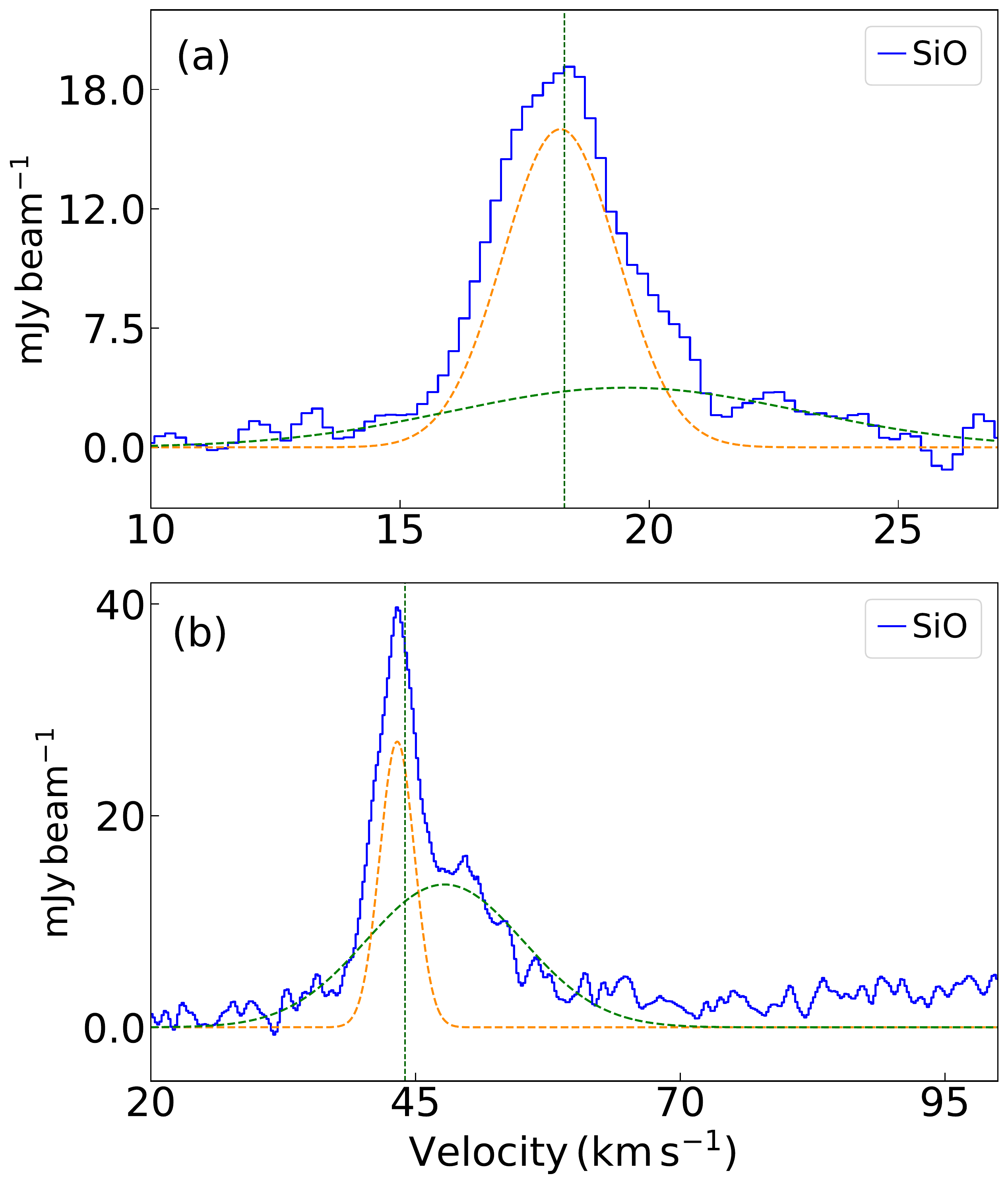}
    \caption{Clump-averaged spectra of SiO (2-1) detected towards G12.42+0.50 and G19.88-0.53 using combined 12-m + 7-m data are shown in panel (a) and (b), respectively. The spectra are averaged over a region enclosed by the 3 $\sigma$ contour continuum emission for G12.42 and G19.88, respectively. The spectra are boxcar smoothed by 4 channels resulting in a velocity resolution of 0.84$\,\rm km\,s^{-1}$. Orange and green dashed lines indicate the decomposed components of SiO (2-1) line. The vertical dashed line in each panel marks the LSR velocity of 18.3 and 44.0 $\,\rm km\,s^{-1}$ for G12.42 and G19.88, respectively.}
    \label{1242+1988_sio_3sigma}
\end{figure}
\subsection*{\it Physical properties of detected cores}

\noindent Assuming the dust emission to be optically thin, the core masses can be estimated using the following expression
\begin{equation}
M_{\rm core} = \frac{F_{\rm \nu}\,D^{2}\,R_{\rm gd}}{B_{\nu}(T_{\rm d})\,\kappa_{\nu}},
\label{M_dust}
\end{equation}
where ${\kappa}_{\nu}={\kappa}_{\nu_0}\bigg(\dfrac{\nu}{\nu_0}\bigg)^{\beta}$ is the dust opacity, $\beta$ is the dust emissivity index taken as 1.75 in our calculations, $M_{\rm core}$ is the mass of the cores, $F_{\nu}$ is the integrated flux density, $D$ is the distance to the source, $R_{\rm gd}$ is the gas to dust ratio and $B_{\nu}$ is the Planck function at dust temperature $T_{\rm d}$. Considering $R_{\rm gd} = 100$ and using ${\kappa}_{\nu_0}$ to be 0.1~cm$^2$ g$^{-1}$ at 1~THz \citep{1990AJ.....99..924B}, ${\kappa}_{\nu}$ is estimated to be 0.18 and 0.21~cm$^2$ g$^{-1}$ for G12.42 (at $\sim$99.93~GHz) and G19.88 (at $\sim$111.0~GHz), respectively. For G12.42, $T_{\rm dust}$ is taken to be 25~K, following the assumption made in \citet{HLLiu21} and consistent with the cold component modelled for the EGO in Issac19. 
It should be noted that a higher temperature would yield lower mass values. For instance if one considers a temperature of the order of 100~K, then the core masses would be a factor of four lower. In comparison, a relatively better estimate of the core mass is possible for the G19.88 protocluster. For this, Issac20 have detected multiple transitions of CH$_{3}$OH molecule towards the cores associated with G19.88. Using it, they generated the rotational temperature diagram (RTD) to estimate CH$_{3}$OH rotational temperature for each core in the range $\sim$47 - 116~K and these are used to calculate core masses. The core MM7, which was not detected in Issac20, however, falls within the aperture used to extract the CH$_{3}$OH spectrum to construct the RTD for MM1, MM3, and MM4 where the rotational temperature was estimated to be 82.7~K. We adopt the same temperature for MM7.
Taking into account the different $\beta$ value used and the slight difference in core aperture retrieved, the  estimated mass of the cores are in agreement with Issac20. 
We have also derived the mass surface density using the expression
\begin{equation}
{\Sigma}_{\rm core} = \dfrac{M_{\rm core}}{\pi R^{2}_{\rm core}},
\end{equation}
where $R_{\rm core}$ is the core radius which is taken to be half of the geometric mean of FWHM$_{\rm maj}$ and FWHM$_{\rm min}$ at the core distance. Masses and mass surface densities of the cores are tabulated in Table \ref{parameter1242} and \ref{parameter1988}. The 3$\sigma$ mass sensitivities for G12.42 and G19.88 are $\sim$1.2 and $\sim$0.4~M$_\odot$, respectively, considering the temperatures used for core mass calculations.
At the above sensitivity, the ALMA continuum maps reveal the most complete population of cores in these two protocluster EGOs.
\begin{figure}
    \centering
    \includegraphics[width=0.47\textwidth]{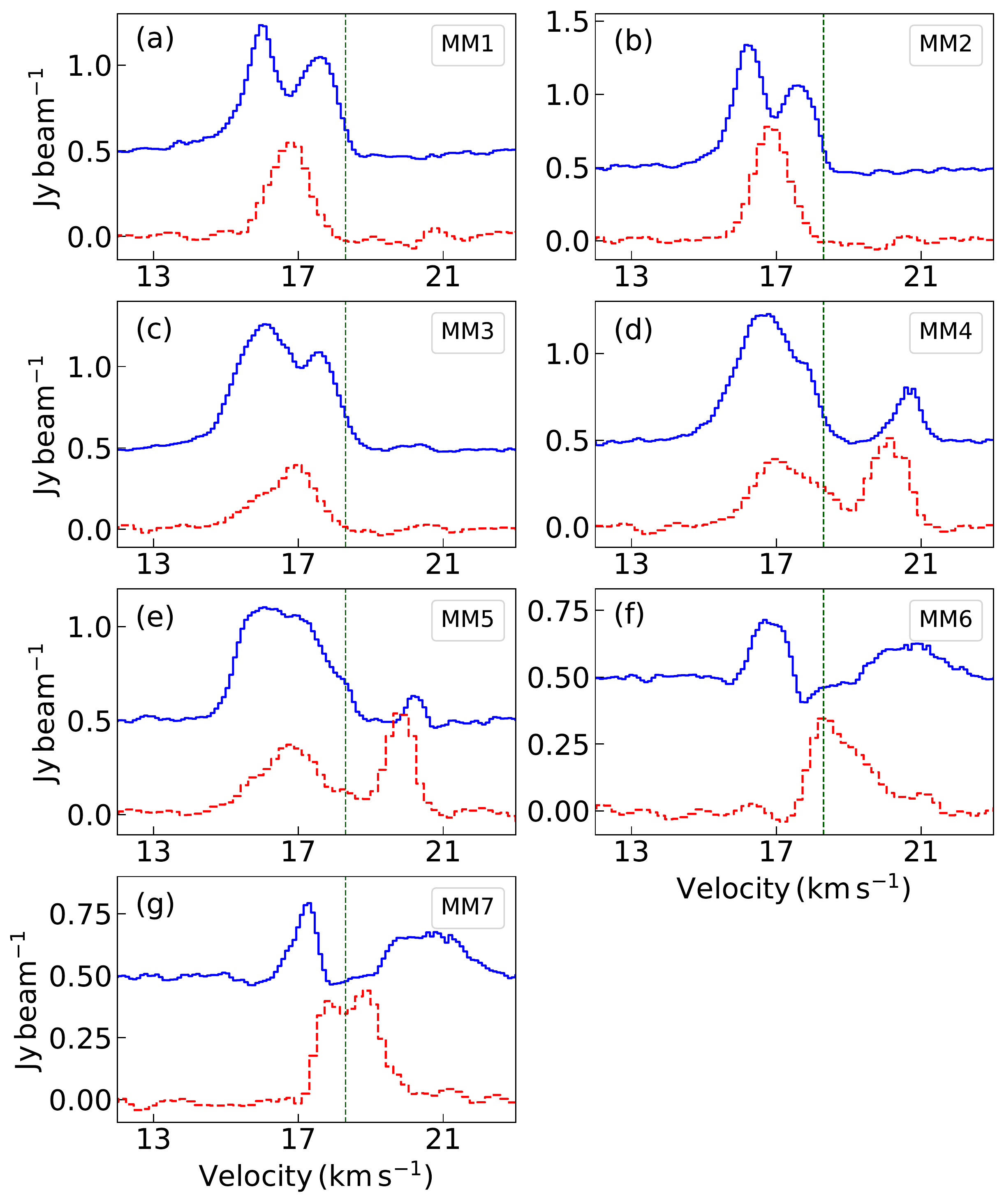}
    \caption{HCO$^{+}$ and H$^{13}$CO$^{+}$ spectra taken towards the cores associated with G12.42+0.50 are shown in blue (solid) and red (dashed) in each panel, respectively using combined 12-m + 7-m data. HCO$^{+}$ spectra are vertically offset by 0.5~$\rm Jy\,beam^{-1}$  and H$^{13}$CO$^{+}$ spectra are scaled-up by a factor of 3 in all the panels. Vertical dashed lines mark the LSR velocity $V_{\rm LSR}$ = 18.3$\,\rm km\,s^{-1}$. Plots (a)-(g) present the core-averaged spectra covering (a) MM1 (b) MM2 (c) MM3 (d) MM4 (e) MM5 (f) MM6 and (g) MM7, respectively.}
    \label{1242_allcores_hco_h13co}
\end{figure}
\begin{figure}
    \centering
    \includegraphics[width=0.45\textwidth]{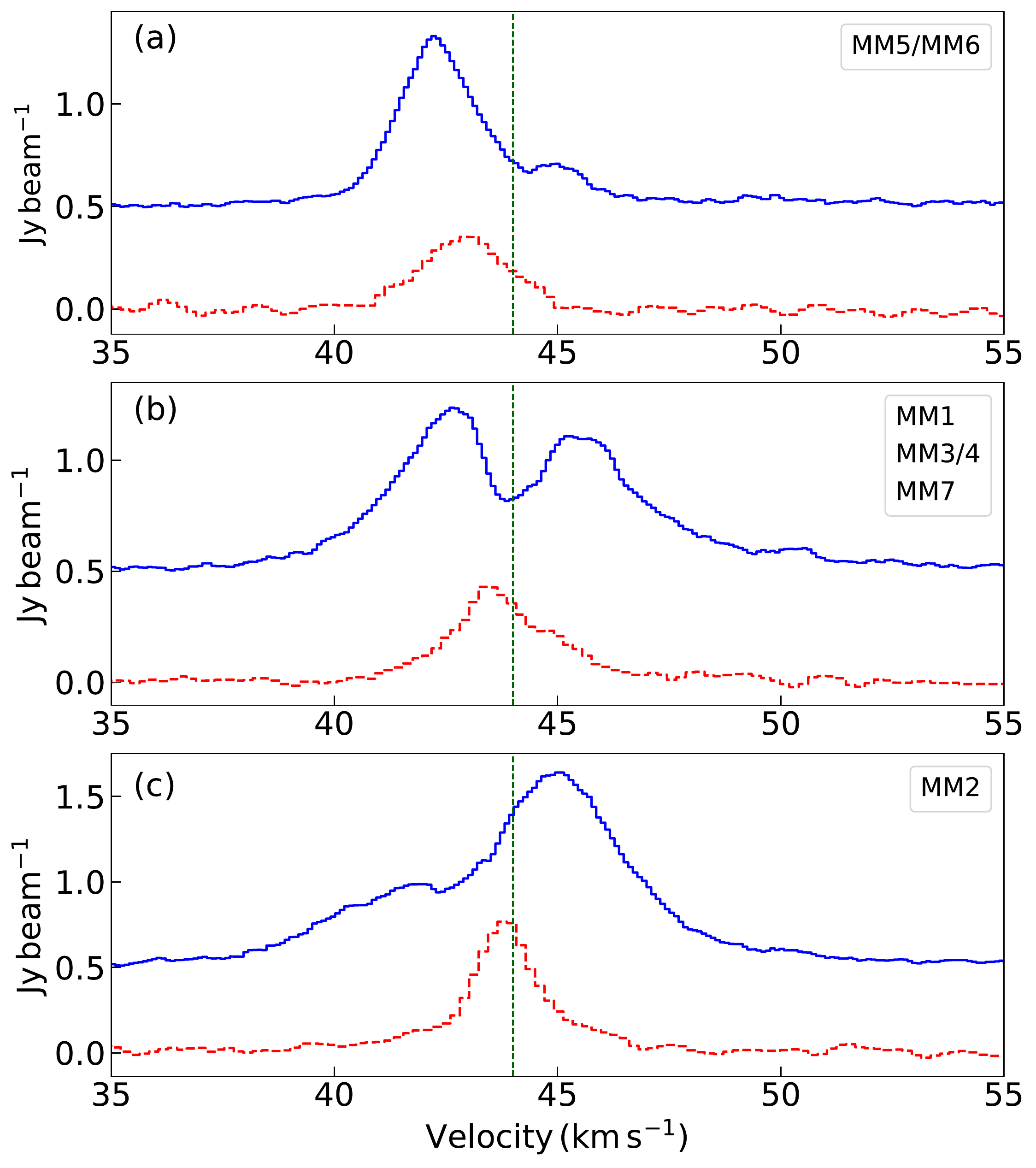}
    \caption{HCO$^{+}$ and H$^{13}$CO$^{+}$ spectra taken towards the cores associated with G19.88-0.53 are shown in blue (solid) and red (dashed) in each panel, respectively using combined 12-m + 7-m data. HCO$^{+}$ spectra are vertically offset by 0.5~$\rm Jy\,beam^{-1}$ and H$^{13}$CO$^{+}$ spectra are scaled-up by a factor of 3 in all the panels. Vertical dashed lines mark the LSR velocity $V_{\rm LSR}$ = 44.0$\,\rm km\,s^{-1}$. Plots (a)-(c) present the spectra averaged over cores (a) MM5 and MM6 (b) MM1, MM3/MM4, and MM7 (c) MM2, respectively}
    \label{1988_allcores_hco_h13co}
\end{figure}
\subsection{Molecular line emission from combined ALMA 12-m array and ACA data}
\label{mol_line_result}
Detailed kinematic information is key to understanding the gas dynamics in the star-forming clouds which enables us to probe the star-formation scenario therein. 
The velocity integrated intensity (moment zero) maps of the molecular transitions observed for these sources in the ATOMS survey, are shown in Figures \ref{all_mole_mom0_1242} and \ref{all_mole_mom0_1988} illustrating the spatial distribution of these species in the two star-forming regions. Multi-peak features are seen in most molecular transitions. There is no $\rm H_{\alpha}$ emission detected. For G12.42, $\rm H^{13}CO^{+}$ map reveals interesting filamentary morphology. This transition is a good tracer of filamentary clouds which is illustrated in the detection of a complex network of narrow filamentary structures toward a massive infrared dark cloud, NGC 6334S \citep{2022ApJ...926..165L}. In both star-forming complexes (G12.42 and G19.88), maps of the outflow/shock tracers (SiO, SO, and CS) reveal elongated and collimated structures. To investigate the global and the core kinematics of G12.42 and G19.88, we have used the spectral line data of H$^{13}$CO$^{+}$, HCO$^{+}$, and SiO transitions. 
In Figures \ref{1242+1988_hco+h13co_3sigma} and \ref{1242+1988_sio_3sigma}, we present the clump-averaged spectra, extracted over the region enclosed within the 3$\sigma$ level of the continuum emission for G12.42 and G19.88. The systemic velocity of 18.3 $\rm km\,s^{-1}$ \citep{2015MNRAS.451.2507Y} and 44.0 $\rm km\,s^{-1}$ \citep{{2007ApJ...654..361Q}, {2002ApJ...566..945B}} for G12.42 and G19.88, respectively, are shown as green dashed vertical lines in the plots. 
As seen in Figure \ref{1242+1988_hco+h13co_3sigma}(a), the line profiles of HCO$^{+}$ (1 - 0) and H$^{13}$CO$^{+}$ (1 - 0) display double peaked, blue-skewed profiles being more prominent in the optically thick transition of HCO$^{+}$ (1 - 0). 
The dip in the line profile for H$^{13}$CO$^{+}$ (1 - 0) is at the systemic velocity (Figure \ref{1242+1988_hco+h13co_3sigma}(a)), while that for HCO$^{+}$ (1 - 0) is redward of it. Extended line wings, seen prominently here in HCO$^{+}$, are the signature of outflows.
The SiO spectra are shown in Figure \ref{1242+1988_sio_3sigma}(a) where the spectrum fits to a combination of a broad and a narrow Gaussian components. The nature of this profile will be further discussed in Section \ref{outflow_activity}.

For G19.88, the relatively optically thin H$^{13}$CO$^{+}$ (1 - 0) spectrum shows a single peak profile with the peak consistent with the LSR velocity, whereas, the HCO$^{+}$ spectrum of G19.88 is double-peaked and marginally blue-skewed with the dip at the systemic velocity where the single component $\rm H^{13}CO^{+}$ line peaks (refer Figure \ref{1242+1988_hco+h13co_3sigma}(b)). Outflow wings are seen in both HCO$^{+}$ and $\rm H^{13}CO^{+}$ transitions. The SiO spectrum (Figure \ref{1242+1988_sio_3sigma}(b)) also displays the double-Gaussian profile as discussed above.

To probe the gas kinematics at the core level, we extract the $\rm HCO^+$ and $\rm H^{13}CO^+$ spectra over circular areas of diameter $\sim$2.5~arcsec (typical beam size of the molecular line data) that adequately covers the detected core apertures in both complexes. 
For G12.42, H$^{13}$CO$^{+}$ and HCO$^{+}$ spectra extracted over an area enclosing MM1, MM2, MM3, MM4, MM5, MM6, and MM7 are shown in Figure \ref{1242_allcores_hco_h13co}(a) - (g), respectively. The H$^{13}$CO$^{+}$ spectra covering cores MM1, MM2 and MM3 display single peaked profiles, whereas the HCO$^{+}$ profiles are blue-skewed, double-peaked with the dip consistent with the peak of H$^{13}$CO$^{+}$ (1-0) emission. Prominent blue line wings are discernible in the HCO$^{+}$ line profiles of MM1 and MM2. For MM3, the blue component is relatively broader, and the outflow wing is marginally visible.
Spectra for the other cores show complex, double-peaked profiles. For the cores MM6 and MM7, the two observed components of HCO$^{+}$ and H$^{13}$CO$^{+}$ do not peak at the same velocity. The peak of H$^{13}$CO$^{+}$ lies in between and close to the dip seen in the HCO$^{+}$ spectra. This could be a possible infall signature.
Given the beam size ($\sim$2.5 arcsec) of the molecular line data, it is difficult to extract the spectra over individual cores in G19.88 as the regions would overlap. Hence, for this protocluster, we extract H$^{13}$CO$^{+}$ and HCO$^{+}$ spectra over apertures covering MM5-MM6, MM1-MM3-MM4, and MM2 which are shown in Figure \ref{1988_allcores_hco_h13co}(a), (b), and (c), respectively. 
In the spectra covering cores MM5 and MM6, the HCO$^{+}$ line profile seem to be a blue-skewed, double-profile though the red component is not very prominent. The H$^{13}$CO$^{+}$ spectrum shows single profile with a bump on the red side. The possibility of this being another component cannot be ruled out. The spectrum for the region covering MM1-MM3-MM4-MM7 shows a blue-skewed, double-peaked HCO$^{+}$ line with a single peaked H$^{13}$CO$^{+}$ profile, the peak of which coincides with the dip seen in the other and matches with the systemic velocity. The line profiles for MM2 are single peaked. All regions show strong signatures of outflow wings, being most pronounced for MM2. Higher resolution molecular line data are essential to probe the kinematics of the individual cores in G19.88.

\section{Gravitational stability and global collapse scenario}
\label{grav_stab_global_coll}
In this section, we investigate the gravitational stability of the two massive, protostellar clumps under study and search for kinematic signatures, if any, of global infall. First, we determine whether the clumps are gravitionally bound. 
For this, we calculate the virial parameter, $\rm \alpha_{vir}$ which is the ratio of the virial mass, $M_{\rm vir}$, to the mass of the clump considered. $M_{\rm vir}$ of a clump is defined as the amount of mass that can be supported against self-gravity by both thermal and turbulent pressure. This is given by  \citep{2016MNRAS.456.2041C}
\begin{equation}
M_{\rm vir} = \frac{5\ r\ \Delta V^2}{8\ {\rm ln}(2)\ a_1\ a_2\ G} \sim 209\ \frac{1}{a_1\ a_2} \left(\frac{\Delta V}{\rm km\ s^{-1}} \right)^2\ \left(\frac{r}{\rm pc}\right) \rm M_{\odot}
\label{M_vir}
\end{equation}
Here, 
$\Delta V$ is the line width of an optically thin line, 
$r$ is the effective radius of the clump, the constant $a_1$ accounts for the correction for power-law density distribution, and is given by $ a_1 = (1-p/3)/(1-2p/5)$, for $ p < 2.5$ \citep{1992ApJ...395..140B} where we adopt $p=1.8$ \citep{2016MNRAS.456.2041C}. The constant $a_2$ takes into account the shape of the clump which we assume here to be spherical and consider $a_2=1$. 
For $\Delta V$, we use the line width of $\rm H^{13}CO^+(1-0)$ line. The $\rm H^{13}CO^+(1-0)$ line is generally found to be optically thin in Infrared dark clouds (IRDCs) and clumps \citep[e.g.][]{2012ApJ...756...60S}. However, in the densest part of clumps (i.e., in the cores), this may not always be the case. Following the approach outlined in  \citet{{2020ApJ...901...31L},{HLLiu22a}} and under the assumptions of local thermodynamic equilibrium, we calculate the optical depth of the detected cores from the ratio of the peak intensities of H$^{13}$CO$^{+}$ and HCO$^{+}$ emission (see Eq. 3 of \citet{2020ApJ...901...31L}). 
The isotopic abundance ratio, [$\rm HCO^+$/$\rm H^{13}CO^+$], is calculated to be 52 and 50 for G12.42 and G19.88, respectively, 
using the relation of $\rm ^{12}CO/^{13}CO$ as a function of the Galactocentric distance \citep{2013A&A...554A.103P}. Considering the peak intensities and the above ratios, the optical depth values are estimated to lie in the range of 0.1-0.7 and 0.1-0.2, for G12.42+0.50 and G19.88-0.53, respectively. This justifies the optically thin assumption for the H$^{13}$CO$^{+}$ line emission in both regions.
\begin{table}
\caption{3~mm clump parameters}
\centering
\begin{tabular}{c c c c c c}
 \hline
Clump & Line width ($\Delta V$) & Radius ($r$) &  $M_{\rm  cl-3mm}$ & $M_{\rm   vir}$ &$\alpha_{\rm  vir}$\\[0.5ex] 
& $\rm km\,s^{-1}$ & (pc) & ($\rm  M_{\odot}$) & ($\rm  M_{\odot}$) & \\[0.5ex] 
\hline
G12.42 & 1.8 & 0.3 &  478 & 126 & 0.3 \\
\hline
G19.88 & 3.0 & 0.4 &  1078 & 475 & 0.4 \\
\hline
\end{tabular}
\label{vir_par}
\end{table}
Considering the ATOMS clump, the effective radii of the clumps, harbouring the identified cores in these two regions, are estimated from the area covering emission above the 3$\sigma$ level in low-resolution 7-m array maps. Further, integrating the fluxes within this area, the masses are determined from Eq. \ref{M_dust}. Here, we have considered mean temperatures of 25~K (Issac19) and 18.6~K (Issac20) for G12.42 and G19.88, respectively. 
The $\rm H^{13}CO^+$ line widths are estimated by Gaussian profile fitting to the spectra presented in Figures \ref{1242+1988_hco+h13co_3sigma}(a) and (b).
For G12.42, where two components are seen, the average of the FWHM obtained by fitting individually to the components is considered. The estimated H$^{13}$CO$^{+}$ line widths, effective radii and mass of the clumps using 3~mm data, $ M_{\rm cl-3mm}$, virial mass, $M_{\rm vir}$, and the virial parameter, $\alpha_{\rm vir} = M_{\rm vir}/M_{\rm cl-3mm}$ are listed in Table \ref{vir_par}. Similar values are obtained if one considers the mass and radius of the associated larger ATLASGAL clumps (Issac19; Issac20).

The virial parameter enables us to understand the stability of the protostellar clumps. Clumps are gravitationally bound and unstable
to collapse if they are supercritical with $\alpha_{\rm vir} <<  \alpha_{\rm cr}$, $\alpha_{\rm cr} $ being a critical virial parameter. If we consider non-magnetized clumps then $\alpha_{\rm cr} \sim 2$ whereas, $\alpha_{\rm cr} << 2$ is possible in the presence of strong magnetic support. For both the protostellar clumps
we estimate $\alpha_{\rm vir} \sim 0.3 - 0.4 $.  
The actual values could be still lower since the presence of infall/outflow in these regions would result in broadened line profiles thus leading to overestimation of the parameter. Such low values of the virial parameter are generally observed towards high-mass star-forming clumps \citep[e.g.][]{{2011A&A...530A.118P},{2013ApJ...779..185K},{2016MNRAS.456.2041C},{2019ApJ...878...10T}}. For G12.42 and G19.88, these estimated low values indicate that these clumps are supercritical and are undergoing strong gravitational collapse if no other supporting mechanism (e.g. magnetic field) exists. Magnetic field effects, if present, have not been considered in the estimation of $\alpha_{\rm vir}$ as no magnetic field study is available for these protoclusters in literature. Magnetic field plays an important role and if there exists additional magnetic support then including the Alfv\'en velocity term would give a larger value for $\alpha_{\rm vir}$ (see Eq. 9 of \citet{2019ApJ...878...10T}).

Several studies have used optically thick molecular line transitions to decipher global gas dynamics in star-forming clumps \citep[e.g.][]{{2015MNRAS.450.1926H},{2012A&A...538A.140K}}. HCO$^{+}$ is an excellent dense gas tracer that is commonly used to reveal global infall signatures. This constitutes a double-peaked, blue asymmetrical line profile with a self-absorption dip at the systemic velocity. To rule out multiple components along the line-of-sight, for a bona fide infall candidate, the relatively less optically thick isotopologue, H$^{13}$CO$^{+}$, is also required to be single peaked coinciding with the above self-absorption dip. 
G12.42 has been identified as an infall candidate by \citet{2015MNRAS.450.1926H} based on the line profiles as discussed above. This inference was also confirmed by Issac19. Both these studies are based on single dish MALT90 molecular line data. Issac19 used the optically thin H$^{13}$CO$^{+}(1-0)$ and the optically thick HCO$^{+} (1-0)$ lines to trace infall. Using similar approach and another transition ($\rm J = 4 - 3$) of the same pair of lines, no infall is inferred for the G19.88 clump \citep{2012A&A...538A.140K}. This is consistent with the conclusion arrived at by another study conducted by \citet{2016ApJS..225...21J} using the HCN and HNC molecular line transitions. Both these results are again based on single dish observations. 
 
\begin{figure*}
     \centering
     \begin{subfigure}[b]{0.48\textwidth}
         \centering
         \includegraphics[width=\textwidth]{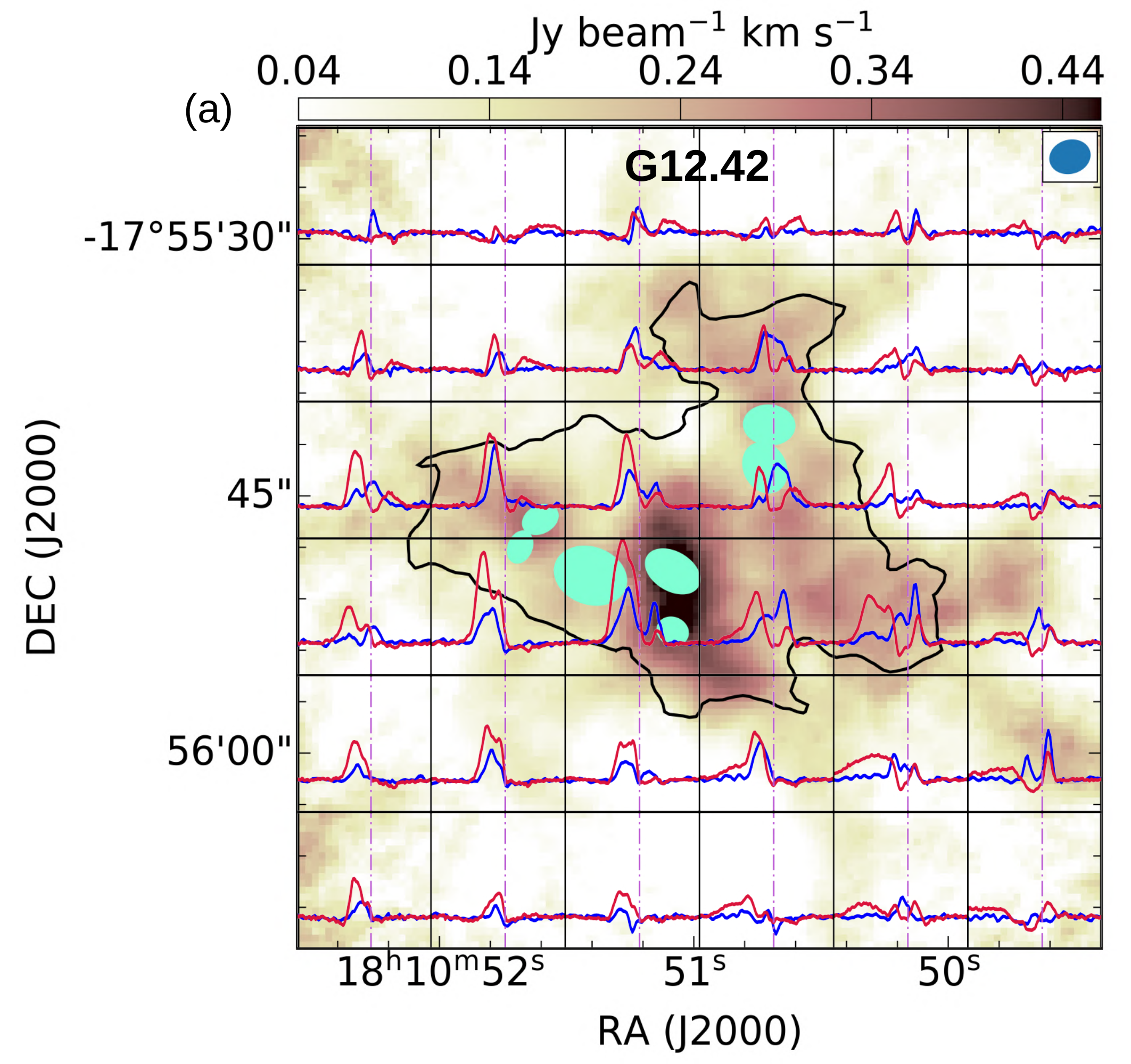}
     \end{subfigure}
     \hfill
     \begin{subfigure}[b]{0.48\textwidth}
         \centering
         \includegraphics[width=\textwidth]{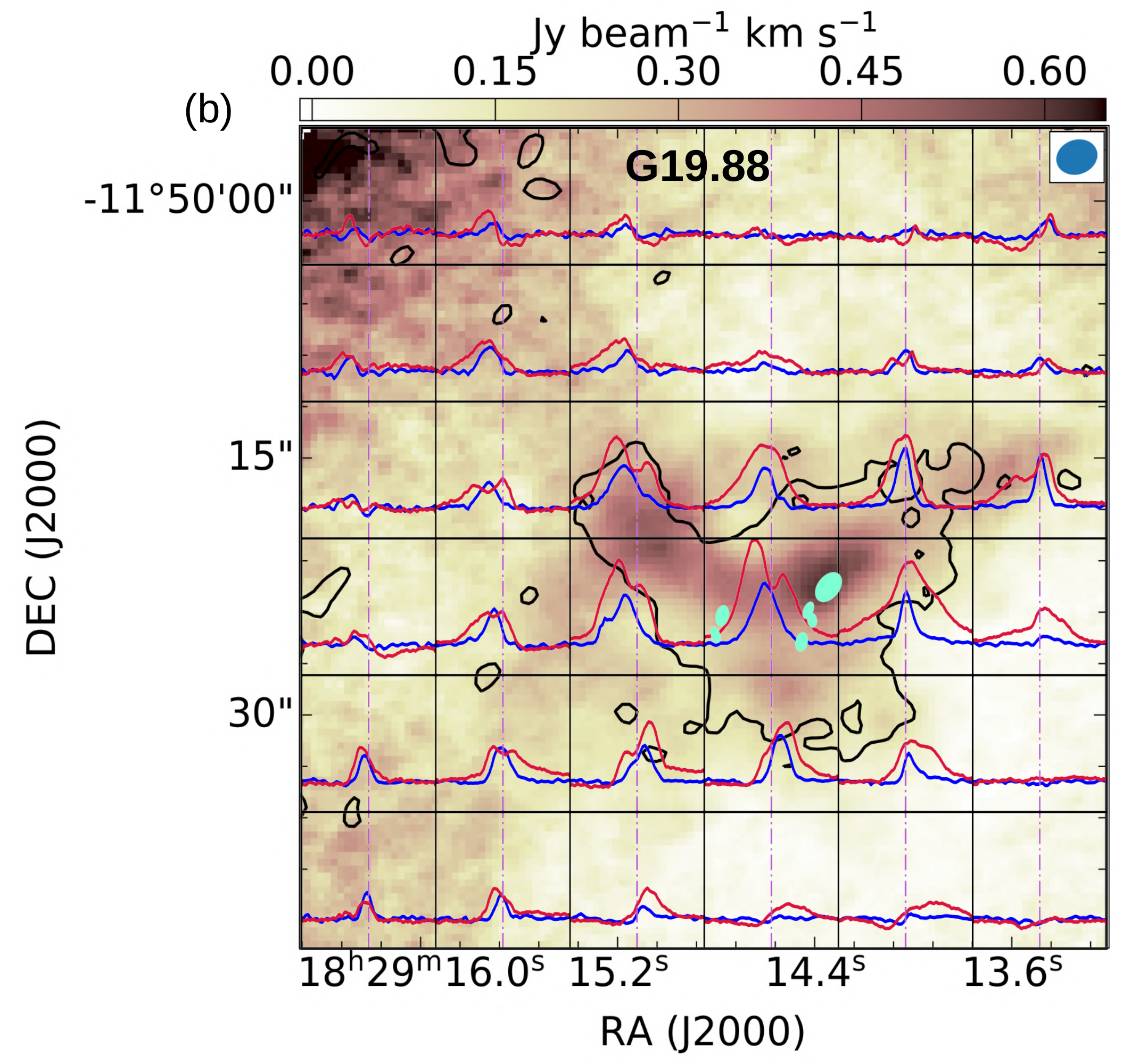}
     \end{subfigure}
     \hfill
        \caption{(a) The colour scale shows the moment 0 map of H$^{13}$CO$^{+}$ for G12.42+0.50. The 3$\sigma$ contours of the 3~mm emission (combined data) is shown in black. The HCO$^{+}$ spectrum shown in red and the H$^{13}$CO$^{+}$ spectrum shown in blue are overlaid. The grid size is $\rm 8 arcsec \times 8 arcsec$ ($\rm 0.09~pc \times 0.09~pc$ at the distance to the source). The velocity range in each  grid is (10$\,\rm km\,s^{-1}$, 25$\,\rm km\,s^{-1}$). (b) Same as (a) but for G19.88-0.53. The velocity range in each  grid is (38$\,\rm km\,s^{-1}$, 50$\,\rm km\,s^{-1}$). The grid size is $\rm 8 arcsec \times 8 arcsec$ ($\rm 0.13~pc \times 0.13~pc$ at the distance to the source). In both the panels the dashed vertical lines indicate the LSR velocity, H$^{13}$CO$^{+}$ spectra are scaled-up by a factor of 3 in all the grids, and filled ellipses represent the identified cores towards the star-forming regions. Beam sizes are indicated at the top right in each figure.}
        \label{grid_plot}
\end{figure*}

We revisit the infall analysis with high-resolution ATOMS-ALMA observation using HCO$^{+}(1-0)$ and H$^{13}$CO$^{+}(1-0)$ lines. 
Figures \ref{1242+1988_hco+h13co_3sigma}(a) and (b) show these line transitions for G12.42 and G19.88, respectively. Considering G12.42, the HCO$^{+}(1-0)$ spectrum shows a double-peaked, blue-skewed profile consistent with the MALT90 spectrum. However, the optically thin emission of H$^{13}$CO$^{+}(1-0)$ also displays a double peak implying two line-of-sight velocity components. To investigate further,  we construct a grid map that is shown in Figure \ref{grid_plot}. Each grid has an area of 8~arcsec $\times$ 8~arcsec and HCO$^{+}$ (red) and H$^{13}$CO$^{+}$ (blue) spectra averaged over each grid area are shown. The moment 0 map of H$^{13}$CO$^{+}$ is shown in colour scale. Both the lines show double peaked profiles in most of the grids though as previously seen, several individual cores display infall signature. Although the virial analysis suggests global collapse, it seems highly unlikely that the blue-skewed clump-averaged profile of HCO$^{+}$ from single-dish data indicates infall. It is possible that the poor signal to noise ratio ($\sim 2$) of the MALT90 H$^{13}$CO$^{+}$ spectrum hindered the identification of two peaks. 
Similar results using the ATOMS survey data have been discussed in \citet{2021MNRAS.tmp.2557Z} for the filamentary complex in G286.21+0.17, previously known to be an infall candidate from single dish data. These authors have invoked the presence of two sub-clumps (in relative motion) to explain the observed line profiles. Rotation of the clump, if present, would manifest as blue asymmetric line profiles on one side of the rotation axis and red asymmetry on the other side. In our case, the grid map shows no such signature and hence rotation could be ruled out.

 However, in case of G19.88, the ATOMS-ALMA clump-averaged HCO$^{+}(1-0)$ spectrum displays a distinct, albeit weak, blue-skewed, double-peaked profile while the optically thin  H$^{13}$CO$^{+}$ line is single peaked at the systemic velocity (see Figure \ref{1242+1988_hco+h13co_3sigma}(b)). The grid map (see Figure \ref{grid_plot}(b)) also shows single line profiles with typical infall signatures towards the centre where the cores are located. To quantify the blue-asymmetry of the HCO$^{+}(1-0)$ line profile, we calculate the asymmetry parameter, $A$, which is given by \citep{1997ApJ...489..719M}.
 \begin{equation}
     A = \dfrac{(V_{\rm thick} - V_{\rm  thin})}{\Delta V_{\rm thin}},
\label{delta_V}
 \end{equation}
where $V_{\rm thick}$ denotes the peak velocity of the optically thick line; $V_{\rm thin}$ and $\Delta V_{\rm thin}$ are the peak velocity and FWHM of the optically thin line. To calculate $A$, we used the value of $V_{\rm thick}$ = 43.09 $\rm km\,s^{-1}$, the velocity corresponding to the peak of blue component of HCO$^{+}(1-0)$ line profile, $V_{\rm thin}$ = 44.0 $\rm km\,s^{-1}$, and $\Delta V_{\rm thin}$ = 3.0 $\rm km\,s^{-1}$, which are the peak velocity and FWHM of H$^{13}$CO$^{+}(1-0)$ spectrum, respectively. We estimate $A$ to be $-0.3$ for the clump, which 
satisfies the adopted threshold value of $A < -0.25$ for a genuine blue-skewed profile \citep{1997ApJ...489..719M}. The value that we obtained may be an overestimate given the co-existence of infall and outflow activity. These would considerably increase the velocity dispersion, and thereby decrease the value of $A$. 

Summarizing the above analysis, we conclude that both clumps associated with G12.42 and G19.88 are gravitationally bound and likely to be undergoing an overall collapse. The clump-averaged, double-peaked, blue-skewed HCO$^{+}(1-0)$ line profile for G12.42 is shown to be not an infall signature but arising due to two velocity components. Several individual cores in the protocluster, however, show distinct infall profiles. The velocity components seen could be a complex combination of large-scale inflow and outflow present in this star-forming complex which will be discussed in a later section. In comparison, for the protocluster, G19.88, clear infall signature can be inferred from the clump-averaged spectra and towards the central cores in the grid map.
Furthermore, in the framework of the {\it global hierarchical collapse} (GHC) theory of massive star formation, the absence of the typical infall signature in the line profiles does not preclude collapse \citep{2019MNRAS.490.3061V}. The accepted infall signature of blue-skewed, double peaked line profiles assume roughly spherical collapse, whereas, under the GHC scenario collapse flows are not isotropic and mostly sets in as longitudinal flows along the filaments. The presence of filaments at various spatial scales in both protoclusters (refer Section \ref{dyn-SF}) suggest the likelihood of longitudinal flows and collapse along the filaments and hence could possibly explain the absence of the characteristic infall line profile in G12.42.

\begin{figure*}
     \centering
     \begin{subfigure}[b]{0.48\textwidth}
         \centering
         \includegraphics[width=\textwidth]{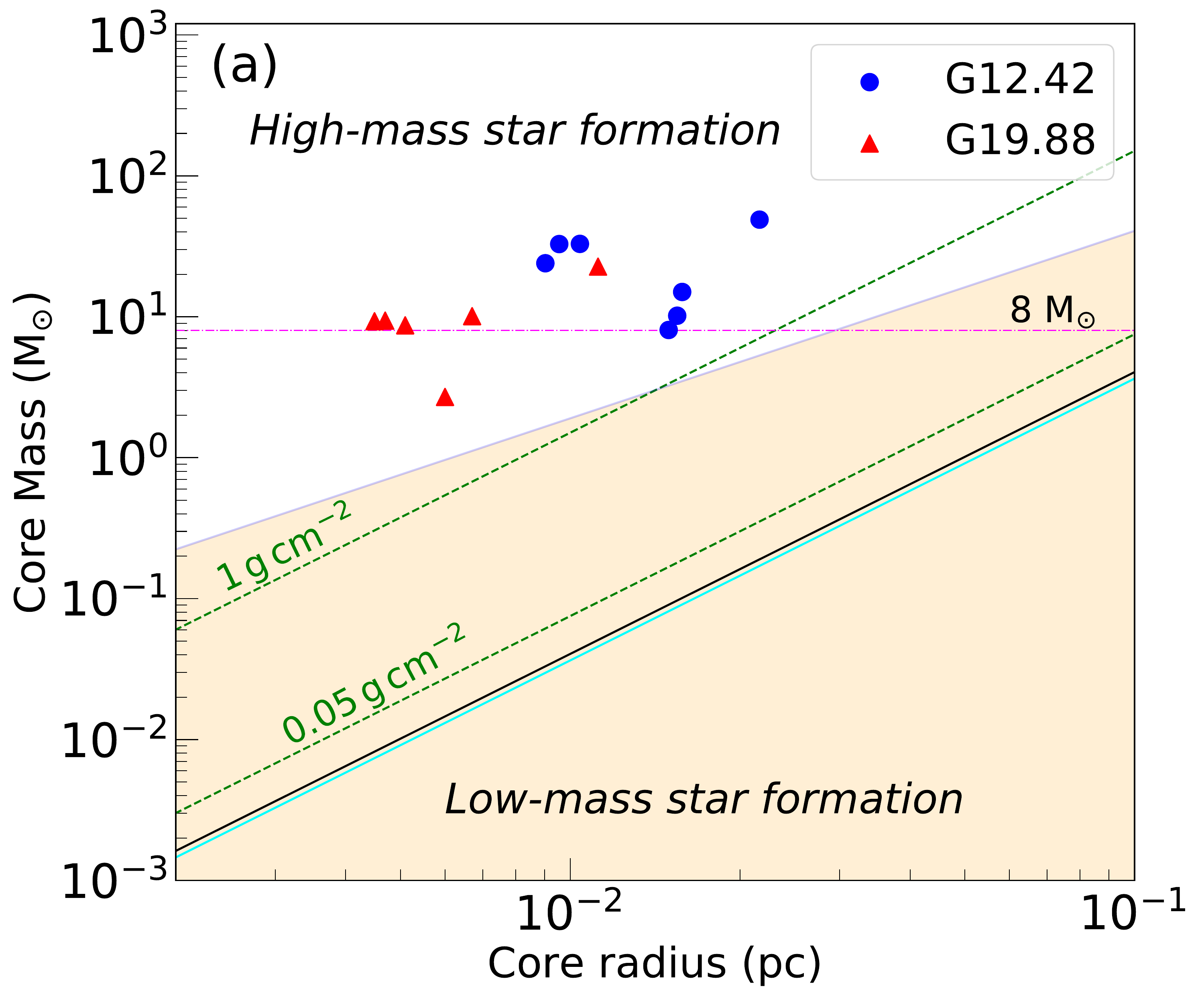}
     \end{subfigure}
     \hfill
     \begin{subfigure}[b]{0.48\textwidth}
         \centering
         \includegraphics[width=\textwidth]{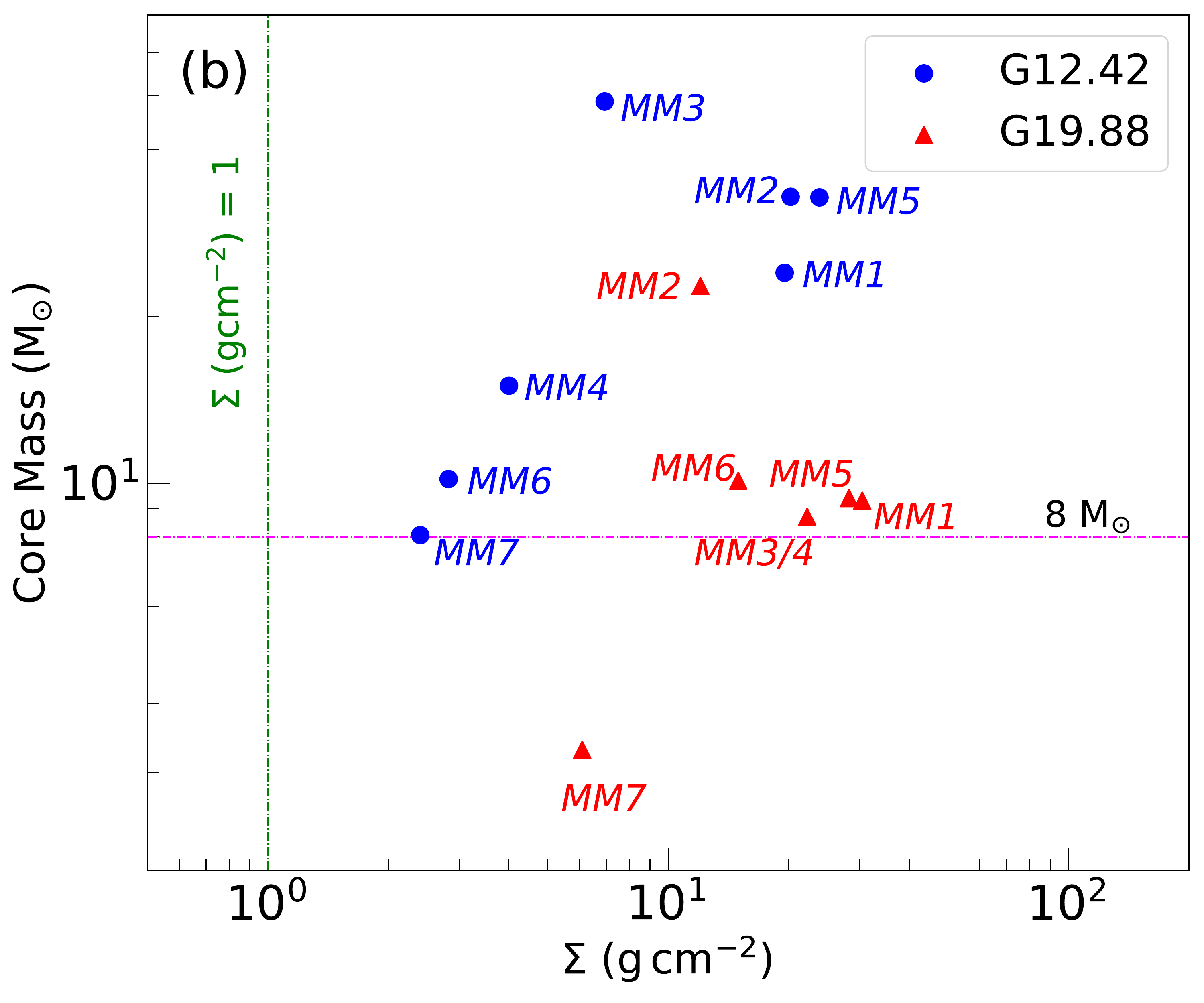}
     \end{subfigure}
     \hfill
        \caption{(a) The mass of the compact cores, identiﬁed using the 3 mm and 2.7 mm maps of G12.42+0.50, and G19.88-0.53, respectively, is plotted as a function of the radius of the cores, and depicted by circles. The cyan and black lines indicate the surface density thresholds of 116 $\rm M_\odot$pc$^{-2}$ ($\sim$ 0.024 \rm g cm$^{-2}$ ) and 129 $\rm M_\odot$pc$^{-2}$ ($\sim$ 0.027 g cm$^{-2}$ ) for active star formation from \citet{2010ApJ...724..687L} and \citet{2010ApJ...723.1019H}, respectively. The shaded region delineates the low-mass star-forming region that does not satisfy the criterion $m(r)$ $>$ 870 $\rm M_\odot$($r$/\rm pc)$^{1.33}$ \citep{2010ApJ...716..433K}.The green dashed lines represent the surface density threshold of 0.05 and 1 g cm$^{-2}$ deﬁned by \citet{2008Natur.451.1082K} and \citet{2014MNRAS.443.1555U}, respectively. (b) Core mass as a function of core mass surface density is shown. The green vertical line indicates a theoretical threshold of 1 g cm$^{-2}$, above which the cores most likely form high-mass stars. The horizontal magenta line in both plots indicates a core mass of 8 $\rm M_\odot$.}
        \label{mass_rad_sdensity}
\end{figure*}
\begin{table*}
\caption{Comparison between observed parameters and estimated Jeans parameters for G12.42 and G19.88. The values in parenthesis denote the median values.}
\centering
\begin{tabular}{c c c c c c c c c }
 \hline
Clump name & $T_{\rm kin}$ & $\rho_{\rm eff}$ & $M^{\rm Th}_{\rm Jeans}$ & $\lambda^{\rm Th}_{\rm Jeans}$ & $M^{\rm turb}_{\rm Jeans}$ & $\lambda^{\rm turb}_{\rm Jeans}$ & $M^{\rm core}_{\rm obs}$ & $\lambda^{\rm core}_{\rm obs}$ \\ [0.5ex] 
 & (K) & (g cm$^{-3}$) & (M$_\odot$) & (pc) & (M$_\odot$) & (pc) & (M$_\odot$) & (pc) \\[0.5ex] 
 \hline
 G12.42 & 24.7 & 2.4$\times$10$^{-20}$ & 14.1 & 0.4 & 162 & 1.0 & 8$-$49 (24) & 0.04$-$0.3 (0.2)\\
 G19.88 & 25.8 & 5.2$\times$10$^{-20}$ & 10.1 & 0.3 & 191 & 0.8 & 3$-$23 (9) & 0.02$-$0.06 (0.03)\\
\hline
 \end{tabular}
\label{Table:Jeans}
\end{table*}
\section{Fragmentation}
\label{frag_Section}
Understanding cloud and clump scale fragmentation is important for deciphering the initial conditions of star formation. The evidence of fragmentation in both protoclusters under study is revealed from the ALMA continuum maps. Thus it is pertinent to investigate the driving mechanism of fragmentation in these regions and to understand whether core formation is governed by the Jeans-type fragmentation process described in \citet{1990sse..book.....K}. 

If the fragmentation is driven by Jeans instability, then the self-gravitating clump is expected to fragment into cores characterized by Jeans mass ($M_{\rm Jeans}$) and Jeans length ($\lambda_{\rm Jeans}$) which are defined as 
\citep{{2014MNRAS.439.3275W},{2014ApJ...785...42P}},
\begin{equation}
    M_{\rm Jeans}= \dfrac{\pi^{5/2}\,c^{3}_{\rm eff}}{6\sqrt{G^{3}\,\rho_{\rm eff}}}
\end{equation}
and
\begin{equation}
    \lambda_{\rm Jeans}=c_{\rm eff} \left ( \dfrac{\pi}{{G\rho_{\rm eff}}} \right)^{1/2},
\end{equation}
where $\rm \rho_{eff}$ is the volume density and $c_{\rm eff}$ is the effective speed of sound. When only thermal support is considered, $c_{\rm eff}$ will be the thermal velocity dispersion ($\sigma_{\rm th}$), given by
\begin{equation}
\sigma_{\rm th} = \sqrt{\dfrac{k_{\rm B} T_{\rm kin}}{\mu\, {m}_{\rm H}}},
\label{sigma_th_equ}
\end{equation}
where $k_{\rm B}$, $T_{\rm kin}$, ${m}_{\rm H}$, and $\mu$ are the Boltzmann constant, the kinetic temperature,  the mass of hydrogen atom and the mean molecular weight per free particle, respectively. We have considered $\mu$ = 2.37 \citep{{2008A&A...487..993K},{2019ApJ...886..102S},{2014MNRAS.439.3275W},{2021arXiv211113869P}}. 

Massive clumps are mostly associated with non-thermal motions dominated by turbulence. In such cases where both thermal and non-thermal (turbulent) support are involved, $c_{\rm eff}$ is replaced by the total velocity dispersion ($\sigma_{\rm tot}$) that includes the non-thermal velocity dispersion of the observed lines ($\rm \sigma_{nt}$) and the thermal velocity dispersion ($\sigma_{\rm th}$) and is expressed as \citep{2017ApJ...849...25L,HLLiu19}
\begin{equation}
  \sigma_{\rm  tot} = (\sigma_{\rm  nt}^2 + \sigma_{\rm th}^2)^{1/2}
\end{equation}
and
\begin{equation}
\sigma_{\rm nt} = \sqrt{(\sigma_{\rm line})^{2} - \dfrac{k_{\rm B}T_{\rm kin}}{m_{\rm species}}},
\label{sigma_nt}
\end{equation}
where $m_{\rm species}$ is the mass of the molecule and $\sigma_{\rm line}$ is the velocity dispersion measured from the spectra of the molecule used.
For interpreting the clump to core fragmentation, we consider the larger ATLASGAL clump since filtering out of emission in the ATOMS 3~mm clump would render the estimate of mass density inaccurate.
Using NH$_3$(1,1) to (3,3) inversion transitions, \citet{2012A&A...544A.146W} derived the average kinetic temperature of 24.7~K and 25.8~K for the clumps associated with G12.42 and G19.88, respectively. We adopted these as the kinetic temperatures of the clumps \citep[][Table 3]{2012A&A...544A.146W} for calculating $\sigma_{\rm th}$. In our calculations, we use the value of mass, and radius of the clumps from Issac19 and Issac20 for G12.42 and G19.88, respectively. These authors have used $\beta =2$ for calculation of masses. To be consistent with the values used in this work, we have recalculated the clump masses with $\beta = 1.75$.
Using $T_{\rm kin}$ = 24.7~K , $M_{\rm clump}$ = 764~$\rm M_\odot$ and $R_{\rm clump}$ = 0.8~pc, we estimate $c_{\rm eff}$ and $\rho_{\rm eff}$ to be 0.29$\,\rm km\,s^{-1}$ and 2.4$\times$10$^{-20}$\rm ~g cm$^{-3}$, respectively for G12.42. Similarly, for G19.88, using $T_{\rm kin}$ = 25.8~K, $M_{\rm clump}$ = 1098~$\rm M_\odot$ and $R_{\rm clump}$ = 0.7~pc, we obtain $c_{\rm eff}$ and $\rho_{\rm eff}$ as 0.3$\,\rm km\,s^{-1}$ and 5.2$\times$10$^{-20}$\rm g cm$^{-3}$, respectively. The non-thermal component, $\sigma_{\rm nt}$, is determined using velocity dispersion ($\sigma_{\rm line}$) measured from the H$^{13}$CO+ spectra. To avoid the influence of current star formation feedback, the velocity dispersions for the fragmentation analysis are estimated from spectra extracted offset from the core cluster and outflow regions. For G12.42 and G19.88, $\sigma_{\rm line}$ is estimated to be 0.6 and 0.8$\,\rm km\,s^{-1}$, respectively.

To understand the driving mechanism of clump to core fragmentation in these two protoclusters, we need to compare the observed masses of the cores and the separation between them with the predicted Jeans parameters. To compute the core separation, we use the minimum  spanning tree (MST) method \citep[e.g.][]{1985MNRAS.216...17B}. The MST algorithm connects a set of nodes with straight lines and calculates the minimum sum of line lengths that connect nodes without any loop in a graph. In our case, the cores are considered as nodes, and the line lengths are the projected spatial separations between the cores. Table \ref{Table:Jeans} summarizes the derived thermal and turbulent Jeans parameters and also lists the range and median values of the observed core masses and separation. 
It is to be kept in mind that the projected separation is on average smaller than the actual separation by a factor of 2/$\pi$ \citep{2019ApJ...886..102S} due to projection effect.

The above analysis enables us to interpret the fragmentation scenario. Similar trend is seen for both protoclusters where the ranges of observed parameters are fairly close to the predicted thermal Jeans parameters. For G12.42, the medians of the observed mass and core separation are within a factor of $\lesssim2$ of the predicted values. In case of G19.88, the median of the observed mass is similar whereas the observed core separation is a factor of 10 smaller. In comparison, the predicted parameters after including turbulent pressure are much larger than the observed values. For G12.42, the predicted turbulent Jeans length (1.0~pc) and mass (162~M$_\odot$) are larger by a factor of $\sim$5 and $\sim$7, respectively, than the median of the observed values. In case of G19.88, the predicted turbulent Jeans length (0.8~pc) and mass (191~M$_\odot$) are larger than the observed values by a factor of $\sim$27 and $\sim$21, respectively. The above analysis suggests that thermal Jeans instability can adequately explain the fragmentation of the massive clumps to cores in both the protoclusters. These results are in good agreement with the results of several statistical and dedicated studies \citep[e.g.][]{{2019ApJ...886..102S},{2021arXiv211113869P},{2013ApJ...762..120P},{2018A&A...617A.100B},{2017ApJ...849...25L}} towards protoclusters where thermal pressure is inferred to be dominant in the hierarchical fragmentation process. Fragmentation where turbulence dominates over thermal pressure has also been inferred in several complexes. In a recent paper based on ATOMS data, \citet{HLLiu22a} discuss the fragmentation process in the filamentary IRDC G034.43+00.24. Consistent with the results presented in several studies \citep[e.g][]{{2009ApJ...696..268Z},{2015ApJ...804..141Z},{2011ApJ...735...64W},{2014MNRAS.439.3275W},{2017ApJ...849...25L}}, the observed hierarchical fragmentation in G034.43+00.24 is seen to be driven by turbulence with the magnetic field also playing a crucial role. 

Of the detected cores, approximately 65 per cent have masses around or below the thermal Jeans mass. The remaining are `super-Jeans' cores likely in different evolutionary phases (Issac19; Issac20 and this current work) ranging from high-mass protostellar objects driving outflows/jets to HMCs to possible UCHII regions. Hence, the more evolved ones could have amassed additional material in their accretion phase.
It is worthwhile to scrutinize these `super-Jeans' cores at higher-resolution (e.g., a few hundred AUs) to investigate if they remain as single cores. If any, the role of turbulence and/or magnetic fields becomes important to explain their stability against fragmentation.

To further understand the nature of the detected cores, we present in  Figure \ref{mass_rad_sdensity}(a) the core mass as a function of the radius. Several commonly adopted thresholds for massive star formation are overlaid in this plot. The identified cores in these protoclusters satisfy the mass-radius requirement ($m(r)$ $>$ 870 $\rm M_\odot$ ${(r/\rm pc)}^{1.33}$) defined by \citet{2010ApJ...716..433K} for high-mass star formation. The plot also displays the thresholds based on the surface mass density for formation of high-mass stars \citep{{2014MNRAS.443.1555U},{2008Natur.451.1082K}} and for efficient star formation \citep{{2010ApJ...724..687L},{2010ApJ...723.1019H}}. In Figure \ref{mass_rad_sdensity}(b), the core mass as a function of the surface mass density is plotted. Cores MM3 in G12.42 and MM2 in G19.88 are the most massive cores detected in these two protoclusters. Similarly, cores MM5 in G12.42 and MM1 in G19.88 are the densest compact cores.

We rule out the effect of mass sensitivity (refer Section \ref{result}) on the observed lack of low-mass cores.
Dearth of low-mass cores in star-forming clumps has also been reported by several authors \citep[e.g.][]{{2015ApJ...804..141Z},{2017ApJ...849...25L}}. Later formation epoch of low-mass stars and feedback from newly formed massive stars suppressing further fragmentation are some of the likely reasons discussed in these papers. In this study, collimated and large scale outflows are identified towards G12.42 and G19.88 (Section \ref{outflow_activity}). Further, Issac19 suggest a co-existence of the UC H\,{\small II} region with an ionised jet in the G12.42 complex. Hence, the presence of multiple outflows and an UC H\,{\small II} region would have possibly inhibited further fragmentation resulting in the observed lack of low-mass protostellar cores. 
%
\begin{figure*}
    \centering
    \includegraphics[width=0.85\textwidth]{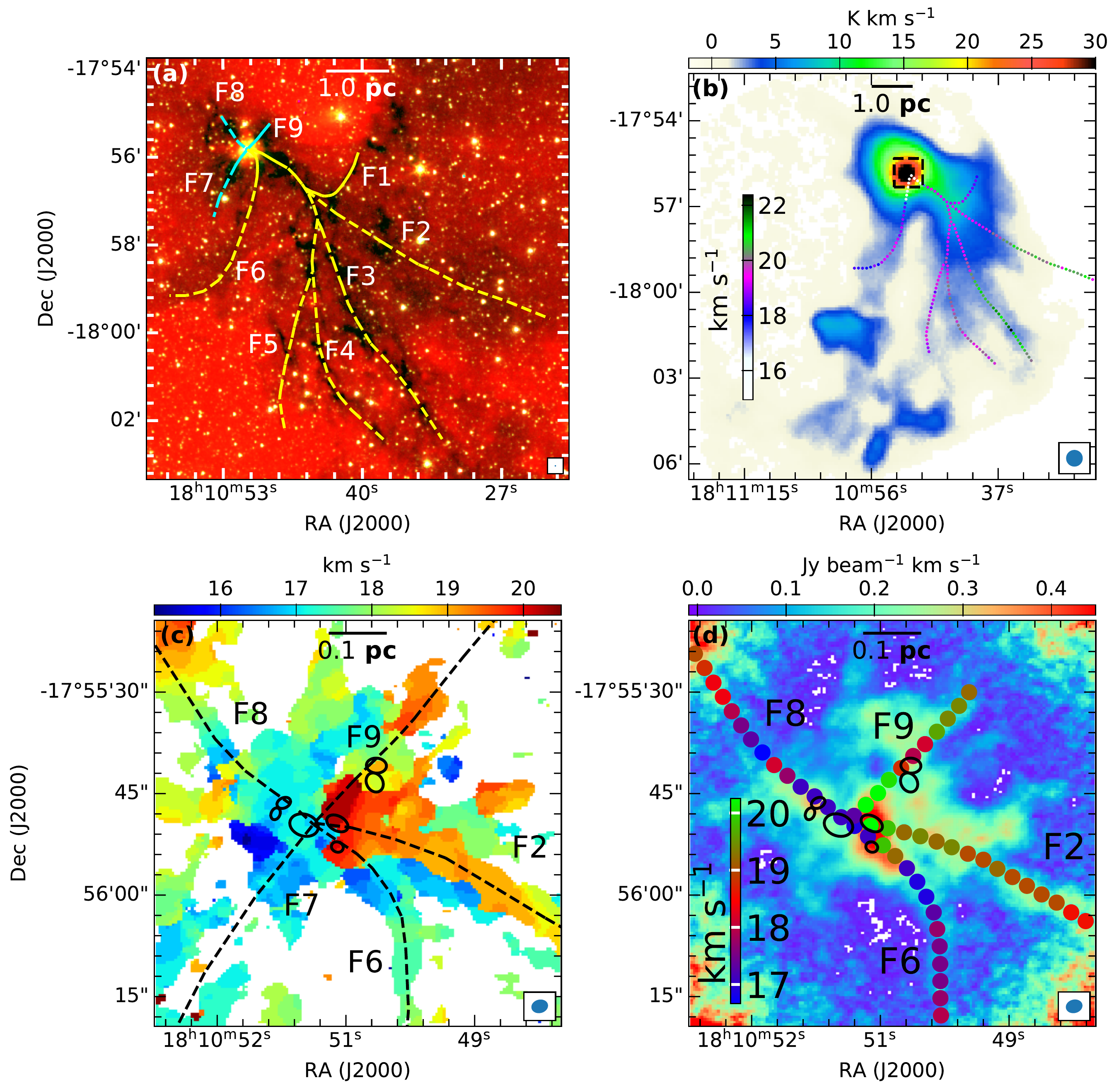}
    \caption{(a) Skeletons of the identified filaments are overlaid on the colour composite image of the region around G12.42 using \textit{Spitzer} 3.6 $\mu$m (blue), 4.5 $\mu$m (green) and 8.0 $\mu$m (red) bands. The yellow dashed curves are adopted from Issac19 and the cyan dashed curves are new filaments identified in this paper (b) Moment zero map of $\rm ^{13}$CO (3 - 2) molecular line data from JCMT is shown in colour scale. The velocity peaks of $\rm ^{12}$CO (3 - 2) adopted from Issac19 are overlaid on it. Dashed square box represents the region of the ATOMS maps shown in panel (c) and (d). (c) Peak velocity map of $\rm H^{13}CO^{+}$ (1 - 0) (obtained using  combined 12-m + 7-m data) overlaid with filaments F2, F6, F7, F8, and F9. (d) The velocity peaks of $\rm H^{13}CO^{+}$ (1 - 0) extracted along the filaments (F2, F6, F8, and F9) are overlaid on the moment zero map of $\rm H^{13}CO^{+}$ (1 - 0) for the region associated with G12.42+0.50 shown in colour scale. The identified 3~mm cores are shown as black ellipses. The beam sizes are shown at the bottom right of each figure.}
    \label{1242-fila}
\end{figure*}
%
\begin{figure}
     \centering
    \includegraphics[width=0.45\textwidth]{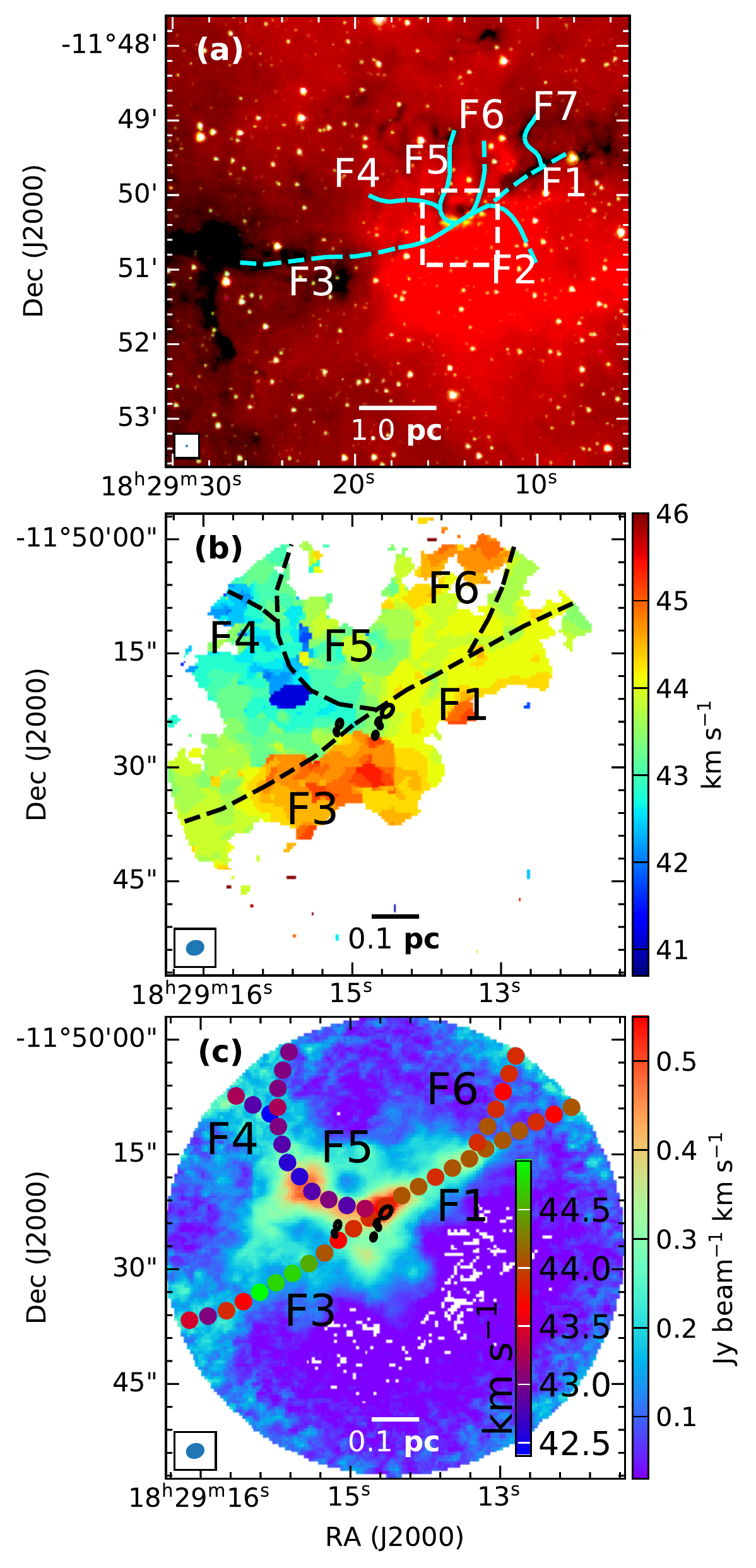}
    \caption{(a) Skeletons of the identified filaments are overlaid on the colour composite image of the region around G19.88 using \textit{Spitzer} 3.6 $\mu$m (blue), 4.5 $\mu$m (green) and 8.0 $\mu$m (red) bands. The dashed square box represents the region of the ATOMS maps shown in panel (b) and (c). (b) Peak velocity map of $\rm H^{13}CO^{+}$ (1 - 0) (obtained using  combined 12-m + 7-m data) overlaid with filaments F1, F3, F4, F5, and F6 is shown in colour scale. (c) The velocity peaks of $\rm H^{13}CO^{+}$ (1 - 0) extracted along the filaments (F1, F3, F4, F5, and F6) are overlaid on the moment zero map of $\rm H^{13}CO^{+}$ (1 - 0) for the region associated with G19.88-0.53 shown in colour scale. Ellipses represent the identified 2.7~mm cores. The beam sizes are shown at the bottom left of each figure.}
    \label{1988-fila}
\end{figure}
%
\begin{figure}
    \centering
    \includegraphics[width=0.48\textwidth]{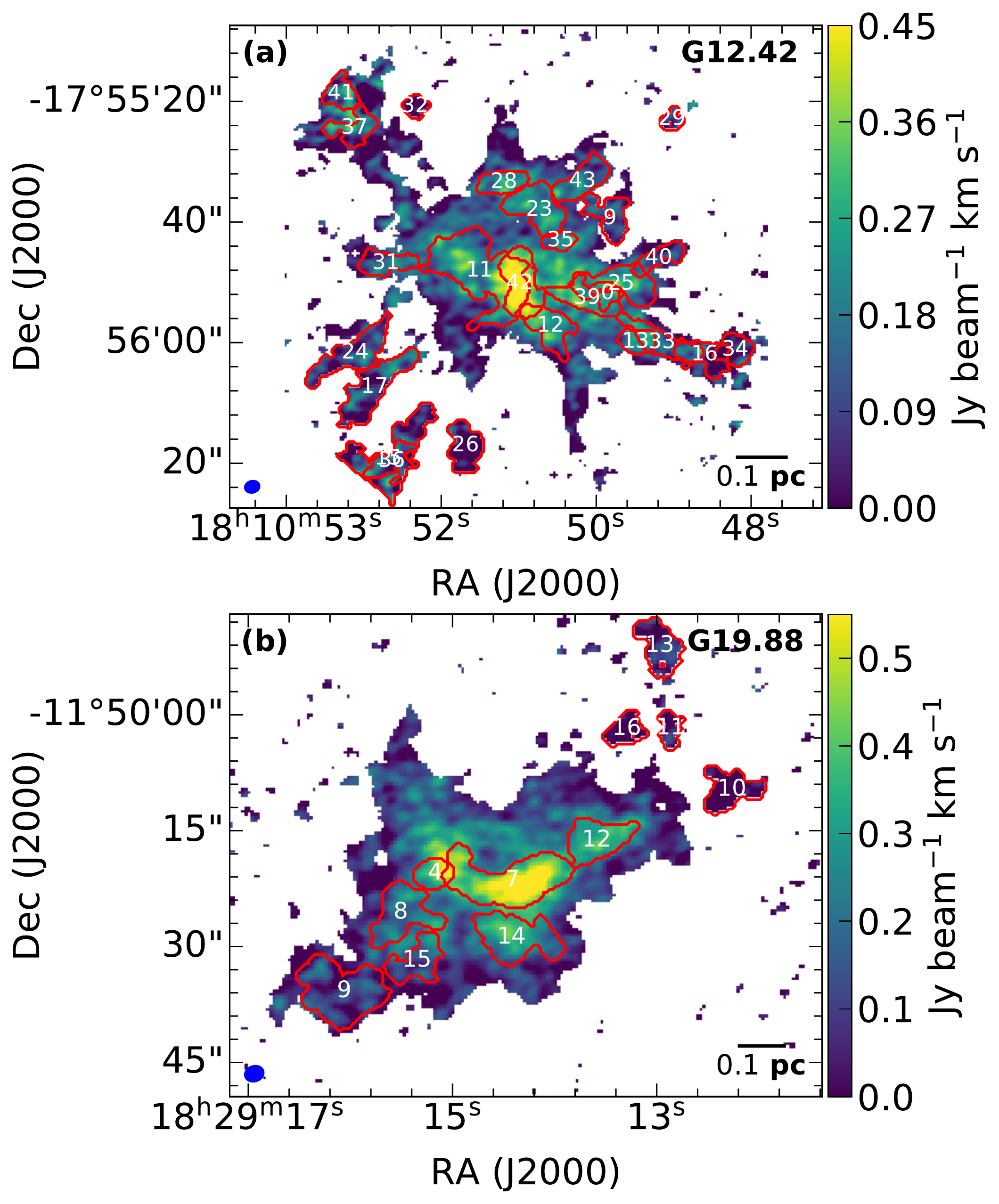}
    \caption{(a) Moment zero map of the modified $\rm H^{13}CO^{+}$ cube where only the pixels having peak intensity (along the spectral axis) greater than five times the noise are shown. The overlaid apertures are two dimensional representations of the three dimensional (PPV) leaf structures identified in G12.42. Each leaf is labelled with its structure ID (refer Table \ref{leaf_params_dendro}). (b) Same as (a) but for G19.88. The beam sizes are shown at the bottom left of each figure.}
    \label{leaf2d}
\end{figure}
%
\section{Gas kinematics in the multi-scale structures associated with the protoclusters}
\label{dyn-SF}
\subsection{Large-scale filamentary structures}
G12.42 has been proposed as a hub-filament system by Issac19 based on the morphology seen in mid-infrared {\it Spitzer} and far-infrared {\it Herschel} images. Using the 8~$\rm \mu m$ {\it Spitzer} image, where the filamentary structures appear as dark (high extinction) lanes, these authors have visually identified six large-scale filamentary structures (F1 -- F6) in the south-west merging towards the central clump. When overlaid on the 350~$\rm \mu m$ {\it Herschel} image, these identified filaments match with the bright, filamentary far-infrared emission (refer Fig. 7 of Issac19). Further, using $\rm ^{12}CO (3-2)$ molecular line data from JCMT, they also reveal bulk motion along the identified filaments and conjecture this as gas inflow. Following similar approach, we revisited the filament identification in G12.42 and detected three additional filaments (F7 -- F9).
In Figure \ref{1242-fila}, we present the morphology and gas kinematics in this hub-filament system. From Figure \ref{1242-fila}(a), two hubs with intertwining filaments can be discerned where the protocluster, G12.42, is associated with the north-east hub.
The large-scale network of filaments is also seen in the JCMT $\rm {}^{13}$CO(3-2) velocity integrated map (Figure \ref{1242-fila}(b)). For better signal-to-noise ratio, the map is convolved to a resolution of 35~arcsec. 

The velocity structure (illustrated as colour coded circles) along the identified MIR filaments by Issac19 are overplotted on the $\rm {}^{13}$CO(3-2) velocity integrated map. The circles depict the  $\rm {}^{12}$CO(3-2) peak velocities obtained within circular regions of radius 3.75~arcsec. Filamentary structures seen in the molecular line map display more extended and complex morphologies.
To avoid optical depth effects  and influence of outflows, it would have been ideal to obtain the velocity structure using the optically thin $\rm {}^{13}$CO(3-2) transition but such an analysis was not possible due to poor signal-to-noise ratio where the peak fluxes within the apertures along the filaments are less than three times the {\it rms} noise.
As illustrated by the colour coded circles, clear velocity gradients $\sim$1$\,\rm km\,s^{-1}{pc}^{-1}$ are seen along the identified filaments with decreasing velocities as one moves towards the hubs. 

The ATOMS survey has probed the inner region of the central hub in G12.42.
We investigate the gas kinematics in this region by constructing the peak velocity map using the high resolution ($\sim$2.5 arcsec) H$^{13}$CO$^{+}$(1-0) ALMA data. In doing so we have considered only the pixels where the peak intensity along the spectral axis is greater than five times \textit{rms}. The map is shown in Figure \ref{1242-fila}(c) where the identified filaments are drawn.
To further elaborate on the velocity structure, we estimate the median values of the peak velocities within circles of radius 1.25~arcsec along the filaments. These are shown in Figure \ref{1242-fila}(d) as color-coded circles overlaid on the moment zero map of H$^{13}$CO$^{+}$. The velocity structure for F7 is not plotted since the peak velocities are masked out (pixel values below five times \textit{rms}) along most part of the filament. Along the filaments F2, F8, and F9, the overall velocity gradient is $\sim$ 2$\,\rm km\,s^{-1}{pc}^{-1}$. 
Considering the filament F6, the velocity field is comparatively complex and shows variation. In the inner region, however, the gradient along the filaments F2, F6, and F9 become steeper. The velocity structure also reveals reversals, jump discontinuities which are difficult to interpret given the complex kinematics associated with the active star-forming cores.

Study of other star-forming regions, like the SDC335 cluster \citep{2013A&A...555A.112P}, Serpens South cluster \citep{2013ApJ...766..115K}, and AFL5142 \citep{2016ApJ...824...31L} has revealed evidence of velocity gradients along ﬁlaments. \citet{2013A&A...555A.112P} and \citet{2013ApJ...766..115K} interpreted velocity gradients of the order of $\sim$1$\,\rm km\,s^{-1}{pc}^{-1}$ as gas inﬂow along ﬁlaments. 
Further, the median values of the velocities along filaments F2 and F9 are $\sim19\,\rm km\,s^{-1}$ and that along F6 and F8 are $\sim17.5\,\rm km\,s^{-1}$, spanning two different ranges. This is consistent with double-peaked profiles seen in the grid presented in Section \ref{grav_stab_global_coll}.

The presence of such large-scale filaments is also seen in G19.88 as illustrated in Figure \ref{1988-fila}(a), where, following the same approach as in G12.42, we have visually identified seven filaments (F1 -- F7) that are seen intersecting at the central hub harbouring the protocluster. Proceeding along the lines discussed for G12.42, the overlay of these filaments on the ATOMS H$^{13}$CO$^{+}$(1-0) peak velocity map is shown in Figure \ref{1988-fila}(b) and the colour-coded velocity structure is shown on the H$^{13}$CO$^{+}$(1-0) moment zero map. The overall velocity gradients along the identified filaments are much lower ($< 0.5\,\rm km\,s^{-1}{pc}^{-1}$) compared to the filaments in G12.42. However, steep velocity fluctuations are seen in F3 and F5. Gas kinematics of the large-scale filaments in this protocluster could not be explored due to unavailability of good quality archival CO molecular line data.
It is worth mentioning here that in an upcoming paper \textit{(Zhou et al. 2022, under review)} of the ATOMS series, well-defined network of filaments have been identified in both these protoclusters using the FILFINDER algorithm on the moment zero maps of  H$^{13}$CO$^{+}$(1-0). These identified filaments appear to match well with the visually identified MIR filaments presented in this work. Further, the multiple outflows detected in both protoclusters (discussed in a later section) influence the kinematics of the inflowing gas along the filaments.


\subsection{Small-scale density structures}
To delve deeper and understand the dynamics that drives the star formation process in the two protoclusters, we perform a Dendrogram analysis along the lines discussed in \citet{HLLiu22b} for the filamentary cloud G034.43+00.24. Density structures at two different scales (branches and leaves) are identified using the H$^{13}$CO$^+$ data cube in position-position-velocity (PPV) space. As mentioned in Section \ref{core_identify} the dendrogram technique decomposes the emission into a hierarchy of sub-structures where leaves are the smallest, bright structures at the top of the tree hierarchy and branches are the faint, extended structures lower down in the tree. 
The local maxima in the data set determine the dendrogram structures defined by distinct isosurfaces around and containing these local maxima. Presence of noisy pixels introduces spurious local maxima. To minimize this, we first generate a noise map where each pixel represents the standard deviation of intensity of that pixel in the line-free channels of the data cube spanning 30 km s$^{-1}$. Using this noise map, a modified data cube of H$^{13}$CO$^+$ is made, where pixels having peak intensity (along the spectral axis) less than five times the noise are masked out.

To generate the dendrogram tree, the following parameters are used: (i) $min\_value = 5 \sigma$, (ii) $min\_delta = 2 \sigma$, and (iii) $min\_npix = N$ pixels. 
$N$ is taken to be five times the synthesized beam area to increase the possibility that the detected structures are resolved both spatially (at least 3 beams) and in velocity (at least 2 channels).
Further details regarding the selection of dendrogram parameters can be found in \citet{2021MNRAS.500.3027D}.
The values of $\sigma$ and $N$ for G12.42 and G19.88 are 0.007, 0.008 $\rm Jy\,beam^{-1}$, and 175, 177, respectively. 
Figure \ref{leaf2d}(a) and \ref{leaf2d}(b) display the two dimensional spatial distribution of the detected leaves overlaid on the moment zero map of the modified data cube of H$^{13}$CO$^+$ for G12.42 and G19.88, respectively. Based on a careful visual scrutiny, we discarded structures 15, 36, and 41 associated with G12.42 for further analysis. These structures are located at the edge and appear truncated.
Following the above procedure, we obtain 23 leaves and 15 branches for G12.42 and for G19.88, 11 leaves and 5 branches are identified. 

The basic parameters of the identified density structures are tabulated in Tables \ref{leaf_params_dendro} and \ref{branch_params_dendro}. These include the coordinates, corrected major and minor axes, the aspect ratio, size, the peak intensity, the velocity range in which the structures span, the mean velocity, <$V_{\rm lsr}$>, of the structures, the velocity variation, $\delta V$ , and the mean velocity dispersion, <$\sigma$> of each identified structures. As discussed and implemented in \citet{HLLiu22b}, a correction factor is multiplied to the major and minor axes. This factor equals the square root of the ratio of the actual area of the density structure in the plane of the sky to the area of the dendrogram extracted ellipse.  
The intensity-weighted first moment map is used to estimate the mean velocity (<$V_{\rm lsr}$>) over the associated velocity range of the structure. The velocity variation $\delta V$ is the standard deviation of $V_{\rm lsr}$ in the structures and the mean velocity dispersion (<$\sigma$>) is determined from the intensity-weighted second moment map. Using Eq. \ref{sigma_nt}, the non-thermal velocity dispersion ($\rm \sigma_{\rm nt}$) is calculated, where $\sigma_{\rm line}$ is replaced by the mean velocity dispersion (<$\sigma$>) values given in Table \ref{leaf_params_dendro} and \ref{branch_params_dendro}. 
The clump averaged $T_{\rm kin}$ from \citep{2012A&A...544A.146W} (refer Section \ref{frag_Section}) is considered in the calculation of $\sigma_{\rm nt}$. 

\begin{figure*}
     \centering
     \begin{subfigure}[b]{0.48\textwidth}
         \centering
         \includegraphics[width=\textwidth]{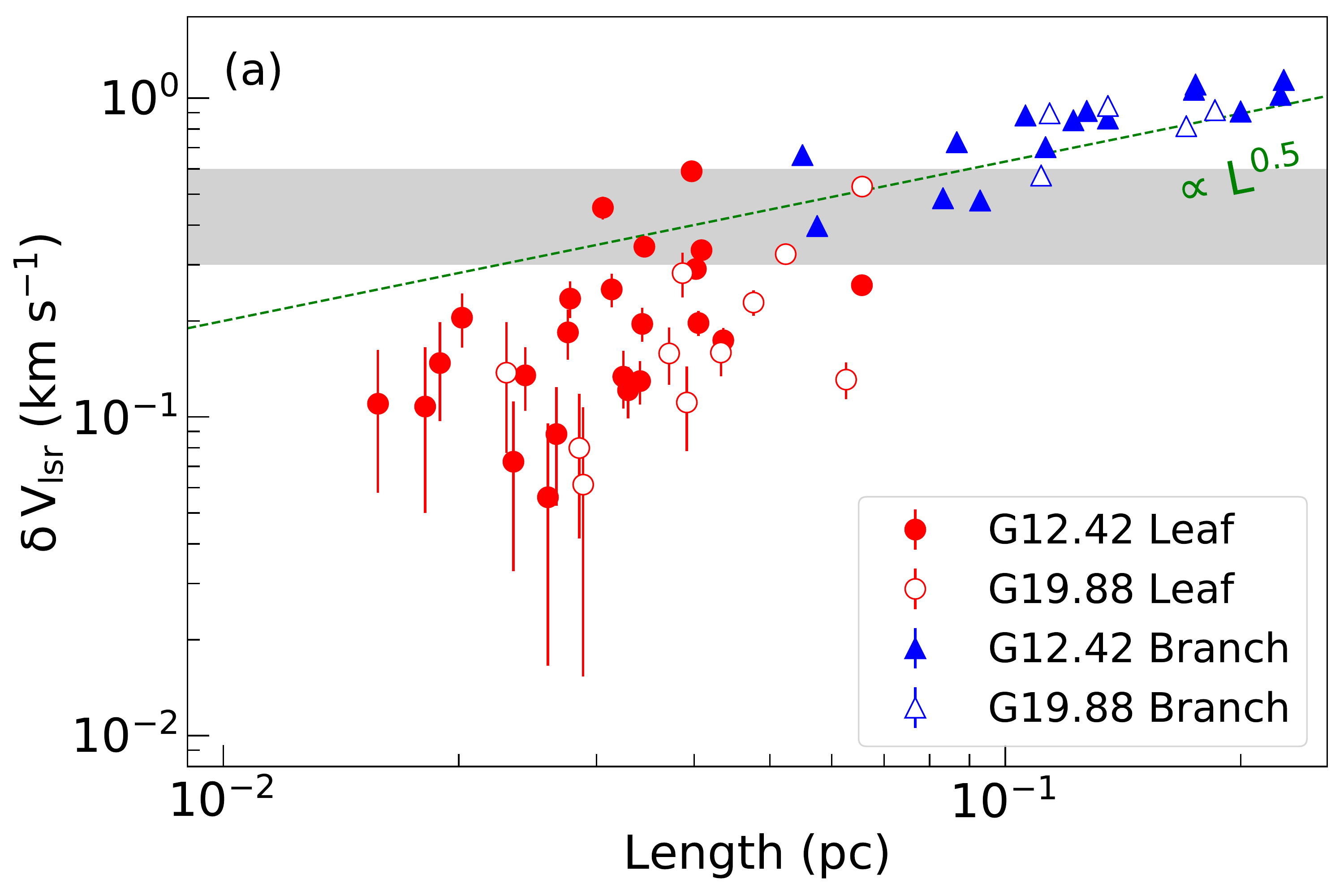}
     \end{subfigure}
     \hfill
     \begin{subfigure}[b]{0.48\textwidth}
         \centering
         \includegraphics[width=\textwidth]{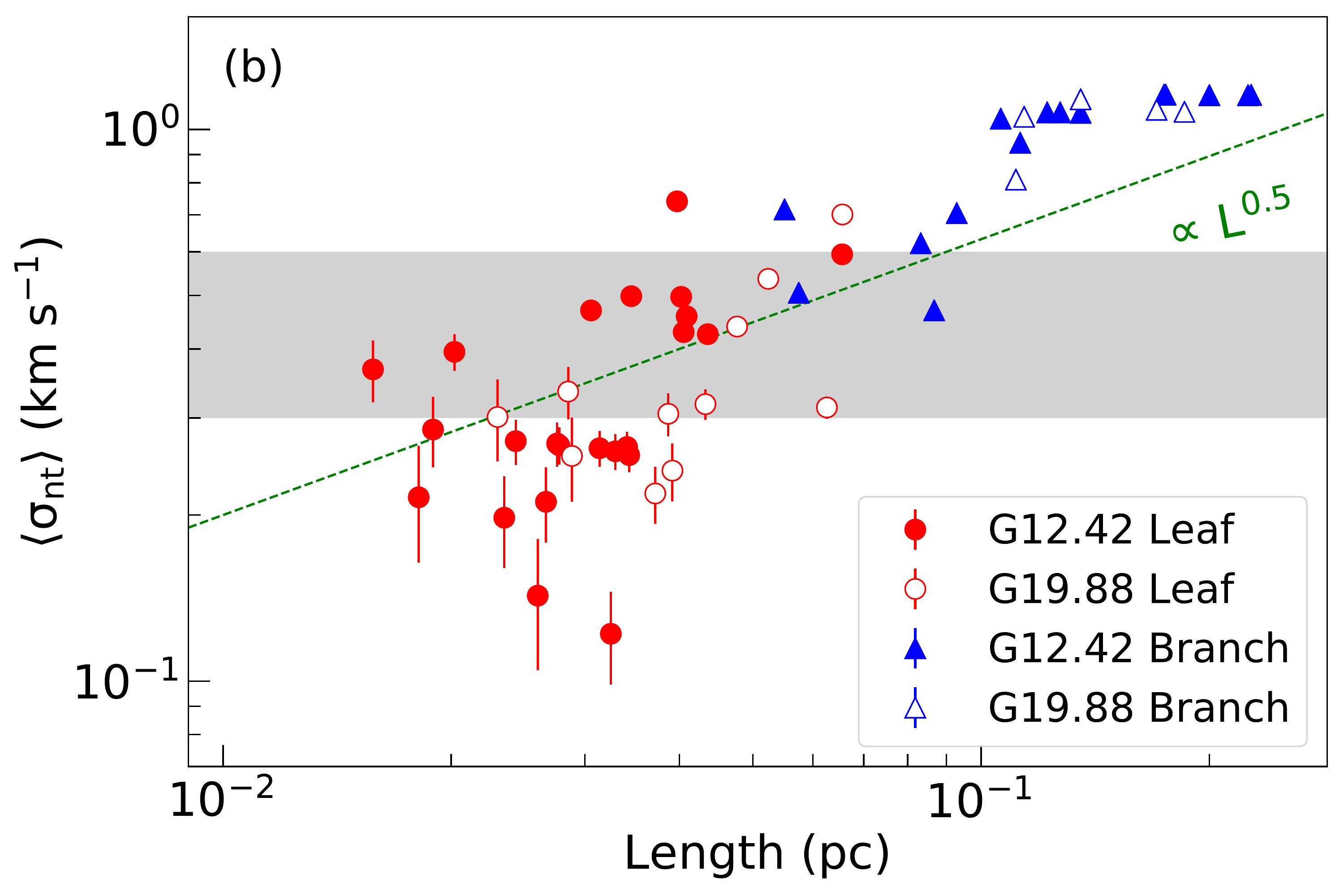}
     \end{subfigure}
     \hfill
        \caption{Velocity size relation for density structures in G12.42 and G19.88. (a) Variation of velocity plotted as a function of size. (b) Non-thermal velocity dispersion as a function of size. In both the panels, circles indicate the leaf structures and triangles show dendrogram branches. Filled and empty markers represent values for G12.42 and G19.88, respectively. In both the panels, the green dashed line shows the Larson relation with $C$ = 2.0 and $\Gamma = 0.5$.}
        \label{larson_plot}
\end{figure*}

Consistent with the Dendrogram tree hierarchy, on average the leaves display larger aspect ratios compared to the branches and have sizes that are smaller by a factor of $\sim4$. The average velocities of the structures reflect the dominant velocity components associated with the protoclusters. It should, however, be noted here that the double velocity component observed in G12.42 will be ingrained in the kinematics of the structures. Compared to the structures identified in G034.43+00.2 \citep{{HLLiu22b}} where the leaves and branches were typically 0.09 and 0.6~pc, respectively, we are probing smaller scale density structures in G12.42 and G19.88 where the median values of leaves and branches are 0.03 and 0.1~pc, respectively. 
To decipher the gas dynamics of these two protoclusters, we investigate the derived values of $\rm \delta V$ and $\rm \sigma_{nt}$ of small and larger-scale density structures detected. The parameter $\delta V$ represents the gas kinematics within the density structure in the plane of the sky and $\sigma_{\rm nt}$ gives the mean gas motion along the line of sight  \citep{{2014ApJ...794..165S},{2014ApJ...797...76L},{HLLiu22b}}. In Figure \ref{larson_plot}, we plot both these parameters as a function of the size-scale, $L$. The branches have larger $\delta V$ and $\sigma_{nt}$ than the leaves, which is consistent with the fact that these cover larger spatial scales and hence likely to reflect large-scale gas motion. Similar velocity trends are seen in the analysis of the density structures in G034.43+00.24 \citep{{HLLiu22b}}. 
The velocity regimes are indicated in the figure. The shaded horizontal band delimitates the sonic ($\leq c_{\rm eff}$) and the transonic ($\leq 2c_{\rm eff}$) levels at 0.3~$\rm km\,s^{-1}$ and 0.6~$\rm km\,s^{-1}$, respectively. The clump-averaged $T_{\rm kin}$ is used in Eq. \ref{sigma_th_equ} to calculate the above levels. The velocity variation, $\delta V$, of majority of the leaves is below the sonic level while most of the branches are above the transonic regime. The distribution in the mean non-thermal velocity dispersion also shows similar behaviour. This shows that gas motions in the larger-scale density structures, the branches, are mostly supersonic while in the smaller-scale structures, the leaves, subsonic or transonic gas motion is observed. 

Internal kinematics of molecular clouds are known to be dominated by supersonic gas motion that is generally attributed to turbulent gas flow, chaotic gravity-driven motions, or orderly large-scale gas inflow \citep[][and references therein]{{2016A&A...587A..97H},{HLLiu22b}}. For turbulent clouds, there exists a correlation between the velocity field and the size which takes a power-law form over a large range of spatial scales probed. This is referred to as Larson's law (velocity dispersion, $\sigma$($\rm km\,s^{-1}$) $\simeq C~ L(\rm pc)^{\Gamma}$, with coefficients $C$ = 1.1 and $\Gamma$ = 0.38) after the pioneering work by \citet{1981MNRAS.194..809L}. Following this, several observational and simulation studies \citep[e.g.][]{{1981MNRAS.194..809L},{1987ApJ...319..730S},{2004ApJ...615L..45H},{2008ApJ...679.1338R}} have shown this power-law scaling in molecular clouds to be consistent with the presence of turbulence, supersonic motions, and shocks due to colliding gas flows. These studies have revealed that the scaling exponent and coefficient show near invariance suggesting the universality of turbulence. Based on several later studies on larger and more homogeneous data sets, the widely accepted values are $C$ = 1.0 and $\Gamma$ = 0.5 \citep{1987ApJ...319..730S}. 

To probe this scaling at the scales of the structures in G12.42 and G19.88, the Larson relation with a power-law exponent of 0.5 is shown in Figure \ref{larson_plot}. The Larson coefficient \citep[1.0;][]{1987ApJ...319..730S} has been offset to the level of the observed velocity parameters. The branches show trends in both plots that are consistent with the canonical Larson's relation slope. The scatter seen in the leaves is larger but there is a definite hint of deviation (i.e. steepening) which is visually more prominent in the $L - \delta V$ plot. Similar studies have been carried out by several authors and in the discussion that follows we give a brief overview of the same. The velocity-size scaling in G034.43+00.24 \citep{HLLiu22b} also shows similar steepening for the leaf structures. Other recent studies based on high-resolution data \citep[e.g.][]{{2016A&A...587A..97H},{2016ApJ...833..204F},{2011MNRAS.411...65B}} have also shown deviation from the Larson law both in the scaling exponent and coefficient. Dendrogram analysis of the linewidth–size relation in Serpens South, using $\rm NH_3$ molecular line data, shows a flat $L - \sigma_{nt}$ distribution \citep{2016ApJ...833..204F}. For Serpens Mains, \citet{2014ApJ...797...76L} implemented similar analysis with $\rm N_2H^+$(1–0) data and present consistent results where $\sigma_{\rm nt}$ remains invariant with size though the 
$L - \delta V$ slope agrees well with the Larson relation. In their analysis of the sonic filament in the Musca cloud, \citet{2016A&A...587A..97H} report a broken power-law fit where a flattening is seen shortward of $\sim$1~pc. 

Summarizing the results of these above mentioned studies and comparing the same with our work, it is possible that local gas kinematics in regions within molecular clouds is different while still preserving the global properties described by the Larson relation. Additionally, the observed deviations reflect the spatial scales at which turbulence ceases to dominate the kinematics of the cloud. The velocity structure seen in the Musca filament \citep{2016A&A...587A..97H} is the first observational evidence of a parsec-scale structure that is fully decoupled from the supersonic turbulent regime of the ISM. Another crucial piece of information that is gathered from the $L - \delta V$ and the $L - \sigma_{\rm nt}$ plots for G12.42 and G19.88 is the transition from supersonic gas motion to transonic velocities at $\sim$0.1pc.  This is in excellent agreement with the observational results presented in \citet{1998ApJ...504..223G} and \citet{2010ApJ...712L.116P} and defines the transitional scale at which turbulent gas motion dissipates. 
From the observed $\rm NH_3$ line width variation, \citet{2008ApJ...672L..33W} have also inferred dissipation of turbulence occurring in the dense regions of the infrared dark cloud G28.34-0.06.
At this juncture, the turbulence induced non-linear density fluctuations proceed towards formation of sonic sub-structures which eventually form the star-forming cores. Hence, for the identified structures within our protoclusters, the deviation from Larson's scaling and the transition scale from supersonic motion corroborate well with each other.

\begin{table*}
\setlength{\tabcolsep}{5pt}
\caption{Physical parameters of leaves for G12.42 and G19.88}
\centering
\begin{tabular}{cccccccccccccccc}
ID & RA & DEC & major & minor & Aspect & Length & $F^{\rm P}$  & Vel. range  & <$V_{\rm lsr}$> & $\delta V$ & <$\sigma$> & <$\sigma_{\rm nt}$> \\
 & (J2000) & (J2000) & (${}^{''}$) & (${}^{''}$) & ratio & pc & $\rm Jy\,beam^{-1}$ & km s$^{-1}$ & km s$^{-1}$ & km s$^{-1}$ & km s$^{-1}$ & km s$^{-1}$\\ 
\hline
     &  & &  & & & & G12.42 &  &   &   &  &  \\
\hline
0     & 18:10:49.67 & -17:55:51.84 & 2.1        & 1.4        & 0.69          & 0.02   & 0.05 & [13.51  , 14.77]  & 14.08 $\pm$ 0.06  & 0.20  $\pm$ 0.04   & 0.40 $\pm$ 0.03  & 0.40 $\pm$ 0.03 \\
9     & 18:10:49.64 & -17:55:39.44 & 3.3        & 2.1        & 0.65          & 0.03   & 0.07 & [15.61  , 17.51]  & 16.44 $\pm$ 0.05  & 0.45  $\pm$ 0.04   & 0.48 $\pm$ 0.02   & 0.47 $\pm$ 0.02 \\
11    & 18:10:51.15 & -17:55:48.07 & 8.3        & 3.8        & 0.46          & 0.07   & 0.30 & [15.83  , 18.15]  & 16.98 $\pm$ 0.01  & 0.26  $\pm$ 0.01   & 0.60 $\pm$ 0.01  & 0.59 $\pm$ 0.01 \\
12    & 18:10:50.33 & -17:55:57.25 & 3.8        & 2.3        & 0.62          & 0.03   & 0.18 & [16.04  , 17.72]  & 16.88 $\pm$ 0.04  & 0.34  $\pm$ 0.03   & 0.50 $\pm$ 0.01  & 0.50 $\pm$ 0.01 \\
13    & 18:10:49.34 & -17:55:59.87 & 2.5        & 1.6        & 0.63          & 0.02   & 0.15 & [16.25  , 16.88]  & 16.58 $\pm$ 0.05  & 0.07  $\pm$ 0.04   & 0.21 $\pm$ 0.04  & 0.20 $\pm$ 0.04 \\
16    & 18:10:48.54 & -17:56:02.02 & 3.8        & 1.9        & 0.5          & 0.03   & 0.10 & [16.46  , 17.72]  & 17.10 $\pm$ 0.04  & 0.25  $\pm$ 0.03   & 0.27 $\pm$ 0.02  & 0.26 $\pm$ 0.02 \\
17    & 18:10:52.37 & -17:56:07.36 & 7.0        & 2.0        & 0.28           & 0.04   & 0.16 & [16.46  , 18.15]  & 17.20 $\pm$ 0.02  & 0.17  $\pm$ 0.02   & 0.43 $\pm$ 0.01  & 0.43 $\pm$ 0.01 \\
23    & 18:10:50.46 & -17:55:37.98 & 4.4        & 2.6        & 0.6          & 0.04   & 0.20 & [16.67  , 19.20]  & 17.94 $\pm$ 0.04  & 0.59  $\pm$ 0.03   & 0.74 $\pm$ 0.01  & 0.74 $\pm$ 0.01 \\
24    & 18:10:52.59 & -17:56:01.70 & 6.5        & 1.9        & 0.29         & 0.04   & 0.15 & [16.88  , 18.36]  & 17.64 $\pm$ 0.02  & 0.20  $\pm$ 0.02   & 0.43 $\pm$ 0.01   & 0.43 $\pm$ 0.01\\
25    & 18:10:49.51 & -17:55:50.32 & 5.8        & 2.1        & 0.36          & 0.04   & 0.17 & [16.88  , 18.57]  & 17.74 $\pm$ 0.03  & 0.29  $\pm$ 0.02   & 0.50 $\pm$ 0.01   & 0.50 $\pm$ 0.01\\
26    & 18:10:51.31 & -17:56:17.06 & 3.6        & 2.2        & 0.59           & 0.03   & 0.13 & [17.30  , 18.57]  & 17.99 $\pm$ 0.04  & 0.13  $\pm$ 0.03   & 0.14 $\pm$ 0.02  & 0.12 $\pm$ 0.02 \\
28    & 18:10:50.87 & -17:55:33.44 & 3.6        & 1.5        & 0.41         & 0.03   & 0.20 & [17.51  , 18.15]  & 17.86 $\pm$ 0.05  & 0.09  $\pm$ 0.04   & 0.22 $\pm$ 0.03  & 0.21 $\pm$ 0.03 \\
29    & 18:10:48.92 & -17:55:22.94 & 1.5        & 1.3        & 0.86           & 0.02   & 0.10 & [17.51  , 18.99]  & 18.23 $\pm$ 0.07  & 0.11  $\pm$ 0.05   & 0.37 $\pm$ 0.05  & 0.37 $\pm$ 0.05 \\
31    & 18:10:52.24 & -17:55:46.81 & 4.4        & 1.3        & 0.29         & 0.03   & 0.16 & [18.15  , 18.99]  & 18.54 $\pm$ 0.05  & 0.18  $\pm$ 0.03   & 0.28 $\pm$ 0.02  & 0.27 $\pm$ 0.02 \\
32    & 18:10:51.89 & -17:55:20.80 & 1.6        & 1.5        & 0.94         & 0.02   & 0.11 & [18.15  , 18.99]  & 18.61 $\pm$ 0.08  & 0.11  $\pm$ 0.06   & 0.23 $\pm$ 0.05  & 0.22 $\pm$ 0.05 \\
33    & 18:10:49.03 & -17:55:59.99 & 5.7        & 1.5        & 0.26          & 0.03   & 0.22 & [18.36  , 19.20]  & 18.85 $\pm$ 0.03  & 0.20  $\pm$ 0.02   & 0.27 $\pm$ 0.02   & 0.26 $\pm$ 0.02 \\
34    & 18:10:48.18 & -17:56:01.22 & 2.4        & 2.1        & 0.85          & 0.03   & 0.19 & [18.36  , 18.78]  & 18.58 $\pm$ 0.05  & 0.06  $\pm$ 0.04   & 0.16 $\pm$ 0.04  & 0.14 $\pm$ 0.04 \\
35    & 18:10:50.21 & -17:55:43.11 & 2.2        & 1.2        & 0.54         & 0.02   & 0.15 & [18.36  , 19.20]  & 18.76 $\pm$ 0.07  & 0.15  $\pm$ 0.05   & 0.29 $\pm$ 0.04  & 0.29 $\pm$ 0.04 \\
37    & 18:10:52.60 & -17:55:24.46 & 3.2        & 1.8        & 0.54         & 0.03   & 0.21 & [18.57  , 19.41]  & 18.98 $\pm$ 0.04  & 0.24  $\pm$ 0.03   & 0.28 $\pm$ 0.02  & 0.27 $\pm$ 0.02 \\
39    & 18:10:49.90 & -17:55:52.62 & 5.1        & 1.7        & 0.33          & 0.03   & 0.19 & [18.78  , 19.62]  & 19.18 $\pm$ 0.03  & 0.13  $\pm$ 0.02   & 0.27 $\pm$ 0.02  & 0.27 $\pm$ 0.02 \\
40    & 18:10:49.07 & -17:55:46.02 & 4.2        & 1.0        & 0.25         & 0.02   & 0.11 & [18.78  , 19.62]  & 19.18 $\pm$ 0.04  & 0.14  $\pm$ 0.03   & 0.28 $\pm$ 0.03  & 0.27 $\pm$ 0.03 \\
42    & 18:10:50.69 & -17:55:50.18 & 5.4        & 2.3        & 0.42          & 0.04   & 0.26 & [18.99  , 20.68]  & 19.84 $\pm$ 0.03  & 0.33  $\pm$ 0.02   & 0.46 $\pm$ 0.01  & 0.46 $\pm$ 0.01 \\
43    & 18:10:49.96 & -17:55:33.27 & 4.8        & 1.7        & 0.35         & 0.03   & 0.17 & [18.99  , 19.83]  & 19.37 $\pm$ 0.03  & 0.12  $\pm$ 0.02   & 0.27 $\pm$ 0.02  & 0.26 $\pm$ 0.02 \\
\hline
     &  & &  & & & & G19.88  &   &   &  &  \\
\hline
4     & 18:29:15.15 & -11:50:20.56 & 2.2        & 1.4        & 0.65         & 0.03   & 0.20 & [40.70  , 41.75]  & 41.22 $\pm$ 0.05  & 0.08  $\pm$ 0.04   & 0.34 $\pm$ 0.04  & 0.33 $\pm$ 0.04 \\
7     & 18:29:14.47 & -11:50:21.46 & 6.7        & 2.5        & 0.37         & 0.07   & 0.33 & [42.17  , 44.71]  & 43.47 $\pm$ 0.03  & 0.53  $\pm$ 0.02   & 0.70 $\pm$ 0.01   & 0.70 $\pm$ 0.01\\
8     & 18:29:15.45 & -11:50:25.58 & 3.4        & 2.6        & 0.74           & 0.05   & 0.16 & [42.60  , 44.07]  & 43.32 $\pm$ 0.03  & 0.23  $\pm$ 0.02   & 0.45 $\pm$ 0.01   & 0.44 $\pm$ 0.01\\
9     & 18:29:15.95 & -11:50:35.79 & 4.8        & 3.2        & 0.67          & 0.06   & 0.14 & [43.23  , 44.28]  & 43.78 $\pm$ 0.02  & 0.13  $\pm$ 0.02   & 0.32 $\pm$ 0.01   & 0.31 $\pm$ 0.01\\
10    & 18:29:12.54 & -11:50:09.69 & 2.9        & 2.0        & 0.69         & 0.04   & 0.12 & [43.23  , 44.49]  & 43.68 $\pm$ 0.06  & 0.28  $\pm$ 0.05   & 0.31 $\pm$ 0.03   & 0.31 $\pm$ 0.03\\
11    & 18:29:13.07 & -11:50:01.76 & 2.0        & 1.0        & 0.52           & 0.02   & 0.08 & [43.23  , 44.28]  & 43.78 $\pm$ 0.08  & 0.14  $\pm$ 0.06   & 0.31 $\pm$ 0.05   & 0.30 $\pm$ 0.05\\
12    & 18:29:13.73 & -11:50:16.25 & 3.6        & 2.0        & 0.55         & 0.04   & 0.26 & [43.44  , 44.49]  & 44.00 $\pm$ 0.04  & 0.16  $\pm$ 0.03   & 0.33 $\pm$ 0.02   & 0.32 $\pm$ 0.02\\
13    & 18:29:13.17 & -11:49:51.05 & 3.2        & 1.9        & 0.58          & 0.04   & 0.13 & [43.65  , 44.71]  & 44.06 $\pm$ 0.05  & 0.11  $\pm$ 0.03   & 0.25 $\pm$ 0.03   & 0.24 $\pm$ 0.03\\
14    & 18:29:14.48 & -11:50:28.82 & 4.3        & 2.5        & 0.57          & 0.05   & 0.20 & [43.86  , 45.76]  & 44.76 $\pm$ 0.03  & 0.32  $\pm$ 0.02   & 0.54 $\pm$ 0.01   & 0.54 $\pm$ 0.01\\
15    & 18:29:15.31 & -11:50:31.83 & 3.2        & 1.7        & 0.52         & 0.04   & 0.11 & [44.28  , 44.92]  & 44.59 $\pm$ 0.05  & 0.16  $\pm$ 0.03   & 0.23 $\pm$ 0.03   & 0.22 $\pm$ 0.03\\
16    & 18:29:13.46 & -11:50:01.83 & 2.1        & 1.5        & 0.71         & 0.03   & 0.10 & [44.28 , 45.13]  & 44.73 $\pm$ 0.06  & 0.06  $\pm$ 0.05   & 0.27 $\pm$ 0.04   & 0.26 $\pm$ 0.04\\
\hline
\end{tabular}
\label{leaf_params_dendro}
\end{table*}
\begin{table*}
\setlength{\tabcolsep}{5pt}
\caption{Physical parameters of branches for G12.42 and G19.88}
\centering
\begin{tabular}{cccccccccccccccc}
ID & RA & DEC & major & minor & Aspect & Length & $F^{\rm P}$  & Vel. range  & <$V_{\rm lsr}$> & $\delta V$ & <$\sigma$> & <$\sigma_{\rm nt}$>\\
 & (J2000) & (J2000) & (${}^{''}$) & (${}^{''}$) & ratio  & pc & $\rm Jy\,beam^{-1}$ & km s$^{-1}$ & km s$^{-1}$ & km s$^{-1}$ & km s$^{-1}$ & km s$^{-1}$\\ 
\hline
     &  & &  & & & & G12.42   &   &   &  &  &\\
\hline
2     & 18:10:50.66  & -17:55:45.87 & 27.92      & 13.64      & 0.49           & 0.23   & 0.30 & [14.77  , 21.10]  & 18.02 $\pm$ 0.01  & 1.14  $\pm$ 0.01   & 1.16 $\pm$ 0.006   & 1.15 $\pm$ 0.006  \\
3     & 18:10:50.67  & -17:55:45.87 & 27.88      & 13.40      & 0.48           & 0.22   & 0.30 & [14.98  , 21.10]  & 18.02 $\pm$ 0.01  & 1.02  $\pm$ 0.01   & 1.16 $\pm$ 0.005   & 1.15 $\pm$ 0.005 \\
4     & 18:10:50.45  & -17:55:48.43 & 21.58      & 13.68      & 0.63           & 0.20   & 0.30 & [14.98  , 21.10]  & 17.94 $\pm$ 0.01  & 0.91  $\pm$ 0.01   & 1.16 $\pm$ 0.005   & 1.15 $\pm$ 0.005 \\
5     & 18:10:50.44  & -17:55:48.37 & 19.42      & 11.66      & 0.60           & 0.18   & 0.30 & [15.19  , 20.89]  & 17.95 $\pm$ 0.01  & 1.10  $\pm$ 0.01   & 1.16 $\pm$ 0.002   & 1.16 $\pm$ 0.002 \\
6     & 18:10:50.45  & -17:55:48.28 & 19.24      & 11.66      & 0.61           & 0.17   & 0.30 & [15.19  , 20.89]  & 17.95 $\pm$ 0.01  & 1.06  $\pm$ 0.01   & 1.16 $\pm$ 0.006   & 1.16 $\pm$ 0.006 \\
7     & 18:10:51.01  & -17:55:50.71 & 15.11      & 6.20       & 0.41           & 0.11   & 0.30 & [15.19  , 19.20]  & 17.43 $\pm$ 0.01  & 0.70  $\pm$ 0.01   & 0.95 $\pm$ 0.003   & 0.95 $\pm$ 0.003\\
8     & 18:10:50.95  & -17:55:50.47 & 13.91      & 4.58       & 0.33           & 0.09   & 0.30 & [15.40  , 18.36]  & 17.09 $\pm$ 0.01  & 0.48  $\pm$ 0.01   & 0.71 $\pm$ 0.004   & 0.70 $\pm$ 0.004\\
14    & 18:10:52.56  & -17:55:23.53 & 9.55       & 5.81       & 0.61           & 0.09   & 0.21 & [16.25  , 20.04]  & 19.07 $\pm$ 0.02  & 0.73  $\pm$ 0.02   & 0.48 $\pm$ 0.005   & 0.47 $\pm$ 0.005 \\
18    & 18:10:50.08 & -17:55:46.79 & 17.42      & 7.76       & 0.45           & 0.14   & 0.26 & [16.46  , 20.89]  & 18.28 $\pm$ 0.01  & 0.86  $\pm$ 0.01   & 1.07 $\pm$ 0.004   & 1.07 $\pm$ 0.004 \\
19    & 18:10:50.14  & -17:55:47.30 & 16.63      & 6.63       & 0.4           & 0.12   & 0.26 & [16.46  , 20.89]  & 18.28 $\pm$ 0.01  & 0.85  $\pm$ 0.01   & 1.08 $\pm$ 0.003   & 1.07 $\pm$ 0.003 \\
20    & 18:10:50.09  & -17:55:47.49 & 17.21      & 6.93       & 0.4           & 0.13   & 0.26 & [16.46  , 20.89]  & 18.28 $\pm$ 0.01  & 0.91  $\pm$ 0.01   & 1.08 $\pm$ 0.004   & 1.07 $\pm$ 0.004 \\
21    & 18:10:50.18  & -17:55:47.66 & 14.16      & 5.87       & 0.41           & 0.11   & 0.26 & [16.67  , 20.68]  & 18.35 $\pm$ 0.02  & 0.88  $\pm$ 0.01   & 1.05 $\pm$ 0.003   & 1.05 $\pm$ 0.003 \\
22    & 18:10:50.52  & -17:55:37.62 & 7.96       & 2.81       & 0.35           & 0.06   & 0.20 & [16.67  , 19.20]  & 17.98 $\pm$ 0.03  & 0.66  $\pm$ 0.02   & 0.72 $\pm$ 0.007   & 0.72 $\pm$ 0.007 \\
30    & 18:10:50.03  & -17:55:52.74 & 14.12      & 3.62       & 0.26           & 0.08   & 0.26 & [18.15  , 20.68]  & 19.18 $\pm$ 0.01  & 0.49  $\pm$ 0.01   & 0.63 $\pm$ 0.005   & 0.62 $\pm$ 0.005 \\
38    & 18:10:50.40  & -17:55:50.64 & 7.20       & 3.39       & 0.47           & 0.06   & 0.26 & [18.78  , 20.68]  & 19.54 $\pm$ 0.02  & 0.40  $\pm$ 0.02   & 0.51 $\pm$ 0.005   & 0.51 $\pm$ 0.005 \\
\hline
     &  & &  & & &  & G19.88 &   &   &  &  &\\
\hline
1     & 18:29:14.57  & -11:50:20.96 & 13.87      & 8.13       & 0.59          & 0.17   & 0.33 & [40.70  , 45.76]  & 43.51 $\pm$ 0.01  & 0.82  $\pm$ 0.01   & 1.09 $\pm$ 0.004   & 1.08 $\pm$ 0.004\\
2     & 18:29:14.61  & -11:50:21.16 & 14.89      & 8.97       & 0.6          & 0.19   & 0.33 & [40.70  , 45.76]  & 43.56 $\pm$ 0.01  & 0.92  $\pm$ 0.01   & 1.08 $\pm$ 0.004   & 1.08 $\pm$ 0.004 \\
3     & 18:29:14.42  & -11:50:20.77 & 11.04      & 6.44       & 0.58          & 0.14   & 0.33 & [40.70  , 45.76]  & 43.54 $\pm$ 0.02  & 0.95  $\pm$ 0.01   & 1.14 $\pm$ 0.004   & 1.13 $\pm$ 0.004 \\
5     & 18:29:14.37  & -11:50:19.65 & 11.94      & 4.22       & 0.35          & 0.11   & 0.33 & [40.70  , 45.13]  & 43.28 $\pm$ 0.02  & 0.90  $\pm$ 0.02   & 1.06 $\pm$ 0.005   & 1.05 $\pm$ 0.005 \\
6     & 18:29:14.34  & -11:50:19.64 & 11.44      & 4.19       & 0.37          & 0.11   & 0.33 & [41.96  , 45.13]  & 43.55 $\pm$ 0.02  & 0.57  $\pm$ 0.01   & 0.81 $\pm$ 0.006   & 0.81 $\pm$ 0.006 \\
\hline
\end{tabular}
\label{branch_params_dendro}
\end{table*}
\subsection{Core kinematics}
\label{core-kinematics}
\subsubsection*{Infall activity}
Following the analysis presented in Section \ref{grav_stab_global_coll}, the virial mass and virial parameter are calculated for all the detected cores and tabulated in Table \ref{vir_par_core}. $\alpha_{\rm vir}$ values are estimated to be less than 2 for all cores. As conjectured for the clumps, the detected cores can also be considered to be supercritical and under collapse. And if that is not the case, then there should exist significant support (e.g. magnetic field) to resist collapse (see Section \ref{grav_stab_global_coll}).

In addition to the above, from the spectra presented in Figures  \ref{1242_allcores_hco_h13co} and \ref{1988_allcores_hco_h13co}, several cores are seen to display the typical infall signature. For G12.42, the cores, MM1, MM2, and MM3 clearly show self-absorbed profiles of $\rm HCO^+$. In case of G19.88, the spectrum extracted over the area covering the cores MM1, MM3, MM4, and MM7 shows this signature. It is difficult in this case to attribute this to any particular core. Using Eq. \ref{delta_V}, we also estimate the asymmetry parameter ($A$) for these core profiles which lie in the range $-0.6$ to $-0.7$. 
These estimated values in combination with the displayed profiles of the $\rm HCO^+$ and the H$^{13}$CO$^{+}$ lines strongly indicate infall activity in these cores. Furthermore, assuming a spherically symmetric core of radius, $R_{\rm core}$, and volume density, $\rho$, we estimate the mass infall rate, $M_{\rm inf}$ = $4\pi R^{2}_{\rm core}\rho V_{\rm inf}$ \citep{2010A&A...517A..66L}, where $V_{\rm inf} = |V_{\rm thick} - V_{\rm thin}| = |V_{\rm H^{13}CO^{+}} - V_{\rm HCO^{+}}|$ is the infall velocity and $\rho = M_{\rm core} \big / \frac{4}{3}\pi R^{3}_{\rm core}$. For the individual cores probed in G12.42, $M\rm_{inf}$ is estimated to be $\sim$6.5$\times$10$^{-3}$ $\rm M_{\odot} yr^{-1}$. For G19.88, the spectra showing infall yield similar values ranging from $\sim$1.4$\times$10$^{-3}$ to  $\sim$6.3$\times$10$^{-3}$ $\rm M_{\odot} yr^{-1}$. 
These are in good agreement with infall rates derived for other high-mass star-forming regions. To cite a few results from literature, \citet{2013MNRAS.436.1335L} estimate the mass infall rates to be $\sim$ 1.5$\times$10$^{-2}$ $\rm M_{\odot} yr^{-1}$, while \citet{2010A&A...517A..66L} and \citet{2010ApJ...710..150C} report  larger ranges of $\sim$1.8$\times$10$^{-3}$ -- 1.3$\times$10$^{-1}$ $\rm M_{\odot} yr^{-1}$ and $\sim$4$\times$10$^{-2}$ -- 1$\times$10$^{-4}$ $\rm M_{\odot} yr^{-1}$, respectively.

\begin{table}
\caption{Virial parameters of cores associated with G12.42+0.50 and G19.88-0.53}
\centering
\begin{tabular}{c c c c}
 \hline
Cores & Line width ($\Delta V$) &  $M_{\rm vir}$ &$\alpha_{\rm vir}$\\[0.5ex] 
& $\rm km\,s^{-1}$ & ($\rm M_{\odot}$) & \\[0.5ex] 
\hline
& G12.42 &\\
\hline
MM1 & 1.2 & 2.0 & 0.1  \\
\hline
MM2 & 1.2 & 2.1 &  0.1  \\
\hline
MM3 & 1.4 & 6.3 &  0.1  \\
\hline
MM4 & 1.7 & 6.3 &  0.4  \\
\hline
MM5 & 1.6 & 3.5 &  0.1  \\
\hline
MM6 & 1.0 & 2.2 &  0.2  \\
\hline
MM7 & 0.9 & 1.9 &  0.2  \\
\hline
& G19.88 &\\
\hline
MM1 & 2.1 & 2.9 &  0.3 \\
\hline
MM2 & 1.6 & 4.1 &   0.2 \\
\hline
MM3 & 2.1 & 3.1 &   0.4 \\
\hline
MM4 & 2.1 & 3.1 &   0.4 \\
\hline
MM5 & 2.0 & 2.9 &   0.3 \\
\hline
MM6 & 2.0 & 4.2 &   0.4 \\
\hline
MM7 & 2.1 & 3.7 &   1.1 \\
\hline
\end{tabular}
\label{vir_par_core}
\end{table}

\subsubsection*{Outflow activity}
\label{outflow_activity}
Jets and outflows are signposts of star formation and prevalent in all mass regimes. The high-velocity and collimated jets, which are indirect tracers of accretion disks, not only carry away excess angular momentum from the rotating disks but also entrain gas and dust from the ambient medium forming large-scale molecular outflows \citep{{2016MNRAS.460.1039P},{2002A&A...383..892B}}. These outflows manifest as broad wings in the line profiles of molecular transitions like HCO$^{+}$, CS, SO, and SiO \citep[e.g.,][]{{2007ApJ...654..361Q},{2013A&A...557A..94S},{2015A&A...581A..85P},{2017ApJ...849...25L},{2019ApJ...878...29L},{2020ApJ...903..119L},{HLLiu21}}. 
Both protoclusters investigated in this study are classified as EGOs and hence are expected to be associated with outflows/jets \citep[e.g.][]{{2008AJ....136.2391C},{2010AJ....140..196D},{ 2012ApJ...748....8T},{ 2013ApJS..206...22C},{2012ApJS..200....2L}}. As discussed earlier, low frequency radio observations with GMRT by Issac19 and Issac20 have shown the presence of ionized jets. These authors have also presented outflows using CO data from JCMT and high-resolution archival ALMA $\rm C^{18}O$ data. Here, we discuss the outflow features observed with the ATOMS survey.  
 
\begin{figure}
    \centering
    \includegraphics[width=0.48\textwidth]{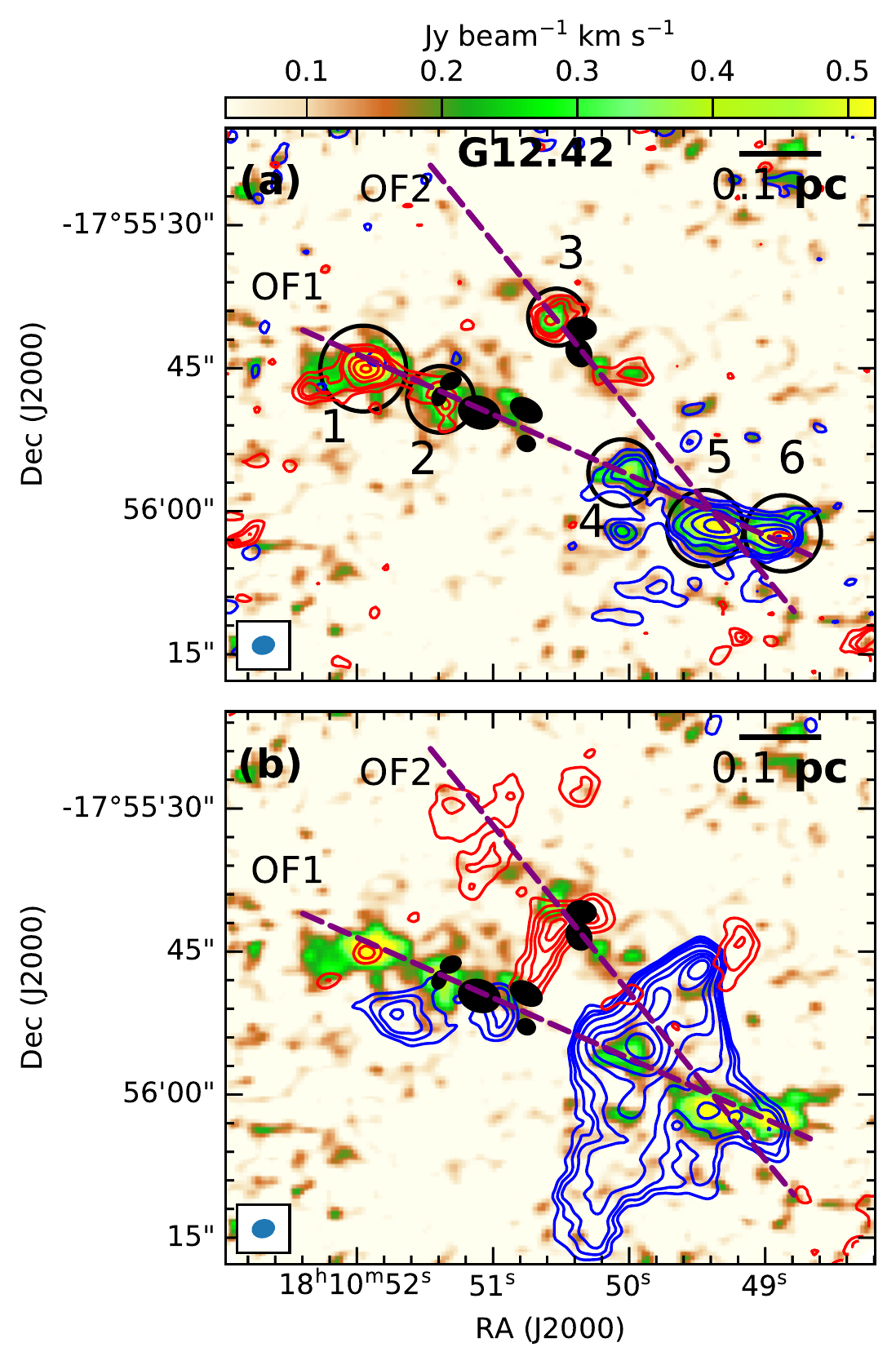}
    \caption{Outflow emission traced by different tracers overlaid on the moment zero map of SiO (2 - 1) for the region associated with G12.42+0.50. Moment zero map for SiO transition obtained by integrating over the full velocity range is shown as colour scale. (a) SiO contours integrated in velocity ranges [0-16.0]$\,\rm km\,s^{-1}$ and [19.8-26.9]$\,\rm km\,s^{-1}$ for blue and red lobes, respectively, are overlaid on the moment zero map of SiO (2 - 1). The contour levels are 3, 5, 7, 11, 16, 22 times $\sigma$ ($\sigma$ = 0.03 $\rm Jy\,beam^{-1}\,km\,s^{-1}$), and 3, 4, 5, 9, 13, 18 times $\sigma$ ($\sigma$ = 0.02 $\rm Jy\,beam^{-1}\,km\,s^{-1}$) for blue and red, respectively. SiO emission only concentrates on several positions labelled as 1 - 6. Black circles represent the area over which the spectra is extracted. (b) Outflow emission traced by HCO$^{+}$ are shown as contours. The contours are integrated in velocity ranges [12-15.0]$\,\rm km\,s^{-1}$ and [21.0-29]$\,\rm km\,s^{-1}$ for blue and red lobes, respectively. The contour levels are 3, 4, 5, 7, 11, 14, 17, 20 times $\sigma$ ($\sigma$ = 0.05 $\rm Jy\,beam^{-1}\,km\,s^{-1}$), and 3, 4, 5, 6, 7, 8 times $\sigma$ ($\sigma$ = 0.04 $\rm Jy\,beam^{-1}\,km\,s^{-1}$) for blue and red, respectively. In both the panels, the dashed lines indicate the possible outflow directions. The identified cores are shown as filled ellipses. The beam sizes are shown at the lower left in each figure.}
    \label{1242_outflow}
\end{figure}
%
\begin{figure}
    \centering
    \includegraphics[width=0.46\textwidth]{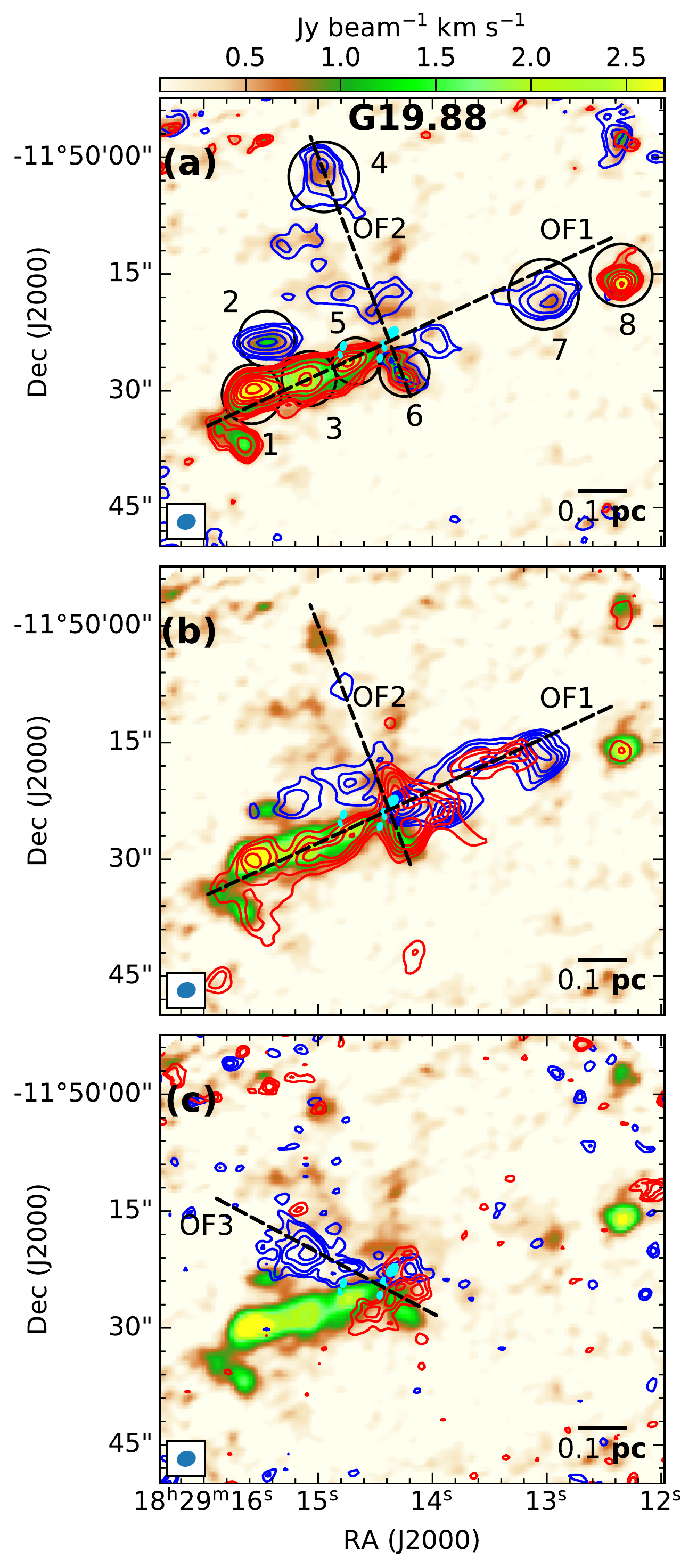}
    \caption{Same as Figure \ref{1242_outflow}, but for G19.88-0.53. Moment 0 map for SiO transition obtained by integrating over the full velocity range is shown as colour scale in all panels. (a) SiO contours integrated in velocity ranges [30.1 - 41.0]$\,\rm km\,s^{-1}$ and [46.5 - 105.0]$\,\rm km\,s^{-1}$ for blue and red lobes, respectively are overlaid. The contour levels are 3, 4, 5, 9, 13, 17, 21 times $\sigma$ ($\sigma$ = 0.033 $ \rm Jy\,beam^{-1}\,\rm km\,s^{-1}$) and 3, 4, 5, 8, 11, 14, 17, 20, 23 times $\sigma$ ($\sigma$ = 0.15 $ \rm Jy\,beam^{-1}\,\rm km\,s^{-1}$) for blue and red, respectively. (b) HCO$^+$ contours integrated in velocity ranges [25.0 - 40.1]$\,\rm km\,s^{-1}$ and [47.4 - 60.0]$\,\rm km\,s^{-1}$ for blue and red lobes, respectively are shown. The contour levels are 3, 4, 5, 6, 7, 8 times $\sigma$, where $\sigma$ = 0.15 and 0.13 $ \rm Jy\,beam^{-1}\,\rm km\,s^{-1}$ for blue and red, respectively. (c) Outflow contours of H$^{13}$CO$^{+}$ (1 - 0) integrated in velocity ranges [29.9 - 41.8]$\,\rm km\,s^{-1}$ and [45.8 - 55.0]$\,\rm km\,s^{-1}$ for blue and red lobes, respectively are presented. The contour levels are 3, 4, 5, 7, 9 times $\sigma$ ($\sigma$ = 0.022 $ \rm Jy\,beam^{-1}\,\rm km\,s^{-1}$) and 3, 4, 5 times $\sigma$ ($\sigma$ = 0.02 $ \rm Jy\,beam^{-1}\,\rm km\,s^{-1}$) for blue and red, respectively. In all the panels, the black dashed lines indicate the possible outflow directions. The identified cores are shown as filled ellipses (cyan). The beam sizes are shown at the lower left in each figure.}
    \label{1988_outflow}
\end{figure}
\begin{figure}
    \centering
    \includegraphics[width=0.45\textwidth]{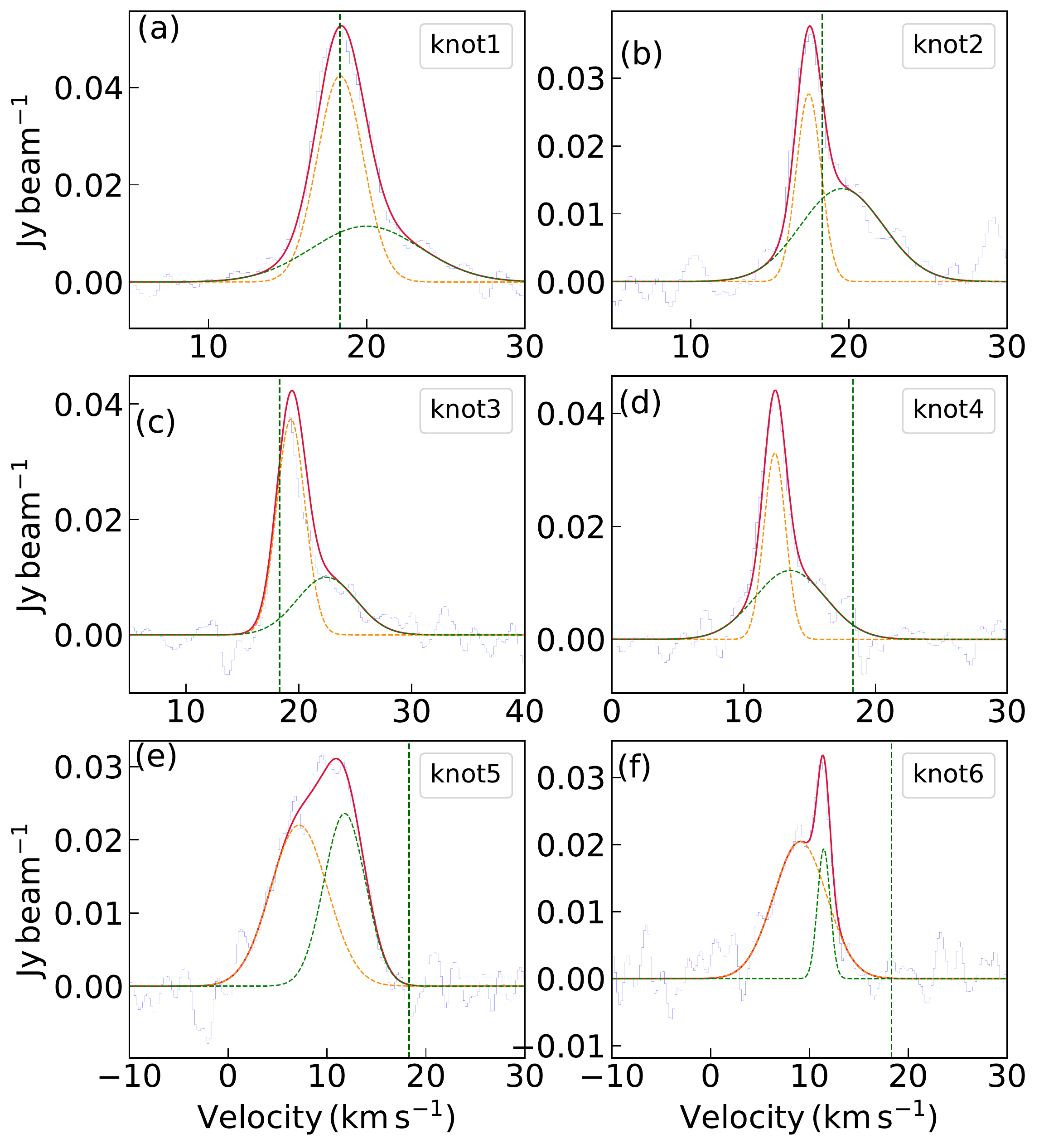}
    \caption{Averaged spectra of six positions marked by black circles in Figure \ref{1242_outflow}(a) presenting the outflow emission traced by different tracers towards G12.42. The SiO (2-1) spectra obtained using  combined 12-m + 7-m data are drawn in blue  lines. The spectra are boxcar smoothed by 4 channels resulting in a velocity resolution of 0.84$\,\rm km\,s^{-1}$. Orange and green dashed lines indicate the decomposed components of SiO (2-1) line and red spectral line is the resultant double Gaussian fitting. Vertical green dashed line marks the systemic velocity (18.3$\,\rm km\,s^{-1}$).}
    \label{1242_sioknot}
\end{figure}
\begin{figure}
    \centering
    \includegraphics[width=0.45\textwidth]{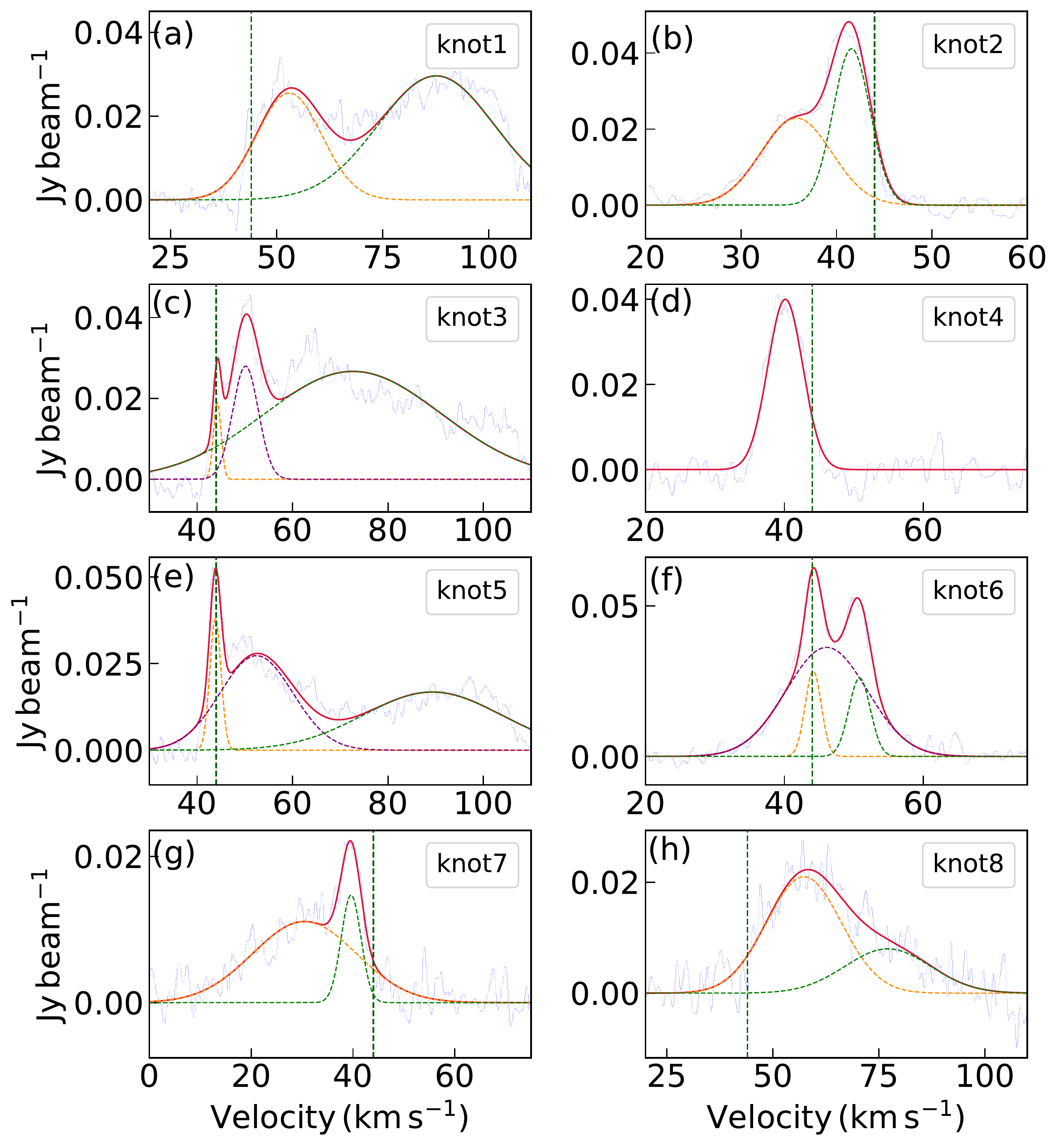}
   \caption{Averaged spectra of eight positions marked by black circles in Figure \ref{1988_outflow}(a) showing the outflow emission traced by different tracers towards G19.88. The SiO (2-1) spectra obtained using  combined 12-m + 7-m data are drawn in blue lines. The spectra are boxcar smoothed by 4 channels resulting in a velocity resolution of 0.84$\,\rm km\,s^{-1}$. Orange, purple and green dashed lines indicate the decomposed components of SiO (2-1) line and red spectral line is the 
   resultant multi Gaussian fitting. Vertical green dashed line marks the systemic velocity (44.0$\,\rm km\,s^{-1}$).}
   \label{1988_sioknot}
\end{figure}

Line profiles of the molecular line spectra presented and discussed in Section \ref{mol_line_result} indicate the presence of outflows driven by one or more of the detected star-forming cores in the two protoclusters. To visualize the orientation and the spatial morphology of these outflows, we present moment 0 maps of SiO with the contours of $\rm HCO^+$ and SiO integrated over the high-velocity wings. Figures \ref{1242_outflow} and \ref{1988_outflow} illustrate the above for G12.42 and G19.88, respectively. For G19.88, we also present the $\rm H^{13}CO^+$ contours. The velocity ranges over which the blue and red-wings of the gas emission is integrated are mentioned in the figure captions.

SiO is considered to be an excellent tracer of shocks \citep{1997A&A...321..293S} and collimated jets in star-forming regions. It is formed when powerful shocks passing through dense molecular gas disrupt dust grains via sputtering or direct photodesorption and vaporization or shattering from the grain-grain collision in shocked environments. This results in enhancement of the gas-phase abundance of SiO by many orders of magnitudes in star-forming regions with molecular outflows as compared to those in quiescent regions \citep[][]{{1987ASSL..134..561I},{1989ApJ...343..201Z},{1992A&A...254..315M},{1999A&A...343..585C}}. 
In the presence of high-velocity gas ($>$ 25$\,\rm km\,s^{-1}$) from shocks, SiO abundance is found to increase up to an abundance of 10$^{-7}$ \citep[e.g.,][]{{1998ApJ...509..768G},{2007A&A...462..163N}}, whereas in dense clouds, abundance is low ( $<$ 10$^{-12}$) \citep[e.g.,][]{1989ApJ...343..201Z}. 
The SiO maps reveal the presence of prominent, bright emission with enhanced knots in the region associated with the protoclusters. 

Pronounced SiO emission is seen along the north-east south-west direction on opposite side of the cluster of cores (MM1, MM2, MM3, MM4, and MM5) in G12.42 possibly tracing an outflow (labelled as OF1 indicated with dashed line in Figure \ref{1242_outflow}). 

Similarly, the presence and orientation of SiO knots indicate a second outflow direction along OF2, likely associated with cores, MM6 and MM7. 
 
In case of G19.88, bright, elongated SiO emission is seen south-east of the core cluster (MM1, MM2, MM3, and MM4), which could be part of an outflow (labelled OF1). A single, bright SiO knot is also seen towards the north-west. 

To confirm the presence of outflows, we inspect the spatial correlation of the observed enhanced SiO emission (as seen in the colour scale in Fig. \ref{1242_outflow} and \ref{1988_outflow}) with that of the high-velocity wings seen in the $\rm HCO^+$ and SiO spectra. These are found to be in good agreement.
 
The contours tracing the high-velocity SiO wings clearly show the blue lobes of OF1 and OF2 in G12.42 (see Figure \ref{1242_outflow}(a)). The pronounced blue-outflow wing of $\rm HCO^+$ appears wide and could possibly be overlapping blue lobes of outflows, OF1 and OF2 (Figure \ref{1242_outflow}(b)). The red lobe along OF2 is wide and clearly seen compared to that of OF1. From the orientation of the outflows and location of the detected lobes, it is likely that OF1 is driven by MM3 and OF2 by MM6 or MM7. Both outflows detected here lie towards the innermost region of the large-scale, wide-angle CO outflow presented by Issac19. 
In the G19.88 complex, OF1 shows the prominent outflow lobes. This outflow is consistent with the ALMA high-resolution CO outflow discussed in Issac20 which probed the inner region. It also matches with the SiO outflow studied by \citet{2007ApJ...654..361Q}. Almost perpendicular to OF1, we detect another possible outflow, OF2, which is also reported by \citet{2007ApJ...654..361Q}. In addition to this, in G19.88, the $\rm H^{13}CO^+$ spectrum also showed outflow wings, the spatial orientation of this shown in Figure \ref{1988_outflow}(c). Labelled as OF3, the direction of this outflow matches very well with the CO outflow mapped by Issac20 using low-resolution ALMA data (refer to Fig. 15(a) of Issac20). There is a cluster of cores 
(MM1, MM2, MM3, and MM4) at the close vicinity of the centroid of the outflows and are likely to be triggered by one or more of them. Based on GMRT radio data, Issac20 identify the presence of an ionized jet coincident with MM2, making it the most promising candidate for driving the SiO outflows given that SiO emission originates from the high-velocity C-type or CJ-type shocks that are associated with protostellar jets \citep{2016A&A...589A..29W}.

Several studies of star-forming regions have shown that the SiO spectra can be better fitted with two Gaussian components, a broad and a narrow one. The broader ones are commonly attributed to high-velocity shocks from protostellar outflows \citep[][]{{2017ApJ...849...25L},{2021MNRAS.tmp.2557Z}}. 
The narrow components are argued to be related to low-velocity shocks possibly arising due to low-velocity outflows \citep{2016A&A...595A.122L}, colliding flows \citep{2020MNRAS.496.2790L}, or could be thermal SiO emission  \citep{{2021MNRAS.tmp.2557Z}}. Additionally, it may be the imprints of a cloud–cloud collision that has triggered star formation in the cloud \citep[][]{{2010MNRAS.406..187J},{2016A&A...586A.149C},{2020MNRAS.499.1666C}}. To understand the nature of the detected SiO knots along the proposed outflow axes, we present in Figure \ref{1242_sioknot} and \ref{1988_sioknot}, spectra extracted from identified knots in G12.42 and G19.88. All the knots in G12.42 display the aforementioned typical profile with the narrow and the broad components. Interestingly, though knots \#1 and \#2 correspond to red outflow lobes (Figure \ref{1242_outflow}), the narrow SiO (2-1) components peak at the systemic velocity and are blueshifted by $\sim$ 0.9$\rm \, km\,s^{-1}$, respectively. Hence it is possible that the narrow components of SiO (2-1) spectra at these knots are not produced by outflow. Knots with similar profiles are also observed in the G286.21+0.17 cloud \citep{2021MNRAS.tmp.2557Z}. Such profiles suggest that the narrow components could either be originating from filamentary collisions or decelerated shocked material in interaction regions of outflows and dense regions in the clump. For G12.42, the filamentary collision picture finds support in the presence of a network of converging filaments detected in the protocluster. 
In the case of G19.88, all except knot \#4 display double component profiles. Knot \#4, that is located along OF2, shows only a single narrow component. The absence of high-velocity wings and the broad component in the spectrum contradicts the association of this SiO knot with an outflow. In this case, one needs to further investigate the veracity of the outflow OF2 since the red SiO lobe of it could be a part of the OF1 outflow. Another noticeable feature in G19.88 is that the red-wing in the spectra of knots \#1, \#3, \#5 and \#8 extends up to very high velocities ($\rm \sim 60~\rm\,km\,s^{-1}$ with respect to the systemic velocity). Similar feature is also seen in Figure \ref{1242+1988_sio_3sigma}(b).

\section{Gravity or turbulence: Deciphering the driving mechanism for massive star formation in the protoclusters}
\label{driving_mech}
Understanding the formation mechanism of massive stars has been one of the major focus areas of stellar astrophysics in the last decades or so. Together with the manyfold improvement in the observational database,  theoretical models including the {\it core-accretion} \citep{{2003ApJ...585..850M},{2005Natur.438..332K},{2006ApJ...638..369K}} and the {\it competitive-accretion} \citep{{2001MNRAS.323..785B},{2004MNRAS.349..735B}} hypotheses have played a key role in unravelling the complex formation processes involved. The core-accretion theory presents a scaled-up version of the well-established theory of low-mass star formation. Here, massive stars form from pre-stellar cores, that are supported by turbulence or the magnetic field, via enhanced disk accretion. While successfully addressing the `radiation pressure' problem, it leaves much to debate about the collapse time-scales and fragmentation of the massive pre-stellar cores. The competitive-accretion model, on the other hand, proposes fragmentation of clumps to low-mass cores and that the central cores chaotically gather mass by competing for it from intraclump gas reservoir. In addition to these pioneering theories, there are the current set of hierarchical fragmentation-based models for massive star formation. The two widely studied ones are the {\it inertial-inflow} \citep[IF;][]{2020ApJ...900...82P} and the GHC \citep{2019MNRAS.490.3061V} models. The IF model advocates for large-scale converging inertial inflow, driven by supersonic turbulence, to channel mass onto the accreting massive star. As proposed by these authors, the mass reservoirs are not controlled by gravity and global collapse is not necessary given the occurence of turbulence driven compression at all scales. The GHC model, proposes a very different uniform framework in which chaotic gravity-driven motions dominate at all scales. This is similar to the hierarchical Hoyle-like gravitational fragmentation taking place in the cloud where non-linear density fluctuations are formed by the initial turbulence. The conflicts and similarities of these hierarchical fragmentation models with the core-accretion and competitive-accretion models have been extensively discussed in \citet{2020ApJ...900...82P} and \citet{2019MNRAS.490.3061V}.  

The detailed analysis of the ATOMS data gives a better insight regarding the interplay between gravity and turbulence in driving star formation in the two protoclusters studied here. Large-scale filamentary gas inflow is seen converging towards the central hub in protocluster G12.42 and evidence of a filamentary structure is also observed in  G19.88. These filaments could be a natural manifestation of supersonic turbulence which finds support in the gas motion of the next spatial scale structure, the branches. These supersonic density structures are consistent with Larson's velocity-size scaling presenting similar power-law exponent ($\sim$0.5). While this supports the IF model, one cannot rule out the possibility of a gravity-driven mode in the filamentary flows which could be coupled to the turbulence cascade. Dissipation of turbulence at $\sim$0.1~pc is observed where the smaller-scale structures show transonic or subsonic gas motions. Beyond this, the leaf structures deviate from Larson's relation, showing a steeper slope. In conjunction with the virial analysis of the detected cores, where gravitational contraction is inferred from the low virial parameters derived, one can conjecture the onset of turbulence decoupling and dominance of gravitational acceleration towards the collapsing cores. The fragmentation analysis lends support to turbulence not being the key player driving clump to core fragmentation. 
Furthermore, \citet{2005Natur.438..332K} have shown that competitive-accretion could be a viable mechanism to form massive stars if the clumps that feed the accreting star are characterized by low virial parameters ($\alpha_{\rm vir} < 1$). Similar low values of $\alpha_{\rm vir}$ are estimated for the dust clumps associated with the protoclusters.
The predictions of the competitive accretion model are also argued to be consistent with that of the GHC model \citep{2019MNRAS.490.3061V}. Furthermore, a large population (more than 60 per cent) of the detected cores have masses $\lesssim M_{\rm Jeans}$, lending further support to the competitive accretion and the GHC scenario \citep{2019ApJ...886..102S}. 
From the analysis presented, it is difficult to favour one model over the other. Based on the GHC and IF models, the overall observed picture of massive star formation could be a result of the scale-dependent combined effect of turbulence and gravity, which dominate on large and small scales, respectively.
From the results presented here and in \citet{HLLiu22a,HLLiu22b}, the need for similar case studies is evident. Probing a large number of star-forming regions would enable us to obtain a better insight on the scale-dependent contribution of gravity and turbulence to the processes involved in high-mass star formation. Another key ingredient is the magnetic field. A complete understanding of these three crucial components is required to decipher the intricacies of massive star formation.

\section{Conclusions}
\label{conclusion}
We have carried out a detailed continuum and multi-spectral line study of the region associated with the protoclusters G12.42+0.50 and 19.88-0.53 using data from the ATOMS survey. The main results are summarized below:
\begin{itemize}
\item[(i)] The continuum maps reveal the presence of seven cores in each of the two protoclusters. Every core qualifies as a massive star forming core satisfying the mass-radius relation and the surface mass density threshold. The observed dearth of low-mass cores implies the influence of stellar feedback on the process of fragmentation. \\[0.5mm]

\item[(ii)]The global collapse scenario of the associated clumps has been addressed. Though previously identified as an infall candidate based on the nature of single dish line profiles, the ATOMS data of G12.42 show the presence of two distinct velocity components. However, several cores display characteristic infall signatures in the $\rm HCO^+$ line profiles supported by single peak $\rm H^{13}CO^+$ spectra. For G19.88, an infall signature is observed from the $\rm HCO^+$ and $\rm H^{13}CO^+$ line profiles. The virial analysis yields low values of the virial parameter ($\alpha_{\rm vir} << 2 $) indicating gravitational collapse in both protoclusters. \\[0.5mm]

\item[(iii)] From the estimates of Jeans parameters (mass and length) and the observed range of mass and separation of the cores, we infer that turbulence does not drive the fragmentation of the clumps and thermal Jeans fragmentation adequately explains the process. \\[0.5mm] 

\item[(iv)] Gas kinematics at different spatial scales has been investigated. 
JCMT $\rm {}^{13}CO$ (3 - 2) data show large scale, converging filamentary gas inflow in the G12.42 complex. For G19.88, such high quality CO molecular line data are required in the future to probe the gas motion in the associated filaments. \\[0.5mm]

\item[(v)] From the dendrogram analysis of the optically thin $\rm H^{13}CO^+$ transition, several multi-scale density structures (branches and leaves) are detected. The characteristic sizes (median values) are 0.1~pc and 0.03~pc for the branches and leaves, respectively. Combining the two protoclusters, 20 larger-scale branch structures and 34 smaller scale leaf structures are identified. Analysis of the velocity variation, $\delta V$ and the non-thermal velocity dispersion, $\sigma_{\rm nt}$ shows that the gas motion in the branch structures is supersonic and the leaves are mostly subsonic. The transition from the supersonic to the transonic regime occurs at spatial scales of $\sim$0.1~pc consistent with the scale at which dissipation of turbulence sets in. \\[0.5mm]

\item[(vi)] Consistent with the above result, in the $\delta V$ -- $L$ and $\sigma_{\rm nt}$ -- $L$ plots, the branches follow the canonical Larson's relation and the leaves display a deviation (steepening) from it. \\[0.5mm]

\item[(vii)] The smallest scale structure, the cores are likely to be in strong gravitational collapse as inferred from the low values of their virial parameters. In addition, several cores display the characteristic infall profiles with mass infall rates ($\sim$1 -- 7$\times$10$^{-3}$ $\rm M_{\odot} yr^{-1}$) being consistent with those found in other high-mass star-forming cores.
Using SiO and $\rm HCO^+$ molecular lines, multiple outflows are identified, which are likely to be driven by MM3 (49$\,\rm M_{\odot}$) in G12.42 and MM2 (23$\,\rm M_{\odot}$) in G19.88. In G19.88, an additional outflow is also identified using the $\rm H^{13}CO^+$ line.\\[0.5mm]

\item[(viii)] Based on the fragmentation analysis, the gas kinematics at various spatial scales, and the predictions of the IF and GHC models, the results obtained from this study indicate that the scale-dependent combined effect of turbulence and gravity is driving the formation of massive stars in the protoclusters.
\end{itemize}

The results of this study impel us to strongly advocate for more detailed case studies to address the exact behaviour of fragmentation and gas kinematics in protoclusters. EGOs associated with filamentary IRDCs present a sample of promising candidates as they are likely to be protoclusters harbouring stars with different masses and at different stages of evolution (Issac20 and references therein). These would provide a strong database for considering scale dependent contribution of turbulence and gravity in theories of massive star formation. 

\section*{Acknowledgements}
H.L.L. is supported by National Natural Science Foundation of China (NSFC) through the grant No.12103045. 
T.L. acknowledges the supports by National Natural Science Foundation of China (NSFC) through grants No.12073061 and No.12122307, the international partnership program of Chinese Academy of Sciences through grant No.114231KYSB20200009, Shanghai Pujiang Program 20PJ1415500 and the science research grants from the China Manned Space Project with no. CMS-CSST-2021-B06. 
C.W.L. is supported by the Basic Science Research Program through the National Research Foundation of Korea (NRF) funded by the Ministry of Education, Science and Technology (NRF-2019R1A2C1010851), and by the Korea Astronomy and Space Science Institute grant funded by the Korea government (MSIT) (Project No.2022-1- 840-05). %
A.S., G.G. and L.B. gratefully acknowledge support by the ANID BASAL projects ACE210002 and FB210003. A.S. also acknowledges funding support through Fondecyt Regular (project codes  1180350 and 1220610) and from the Chilean Centro de Excelencia en Astrofísica y Tecnologías Afines (CATA) BASAL grant AFB-170002. 
This research was carried out in part at the Jet Propulsion Laboratory, which is operated by the California Institute of Technology under a contract with the National Aeronautics and Space Administration (80NM0018D0004). 
S.L.Q. is supported by the National Natural Science Foundation of China (grant No. 12033005). 
K.W. acknowledges support by the National Key Research and Development Program of China (2017YFA0402702, 2019YFA0405100), and the National Science Foundation of China (11973013, 11721303). 
C.E. acknowledges the financial support from grant RJF/2020/000071 as a part of Ramanujan Fellowship awarded by Science and Engineering Research Board (SERB), Department of Science and Technology (DST), Govt. of India. 
This paper makes use of the following ALMA data: ADS/JAO.ALMA\#2019.1.00685.S and \#2017.1.00377.S. ALMA is a partnership of ESO (representing its member states), NSF (USA), and NINS (Japan), together with NRC (Canada), MOST and ASIAA (Taiwan), and KASI (Republic of Korea), in cooperation with the Republic of Chile. The Joint ALMA Observatory is operated by ESO, AUI/NRAO, and NAOJ. This research made use of astrodendro, a Python package to compute dendrograms of Astronomical data ({\url{http://www.dendrograms.org/}}). This research made use of Astropy, a community-developed core Python package for Astronomy (Astropy Collaboration, 2018).

\section*{Data Availability}
The data underlying this article will be shared on reasonable request
to the corresponding author.

\bibliographystyle{mnras}
\bibliography{reference} 

\vspace{5mm}
\noindent
Author affiliations:\\

\noindent 
$^{1}$Indian Institute of Space Science and Technology, Thiruvananthapuram 695 547, Kerala, India\\
$^{2}$Department of Astronomy, Yunnan University, Kunming, 650091, PR China \\
$^{3}$Shanghai Astronomical Observatory, Chinese Academy of Sciences, 80 Nandan Road, Shanghai 200030, Peoples Republic of China \\
$^{4}$Key Laboratory for Research in Galaxies and Cosmology, Shanghai Astronomical Observatory, Chinese Academy of Sciences, 80 Nandan Road, Shanghai 200030, Peoples Republic of China \\
$^{5}$Indian Institute of Astrophysics, Koramangala II Block, Bangalore 560 034, India\\
$^{6}$University of Science and Technology, Korea (UST), 217 Gajeong-ro, Yuseong-gu, Daejeon 34113, Republic of Korea\\
$^{7}$Korea Astronomy and Space Science Institute, 776 Daedeokdaero, Yuseong-gu, Daejeon 34055, Republic of Korea\\
$^8$Departamento de Astronom\'{\i}a, Universidad de Chile, Las Condes, Santiago 7550000, Chile\\
$^9$Jet Propulsion Laboratory, California Institute of Technology, 4800 Oak Grove Drive, Pasadena, CA 91109, USA\\
$^{10}$Department of Physics, P.O. box 64, FI- 00014, University of Helsinki, Finland\\
$^{11}$Departamento de Astronom\'ia, Universidad de Concepci\'on, Av. Esteban Iturra s/n, Distrito Universitario, 160-C, Chile \\
$^{12}$Max-Planck-Institute for Astronomy, K\"{o}nigstuhl 17, 69117 Heidelberg, Germany \\
$^{13}$Kavli Institute for Astronomy and Astrophysics, Peking University, 5 Yiheyuan Road, Haidian District, Beijing 100871, People's Republic of China\\
$^{14}$Department of Astronomy, Peking University, 100871, Beijing, People's Republic of China\\
$^{15}$Satyendra Nath Bose National Centre for Basic Sciences, Block-JD, Sector-III, Salt Lake, Kolkata-700 106 \\
$^{16}$School of Physics and Astronomy, Sun Yat-sen University, 2 Daxue Road, Zhuhai, Guangdong, 519082, People's Republic of China\\
$^{17}$Indian Institute of Science Education and Research Tirupati, Rami Reddy Nagar, Karakambadi Road, Mangalam (P.O.), Tirupati 517 507, India\\

\label{lastpage}
\end{document}